\documentclass[floatfix,a4paper,aps,prd,twocolumn,preprintnumbers]{revtex4} 
\usepackage{epsfig}
\usepackage{axodraw}
\usepackage{amsmath}
\usepackage{amssymb}
\usepackage{amsfonts}
\usepackage{graphicx}
\usepackage{graphics,subfigure}
\usepackage{axodraw}
\usepackage{float}
\usepackage{booktabs}
\usepackage{color}
\usepackage{geometry}
\usepackage{rotating}
\usepackage{atlasphysics}
\usepackage{multirow}

\usepackage{array}
\usepackage{dcolumn}
\usepackage{rotating}

\newbox\charbox
\newbox\slabox
\def\s#1{{      
    \setbox\charbox=\hbox{$#1$}
    \setbox\slabox=\hbox{$/$}
    \dimen\charbox=\ht\slabox
    \advance\dimen\charbox by -\dp\slabox
    \advance\dimen\charbox by -\ht\charbox
    \advance\dimen\charbox by \dp\charbox
    \divide\dimen\charbox by 2
    \raise-\dimen\charbox\hbox to \wd\charbox{\hss/\hss}
    \llap{$#1$}
}}

\newcommand{\mymed}{\vspace{0.12cm}}
\newcommand{\Etmiss}{\met}

\newcommand{\newc}{\newcommand}
\newc{\ssup}{\tilde{u}}
\newc{\ssdown}{\tilde{d}}
\newc{\ssstrange}{\tilde{s}}
\newc{\sscharm}{\tilde{c}}
\newc{\sstop}{\tilde{t}}
\newc{\ssbottom}{\tilde{b}}
\newc{\sse}{\sel}
\newc{\ssmu}{\smu}
\newc{\sstau}{\stau}
\newc{\ssnue}{\tilde{\nu}_{e}}
\newc{\ssnumu}{{\tilde{\nu}_{\mu}}}
\newc{\ssnutau}{{\tilde{\nu}_{\tau}}}
\newc{\ssbnue}{\tilde{\nu}_{e}^*}
\newc{\ssbnumu}{\tilde{\nu}_{\mu}^*}
\newc{\ssbnutau}{\tilde{\nu}_{\tau}^*}
\newc{\lsim}{\stackrel{<}{\sim}}
\newc{\gsim}{\stackrel{>}{\sim}}
\newc{\lam}{\lambda}
\newc{\kap}{\kappa}

\def\rpv{\mbox{\ensuremath{\, \slash\kern-.6emR_{p}}}}

\begin{document}
 
\title{Stau as the Lightest Supersymmetric Particle in R-Parity Violating SUSY Models:\\
Discovery Potential with Early LHC Data}

\author{K.~Desch}
\email[]{desch@physik.uni-bonn.de}
\affiliation{Physics Institute, University of Bonn, Bonn, Germany}

\author{H.~K.~Dreiner}
\email[]{dreiner@th.physik.uni-bonn.de}
\affiliation{Bethe Center for Theoretical Physics and Physics Institute, University of Bonn, Bonn, Germany}
\affiliation{SCIPP, University of California, Santa Cruz, CA 95064, USA}

\author{S.~Fleischmann}
\email[]{Sebastian.Fleischmann@cern.ch}
\affiliation{Physics Institute, University of Bonn, Bonn, Germany}

\author{S.~Grab}
\email[]{sgrab@scipp.ucsc.edu}
\affiliation{SCIPP, University of California, Santa Cruz, CA 95064, USA}

\author{P.~Wienemann}
\email[]{wienemann@physik.uni-bonn.de}
\affiliation{Physics Institute, University of Bonn, Bonn, Germany}

\begin{abstract}
We investigate the discovery potential of the LHC experiments for
$R$-parity violating supersymmetric models with a stau as the lightest
supersymmetric particle (LSP) in the framework of minimal supergravity. We classify
the final states according to their phenomenology for different
$R$-parity violating decays of the LSP. We then develop event
selection cuts for a specific benchmark scenario with promising
signatures for the first beyond the Standard Model discoveries at the
LHC. For the first time in this model, we perform a detailed 
signal over background analysis. We use fast detector
simulations to estimate the discovery significance taking the most
important Standard Model backgrounds into account. Assuming an
integrated luminosity of 1~$\ifb$ at a center-of-mass energy of
$\sqrt{s}=7\,\TeV$, we perform scans in the parameter space around
the benchmark scenario we consider.  We then study
the feasibility to estimate the mass of the stau-LSP. We 
briefly discuss difficulties, which arise in the identification
of hadronic tau decays due to small tau momenta and 
large particle multiplicities in our scenarios. 
\end{abstract}

\preprint{BONN-TH-2010-04, SCIPP 10/09}

\maketitle
 
\section{Introduction}
One of the main objectives of the experiments at the Large Hadron
Collider (LHC) is to search for new phenomena beyond the Standard
Model of particle physics (BSM) at and above the TeV energy scale. A
well motivated scenario is the $R$-parity violating minimal
supergravity model (mSUGRA) with baryon triality (B$_3$)
\cite{Hempfling:1995wj,Diaz:1997xc,Allanach:2003eb,Dreiner:2005rd,Lee:2010vj}.
Contrary to the $R$-parity conserving mSUGRA model
(also called the constrained MSSM) \cite{Nilles:1983ge,Martin:1997ns}, lepton number is violated. This
leads to naturally light neutrino masses without introducing either a
new see--saw energy scale at $M_X=\mathcal{O} (10^{10}\,\mathrm{GeV})$
or new gauge--singlet superfields
\cite{see-saw,Allanach:2003eb,Allanach:2007qc,Dreiner:2010ye,Dreiner:2007uj}.
However, within this model the lightest supersymmetric particle (LSP)
is not stable. Thus the lightest neutralino is no longer a viable dark
matter particle 
\footnote{If the lifetime of the neutralino LSP is much larger than
the age of the universe, it is still a good dark matter candidate.
However, the trilinear $R$-parity violating couplings need in this 
case to be smaller than $\mathcal{O}(10^{-20})$ \cite{Barbier:2004ez}.
Note that such $R$-parity violating scenarios are indistinguishable
from $R$-parity conserving scenarios at colliders, because the neutralino LSP 
can escape detection.}
and other candidates must be considered, such as the
axino \cite{Chun:1999cq,Choi:2001cm,Kim:2001sh,Chun:2006ss}, gravitino
\cite{Buchmuller:2007ui,Covi:2009pq,Bobrovskyi:2010ps}, the lightest $U$-parity particle
\cite{Lee:2007fw,Lee:2007qx} or slightly modified models
like the NMSSM with a gravitino LSP
\cite{JeanLouis:2009du}. More importantly for the focus of this paper,
in this case, the LSP is no longer bounded by cosmology to be the
lightest neutralino \cite{Ellis:1983ew}. Any other supersymmetric
particle is possible \cite{Dreiner:1991pe,Dreiner:1997uz}. This can
lead to dramatically different signatures at colliders
\cite{Bernhardt:2008jz,Dreiner:2008rv,Dreiner:2006sv}. Within the
B$_3$ mSUGRA model, certain non--neutralino LSP candidates are
preferred \cite{Bernhardt:2008jz,Dreiner:2008ca,Dreiner:2009fi}. In
particular for small $R$-parity violating couplings the lightest
scalar tau, the stau, is a possible LSP in large regions of the mSUGRA
parameter space. This has been known for the $R$-parity conserving
case for a long time, but has been discarded for the above
cosmological reasons. In the $R$-parity violating case the stau is a natural
LSP. We note however that the stau can also be the lightest
supersymmetric particle within the ($R$-parity conserving) MSSM spectrum if one adds 
another (lighter) particle to the spectrum, like the gravitino. 
The stau then decays into this new particle and 
can be again consistent with cosmological observations; see e.g.
Ref.~\cite{Giudice:1998bp}.

\mymed

In this paper, we consider in detail the discovery potential at the
LHC for the B$_3$ mSUGRA model with a stau LSP. We focus on the case
of early data at 7~TeV center-of-mass energy.
In order to be specific, we mainly restrict
ourselves to one of the B$_3$ mSUGRA benchmark points discussed in
Ref.~\cite{Allanach:2006st}, as well as related scenarios. As we shall
see, besides the good theoretical motivation, the stau LSP scenarios
are also readily testable with early LHC data, because they can lead
to a multi--lepton signature that is hard to achieve with the SM
interactions.

\mymed

The outline of the paper is as follows. In Sec.~\ref{B3-mSUGRA}, we
present the model, which we investigate and then discuss in detail how
a stau LSP can arise. We also review the benchmark scenario that is
relevant for this work.  An overview of all expected stau LSP
signatures at the LHC is given in Sec.~\ref{Sec:stau-lsp-signatures},
\ie~we consider the different dominating $R$-parity violating
operators. We then present in Sec.~\ref{signal_vs_bkg_BC1} in detail
the particle multiplicities and kinematic properties of our benchmark
scenario and for the most important SM backgrounds.  Based on this, we
develop in Sec.~\ref{Sec:Dicov-Potential} a set of cuts that allow a
discovery of these stau LSP scenarios with early LHC data. We
employ a fast detector simulation
and estimate systematic uncertainties in our analysis.
We also investigate the possibility of
reconstructing the stau LSP mass. Finally, we comment in
Sec.~\ref{tauID} on potential difficulties with tau identification due
to the large particle multiplicities in our scenarios. We summarize
and conclude in Sec.~\ref{Sec:summary}.

\mymed 

In App.~\ref{App:properties-BC12}, we review some basic properties
of the investigated benchmark scenario. 
We give an overview over different definitions of significances in
App.~\ref{diff_significances}.

\section{The $R$-parity Violating mSUGRA Model}
\label{B3-mSUGRA}
\subsection{Motivation}
In the B$_3$ mSUGRA model \cite{Nilles:1983ge,Martin:1997ns,Allanach:2003eb} 
there are six free parameters at the unification scale ($M_{\rm GUT}$)
\begin{equation}
M_0,\,M_{1/2},\,A_0,\,\tan\beta,\,\mathrm{sgn}(\mu),\,\mathbf{\Lambda}\,.
\label{msugra-parameters}
\end{equation}
Here $M_0,\,M_{1/2}$ are the universal supersymmetry breaking scalar
and gaugino masses, respectively. $A_0$ is the universal
supersymmetry breaking scalar trilinear coupling and $\tan\beta$ is
the ratio of the two vacuum expectation values. $\mathrm{sgn}(\mu)=
\pm1$ is the sign of the Higgs mixing parameter. 
The magnitude of $\mu$ as well as the respective bilinear
scalar coupling $B_0$ are fixed by radiative electroweak symmetry breaking
\cite{Ibanez:1982fr}.
In the B$_3$ mSUGRA
model the superpotential is extended beyond the MSSM by the following
terms \cite{Dreiner:1997uz}
\begin{equation}
W_{\mathrm{B}_3}=\lam_{ijk} L_iL_j\bar E_k + \lam_{ijk}^\prime 
L_iQ_j\bar D_k + \kap_{i} L_i H_2\,.
\label{B3-superpot}
\end{equation}
Here $L_i,\,Q_i$ denote the lepton and quark $SU(2)$ doublet
superfields. $H_2$ denotes the Higgs $SU(2)$ doublet superfield which
couples to the up--like quarks. $\bar E_i,\,\bar D_i$
denote the lepton and quark $SU(2)$ singlet superfields, respectively.
$i,j,k\in\{1,2,3\}$ are generation indices. $\lam_{ijk}$ denote nine
(anti--symmetric in $i\leftrightarrow j$), $\lam_{ijk}^\prime$
twenty--seven dimensionless couplings. $\kap_i$ are three
dimensionful parameters, which vanish at the unification scale
\cite{Allanach:2003eb}. All the operators in Eq.~(\ref{B3-superpot})
violate lepton number.

\mymed

The parameter $\mathbf{\Lambda}$ in Eq.~(\ref{msugra-parameters}) goes
beyond the $R$-parity conserving mSUGRA model. It refers to a choice of
\textit{exactly one} of the thirty--six dimensionless couplings in
Eq.~(\ref{B3-superpot}) at a time.
\begin{equation}
\mathbf{\Lambda}\in\{\lam_{ijk},\,\lam'_{ijk}\}\,, \quad i,j,k=1,2,3\,.
\end{equation}
Given one coupling at the unification scale, through the
renormalization group equations (RGEs) other couplings are generated
at the weak--scale ($M_{EW}$) \cite{Barger:1995qe,de
Carlos:1996du,Allanach:1999mh}.

\mymed

In principle, one can also choose more than one $R$-parity violating
coupling at the unification scale. We mainly restrict ourselves to
only one coupling, because it makes the investigation of the parameter
space more manageable. However, our approach is also motivated from a
phenomenological point of view, because experimental constraints on products of different
$R$-parity violating couplings are in general more restrictive than on 
only a single coupling \cite{Barbier:2004ez}. Furthermore, in the SM the top Yukawa is
at least an order of magnitude larger than the other Yukawa couplings.

\subsection{Stau LSP}
\label{stau-LSP}
As mentioned above, in the B$_3$ mSUGRA model the LSP is no longer
constrained to be the lightest neutralino, $\ninoone$.  Since the RGEs
are modified compared to the $R$-parity conserving mSUGRA model
by the interactions in Eq.~(\ref{B3-superpot}), as well as the
corresponding soft supersymmetry breaking terms, in principle new LSPs
may arise as a function of the input parameters in
Eq.~(\ref{msugra-parameters}).  For large values of specific
couplings, {\it i.e.} $\mathbf{\Lambda} \gsim \mathcal{O}(10^{-2})$ , this does in fact happen
\begin{eqnarray}
\tilde\mu_R \quad \text{for large} \;\; \lam_{132} \, , \quad \quad \quad \quad \; \, \nonumber \\
\tilde\nu_i \quad \text{for large} \;\; \lam_{ijk}' \, , \quad \quad \quad \quad \; \; \nonumber \\
\tilde e_R \quad \text{for large} \;\; \lam_{121}, \lam_{131}, \lam_{231} \, .
\end{eqnarray}
as discussed in detail in
Ref.~\cite{Bernhardt:2008jz,Dreiner:2009fi,Dreiner:2008ca}. The
relevant couplings are given here in parentheses.  However, even for
small values of arbitrary couplings there are large regions of
parameter space, where the lightest stau, $\tilde\tau_1$, is the
LSP. This occurs for large values of $M_{1/2}$ (which drives up the
$\ninoone$ mass faster than the stau mass) and small values of
$M_0$; see App.~\ref{masses_stau} for the $\tan \beta$
dependence. For a fixed value of $M_0$ there is always a value of
$M_{1/2}$ above which the stau is the LSP \cite{Allanach:2006st}. For
example, at $A_0=-100\,$GeV, $\tan\beta=10$ and $\mathrm{sgn}(\mu)=+1$
the region is approximately given by
\begin{equation}
\tilde\tau_1-\mathrm{LSP}: \quad M_{1/2} \gsim 3.8\cdot M_0 + 175\,
\mathrm{GeV}\,.
\end{equation}
It is the purpose of this paper to study in detail the detectability
of stau LSP models with early LHC data.

\mymed

There is a series of papers in the literature on a stau LSP
\cite{Akeroyd:1997iq,deGouvea:1998yp,Akeroyd:2001pm,Hirsch:2002ys,Abbiendi:2003rn,
Heister:2002jc,Abdallah:2003xc,Allanach:2003eb,Bartl:2003uq,Allanach:2006st,
deCampos:2007bn,Allanach:2007vi,Dreiner:2007uj,Dreiner:2008rv,Bernhardt:2008mz,
Ghosh:2010tp,Mukhopadhyaya:2010qf}.
To the best of our knowledge a stau LSP was first considered in
Ref.~\cite{Akeroyd:1997iq} in a bilinear $R$-parity breaking model
with a focus on charged Higgs phenomenology \footnote{Another early
paper is Ref.\cite{deGouvea:1998yp}, however, here $R$-parity is
conserved.}.  There has been further work on a stau LSP in bilinear
$R$-parity violating models
\cite{Akeroyd:2001pm,Hirsch:2002ys,deCampos:2007bn}. In
Ref.~\cite{deCampos:2007bn}, LHC phenomenology is considered in some
detail, however the focus is on a $\tilde\chi^0_1$ LSP, with only
peripheral mention of the stau LSP discovery reach.  In
Ref.~\cite{Bartl:2003uq} the decay lengths of slepton LSPs are
considered in both bilinear and trilinear $R$-parity violating models.
LEP II searches for a stau LSP were performed in
Refs~\cite{Abbiendi:2003rn,Heister:2002jc,Abdallah:2003xc}, allowing
only for two--body $R$-parity violating decays of the stau.
Recently, Ref.~\cite{Mukhopadhyaya:2010qf} investigated stau LSP scenarios, where
the stau decays leptonically via 2-body decays. It was shown 
that these scenarios might already be discovered in early LHC data 
via same-sign trilepton events.

\mymed

In Ref.~\cite{Allanach:2003eb} the B$_3$ mSUGRA model was constructed.
It was shown that extensive regions of parameter space lead to a stau
LSP.  Furthermore it was shown that in some regions of parameter space
the stau dominantly decays via 2--body decays and in others via
4--body decays. In Ref.~\cite{Allanach:2006st} benchmarks were defined
for exemplary phenomenological and experimental analyses, for both the
2--body and 4--body decays. These are discussed below.  In
Ref.~\cite{Dreiner:2008rv}, the origin of the 2--body and 4--body
decays from the RGEs was studied in detail. The example of resonant
slepton production was then analyzed in the case of a stau LSP.

\mymed

We here present for the first time a comprehensive analysis at the LHC
of stau LSP scenarios. We focus mainly on one specific benchmark point
BC1 \cite{Allanach:2006st}. For this we investigate the signal and the
background in great detail. We include a fast detector simulation
using the program package {\tt Delphes} \cite{Delphes}.

\subsection{Benchmark Scenarios}
\label{benchmarks}
In Ref.~\cite{Allanach:2006st} four Bonn--Cambridge (BC) benchmark
points for studying $R$-parity violating models in detail were defined.
BC1--BC3 are B$_3$ mSUGRA models, BC4 involves a baryon number
violating operator. The main properties of the points BC1--BC3 are 
summarized in Table~\ref{BCs}.
\begin{table}
\begin{tabular}{c||ccc}
&\textbf{BC1}&\textbf{BC2}&\textbf{BC3} \\ \hline\hline &&&\\
Operator & $\lam_{121}L_1L_2\bar E_1$ & $\lam'_{311}L_3Q_1\bar D_1$& $\lam'_{331}L_3Q_3\bar D_1$\\[1mm] \hline&&&\\
\parbox[t]{1.7cm}{Coupling\\ (@$M_{GUT}$)} &$0.032$&$3.5\cdot10^{-7}$&$0.122$\\[1mm]\hline&&&\\
\parbox[t]{1.7cm}{Coupling\\ (@$M_{EW}$)} &$0.048$&$1.1\cdot10^{-6}$&$0.344$\\[1mm]\hline&&&\\
LSP & $\tilde\tau_1$& $\tilde\tau_1$ & $\tilde\nu_\tau$ \\[1mm]\hline&&&\\
$M_0\,$(GeV) &  0 & 0&100\\[1mm] \hline&&&\\
$M_{1/2}\,$(GeV) & 400 & 400 & 250\\[1mm]\hline&&&\\
$\tan\beta$ &13&13&10\\[1mm]\hline&&&\\
$A_0\,$(GeV) &0&0&-100
\end{tabular}
\caption{The main properties of the B$_3$ mSUGRA benchmark points.}
\label{BCs}
\end{table}
BC1 and BC2 have a stau LSP, the focus of our investigation here. In
contrast to BC2, BC1 involves an additional purely leptonic operator,
$\lam_{121}L_1L_2\bar E_1$. Thus the $\ninoone$ and stau decay purely
leptonically. In the production and cascade decay of squarks and
gluinos at the LHC this leads to many leptons in the final
state. In BC2 the stau LSP decays purely hadronically, leading to
significantly less final state leptons. Here we focus
on the BC1 scenario, due to the better prospects of early discovery at
the LHC. The mass spectrum as well as the dominant decay branching
fractions for BC1 are given in App.~\ref{App:properties-BC12}.

\section{Stau LSP Signatures at the LHC}
\label{Sec:stau-lsp-signatures}

As a first step of our analysis, we classify the main LHC signatures,
assuming that the stau decay is dominated by only one $R$-parity
violating operator, \cf~Eq.~(\ref{B3-superpot}). For simplicity
we focus on the cascade process
\begin{equation}
qq/gg \rightarrow \tilde{q} \tilde{q} \rightarrow jj \ninoone \ninoone 
\rightarrow jj \tau \tau \stau_1 \stau_1 \, ,
\label{basic_process}
\end{equation}
where $\tilde{q}$ is a squark, and $j$ denotes a jet. The
final--state staus can only decay via $R$-parity violating operators.
We then classify the signatures according to the possible stau
decays. The results are
\begin{table}
\begin{ruledtabular}
\begin{tabular}{c|c|c} 
 coupling & $\tilde{\tau}_1^+$ decay & LHC signature \\
 \hline
$\lambda_{121}=-\lambda_{211}$ & $\tau^+ \mu^+ e^- \bar \nu_e$ &  \\
 & $\tau^+ \mu^- e^+ \nu_e$ & \\
 & $\tau^+ e^+ e^- \bar \nu_\mu$ & \\
 & $\tau^+ e^- e^+ \nu_\mu$ & $2 j + 4 \tau + 4 \ell + {\met}$ \\
 \cline{1-2}\rule[-2mm]{0mm}{6mm}
$\lambda_{122}=-\lambda_{212}$ & $\tau^+ \mu^+ \mu^- \bar \nu_e$ &\\
& $\tau^+ \mu^- \mu^+ \nu_e$ & with $\ell=e, \mu$\\
 & $\tau^+ e^+ \mu^- \bar \nu_\mu$ & \\
 & $\tau^+ e^- \mu^+ \nu_\mu$ & \\
 \hline
 \hline
$\lambda_{131}=-\lambda_{311}$ & $e^+ \nu_e$ &  \\
\cline{1-2}\rule[-2mm]{0mm}{6mm}
$\lambda_{132}=-\lambda_{312}$ & $\mu^+ \nu_e$ & \\
\cline{1-2}\rule[-2mm]{0mm}{6mm}
$\lambda_{231}=-\lambda_{321}$ & $e^+ \nu_\mu$ & $2 j + 2 \tau + 2 \ell + {\met}$ \\
\cline{1-2}\rule[-2mm]{0mm}{6mm}
$\lambda_{232}=-\lambda_{322}$ & $\mu^+ \nu_\mu$ & \\
\cline{1-2}\rule[-2mm]{0mm}{6mm}
$\lambda_{123}=-\lambda_{213}$ & $\mu^+ \bar \nu_e$ & \\
 & $e^+ \bar \nu_\mu$ & \\
 \hline
 \hline
$\lambda_{133}=-\lambda_{313}$ & $e^+ \bar \nu_\tau$ &  \\
 & $\tau^+ \bar \nu_e$ & $2 j + 2 \tau + 2 \ell + {\met}$ \\
  & $\tau^+ \nu_e$ & $2 j + 3 \tau + 1 \ell + {\met}$ \\
\cline{1-2}\rule[-2mm]{0mm}{6mm}
$\lambda_{233}=-\lambda_{323}$ & $\mu^+ \bar \nu_\tau$ & $2 j + 4 \tau + {\met}$ \\
 & $\tau^+ \bar \nu_\mu$ & \\
  & $\tau^+ \nu_\mu$ & \\
\end{tabular}
\end{ruledtabular}
\caption{\label{tab_stau_LSP_sig_LLE}
Scenarios assuming one non-vanishing $L_i L_j \bar E_k$ operator.  The
column in the middle shows the possible stau LSP decays, for
the $\lambda_{ijk}$ couplings in the left column. The right column
shows the resulting LHC signatures. The SUSY cascade $qq/gg
\rightarrow \tilde{q}\tilde{q} \rightarrow jj \tilde{\chi}_1^0
\tilde{\chi}_1^0 \rightarrow jj \tau \tau \tilde{\tau}_1
\tilde{\tau}_1$ has been assumed. Note that gluino, 
$\tilde{g}$, pair production instead of squark pair production 
will usually give two additional jets; for example via the 
decay $\tilde{g}\rightarrow j \tilde{q}$.}
\end{table}
\begin{table}
\begin{ruledtabular}
\begin{tabular}{c|c|c} 
 coupling & $\tilde{\tau}_1^+$ decay & LHC signature \\
 \hline
$\lambda'_{1jk}$ & $\tau^+ \bar u_j d_k e^+$ &  \\
 & $\tau^+ u_j \bar d_k  e^- $ & \\
 & $\tau^+ \bar d_j d_k \bar \nu_e$ & $6 j + 4 \tau + \ell \ell$\\
 & $\tau^+ d_j \bar d_k \nu_e$ & $6 j + 4 \tau + \ell + {\met}$ \\
 \cline{1-2}\rule[-2mm]{0mm}{6mm}
$\lambda'_{2jk}$ & $\tau^+ \bar u_j d_k \mu^+$ &  $6 j + 4 \tau + {\met}$\\
 & $\tau^+ u_j \bar d_k  \mu^- $ & \\
 & $\tau^+ \bar d_j d_k \bar \nu_\mu$ & \\
 & $\tau^+ d_j \bar d_k \nu_\mu$ & \\
 \hline
 \hline
$\lambda'_{3jk}$ & $u_j \bar d_k$ & $6 j + 2 \tau$\\ 
\end{tabular}
\end{ruledtabular}
\caption{\label{tab_stau_LSP_sig_LQD}%
Same as Tab.~\ref{tab_stau_LSP_sig_LLE} but for one non-vanishing 
$L_i Q_j \bar D_k$ operator.}
\end{table}
\begin{table}
\begin{ruledtabular}
\begin{tabular}{c|c|c} 
 coupling & $\tilde{\tau}_1^+$ decay & LHC signature \\
 \hline
$\lambda''_{ijk}$ & $\tau^+ u_i d_j d_k$ & $8 j+ 2 \tau$ \\
 & $\tau^+ \bar u_i \bar d_j \bar d_k$ & \\
\end{tabular}
\end{ruledtabular}
\caption{\label{tab_stau_LSP_sig_UDD}%
Same as Tab.~\ref{tab_stau_LSP_sig_LLE} but for one non-vanishing 
$\bar U_i \bar D_j \bar D_k$ operator.}
\end{table}
summarized in Tables~\ref{tab_stau_LSP_sig_LLE},
\ref{tab_stau_LSP_sig_LQD}, and \ref{tab_stau_LSP_sig_UDD}, for the
operators $L_i L_j \bar E_k$, $L_i Q_j \bar D_k$, or $\bar U_i \bar
D_j \bar D_k$
\footnote{For completeness, we give in this section also the
signatures for a non-vanishing $\bar U_i \bar D_j \bar D_k$ operator,
where $\bar U_i$ denotes an up-type quark $SU(2)$ singlet
superfield. Note that $\bar U_i \bar D_j \bar D_k$ violates $B_3$;
\cf~for example Ref.~\cite{Allanach:2003eb}.}, respectively. Here we
assume that always only one operator is dominant
\footnote{In principle, there can be additional stau LSP decays
via $R$-parity violating couplings which are generated via RGE running.
This is particularly important if the $R$-parity violating operator
at the unification scale leads to a 4-body decay, whereas the generated
one allows a 2-body decay; see Ref.~\cite{Dreiner:2008rv} for a detailed discussion 
and for explicit examples. For the benchmark point BC1, for example,
$\lam_{323}$ can be generated out of $\lam_{121}$ allowing
a 2-body stau LSP decay, {\it i.e.} $\tilde{\tau}_1^+ \rightarrow \mu^+ \bar \nu_\tau$. 
However, this decay plays no significant role in this work, although it might
be important in other regions of the parameter space. For the parameter scan 
in Sect.~\ref{discovery_potential}, the 2-body decay branching ratio is always $<1\%$ ($\lesssim 10 \%$)
for $\tan \beta < 25$ ($\tan \beta > 25$).}. In the left column
we denote the dominant coupling, in the middle column, the
corresponding dominant 2-- or 4--body stau decays. The resulting final
state signatures for the cascade of Eq.~(\ref{basic_process}) are
given in the right column.  Here $\ell$ denotes an electron or muon,
$\tau$ a tau and ${\met}$ missing transverse energy due to neutrinos
in the final state. We have not included neutrinos from tau decays.

\begin{figure*}[t!]
    \includegraphics[width=1.0\textwidth]{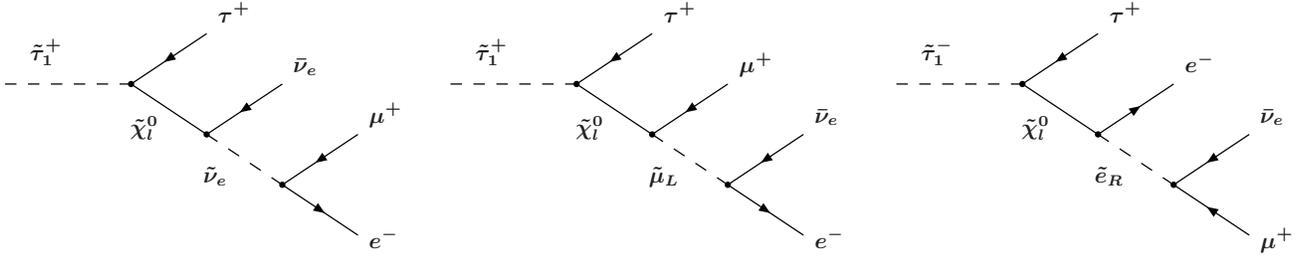}
    \caption{\label{Fig:4body_decay} Feynman diagrams contributing to the 4-body stau LSP decay 
$\tilde{\tau}_1^+ \rightarrow \tau^+ \mu^+ e^- \bar \nu_e$ via $\lambda_{121}$. In this example,
the stau LSP decays via a virtual neutralino $\tilde{\chi}_l^0$ ($l=1\dots 4$). In addition, the
first, second and third diagram involve a virtual electron sneutrino, $\tilde{\nu}_e$, a virtual
(left-handed) smuon, $\tilde{\mu}_L$, and a virtual right-handed selectron, $\tilde{e}_R$,
respectively.}
\end{figure*}

We show in Fig.~\ref{Fig:4body_decay} as an example the 4-body stau LSP
decay into $\tau^+ \mu^+ e^- \bar \nu_e$ via $\lambda_{121}$, {\it cf.} the first decay in Tab.~\ref{tab_stau_LSP_sig_LLE}.
The stau does not directly couple to the $L_1L_2\bar E_1$ operator and thus first couples 
to a virtual neutralino radiating off a tau lepton. The neutralino then couples to, for example,
an electron anti-neutrino and a virtual electron sneutrino, $\tilde{\nu}_e$, which decays via $\lambda_{121}$ 
into an electron and a muon. We end up with a 4-body decay of the stau LSP. 

We observe that the partial width, $\Gamma_4$, for the 4-body decay gets suppressed if 
the sfermion (gaugino) masses increase, {\it i.e.} the behavior is
approximately $\Gamma_4 \sim  m_{\tilde{f}}^{-4}$~($m_{\tilde{\chi}_l^0}^{-2}$) \cite{Allanach:2003eb}. 
Furthermore, we see that for the scenario BC1,
the decay is mainly mediated by the $\tilde{\chi}_1^0$. The
(mainly right-handed stau LSP) couples much stronger to the (bino-like)
$\tilde{\chi}_1^0$ than to the heavier (wino- and higgsino-like) neutralinos.
If we also take into account the Majorana nature of the neutralinos we obtain
from Fig.~\ref{Fig:4body_decay} the second stau LSP decay mode of 
Tab.~\ref{tab_stau_LSP_sig_LLE}.

The stau LSP can in principle also decay via a virtual chargino instead of a virtual
neutralino. However, these decays are suppressed by several orders of
magnitude. On the one hand, they are suppressed due to heavy
propagators. On the other hand, the (mainly right-handed) stau couples
in mSUGRA more strongly to the (bino-like) $\ninoone$ than to the
(wino-like) $\tilde{\chi}_1^+$.
  
Note, that due to the Majorana nature of the $\ninoone$, every charge
combination of the two staus is possible. The flavor and charge of the
leptons and quarks in the final state, are determined via the
different decay modes of the stau LSP. 

\mymed

In general, more complicated decay chains than
Eq.~(\ref{basic_process}) can occur, leading typically to
additional final state particles. The most important are:
\begin{itemize}
\item Additional jets from the production of gluinos 
and their subsequent decays into squarks and quarks.
\item Additional leptons from decays involving heavy neutralinos and
charginos. For example, a left-handed squark might decay to a
$\tilde\chi^0_2$ or $\tilde{\chi}_1^+$, which then decays to a lepton
and a slepton.
\item Additional leptons from the decay of a $\ninoone$
next-to-next-to-next-to LSP (NNNLSP) into right-handed selectrons or
smuons, \ie~$\ninoone \rightarrow \ell^\pm \tilde{\ell}^\mp_R$
followed by $\tilde{\ell}_R^+ \rightarrow \ell^+ \tau^\pm
\stau_1^\mp$.
\end{itemize}
The last process is special for stau LSP scenarios. Within mSUGRA, it
is kinematically only allowed if $M_{1/2} \gg M_0$, \eg~for $M_{1/2}
> 400$ GeV for $M_0=0$ GeV. Note, that $M_{1/2}$ increases the mass of
the (bino-like) $\ninoone$ faster than the mass of the
$\tilde{\ell}_R$ \cite{Drees:1995hj}. If these new decay channels are
open, we can have two additional taus and four additional electrons or
muons in the final state. See Ref.~\cite{Dreiner:2008rv} for explicit
examples.

\mymed

The multi charged lepton final states (especially electrons and
muons) are the most promising signatures to be tested with early LHC
data. It is relatively easy to identify electrons and muons in the
detector and for high multiplicities the SM background is very low
\cite{Aad:2009wy}. In the next section, we therefore investigate the
discovery potential of stau LSP scenarios, where the stau decays in
the 4--body decay mode via the operator $\lambda_{121}L_1 L_2\bar
E_1$, leading to a maximal number of muons and electrons in the final
state, \cf~Tab. \ref{tab_stau_LSP_sig_LLE}. We also analyze
the potential of reconstructing the mass of the stau LSP, once a
discovery has been made.

\mymed   

Note, that scenarios, where the stau decays via a 2-body $L_i L_j \bar
E_k$ mode ($i,j$ or $k=3$) might have a similar discovery potential,
because the charged leptons have on average larger momenta. We also
expect larger values of $\Etmiss$, because a 2-body stau decay via
$L_i L_j \bar E_k$ involves always one neutrino,
\cf~Tab.~\ref{tab_stau_LSP_sig_LLE}.

\mymed  

If the stau LSP couples mainly to a $L_i Q_j \bar D_k$ operator,
there are (at parton level) at least six jets and two taus in
the final state, \cf~Tab.~\ref{tab_stau_LSP_sig_LQD}.
Furthermore, for $i=1,2$, there are possibly one or two
additional charged electrons or muons as well as four additional taus.
Again, due to the Majorana nature of the $\ninoone$, all charge
combinations for these electrons and muons are possible, leading for
example to like sign dilepton events. Unlike the scenarios
shown in Tab.~\ref{tab_stau_LSP_sig_LLE} ($L_i L_j \bar E_k$), the
scenarios in Tab.~\ref{tab_stau_LSP_sig_LQD} ($L_i Q_j \bar D_k$) do
not necessarily lead to $\met$ (apart from neutrinos from tau decays).
It should therefore be possible to directly reconstruct the mass peaks
of sparticles, especially the stau LSP. However, combinatorial
backgrounds due to the many jets in the final state complicate this
task.

\mymed

A special case arises if the $L_i Q_j \bar D_k$ operator involves
quark superfields of the third generation, \ie~$j$ or
$k=3$. For $k=3$ two jets in the final states are $b$-jets, one from
each stau decay. For $i=1,2$ and $j=3$ the situation is similar. We
also obtain one $b$-jet from nearly each stau LSP decay. However,
decays which involve an electron or muon are now kinematically
suppressed or forbidden due to a top quark in the final state, 
\cf~Tab.~\ref{tab_stau_LSP_sig_LQD}.  Finally, for a non-vanishing
coupling $\lambda'_{33k}$, the stau might only decay via a 3-body
mode and a virtual top-quark. See Ref.~\cite{Dreiner:2008rv}
for details.

\mymed

In this paper, we focus on stau LSP scenarios with a lepton-number
violating $L_i L_j \bar E_k$ operator. However, we briefly discuss
signatures, which arise from a non-vanishing baryon-number violating
operator $\bar U_i \bar D_j \bar D_k$.

\mymed

We see in Tab.~\ref{tab_stau_LSP_sig_UDD} that a non-vanishing
$\lambda''_{ijk}$ coupling leads mainly to a 4-body decay of the stau
LSP, resulting in at least eight parton level jets and four taus in
the final state. Due to the many jets, we expect tau identification
via its hadronic decay modes to be very difficult; see
Sec.~\ref{tauID}. A jet from a cascade decay might overlap with the
tau jet. Furthermore, the three jets in a stau decay can be
boosted such that they might appear as only one jet. However, these
jets might still be revealed by investigating the substructure of
jets. See Ref.~\cite{Butterworth:2009qa} for $\ninoone$ LSP scenarios,
with purely hadronic ($\bar U_i\bar D_j\bar D_k$) LSP decays. See also
Refs.~\cite{Butterworth:2002tt,Butterworth:2007ke,Butterworth:2008iy,
Aad:2009wy,Salam:2009jx} for related work. We expect that similar
techniques will work for stau LSP scenarios.

\mymed

For the special case of $j$ or $k=3$, the baryon-number violating 
stau decays lead to two $b$-jets in the final state. For $i=3$,
the final state up-type quark in Tab.~\ref{tab_stau_LSP_sig_UDD} 
will be a top-quark. If the decay into a top quark is kinematically forbidden, 
the stau LSP might decay in a 5-body decay via a virtual top.

\section{Simulation of Signal and Background}
\label{signal_vs_bkg_BC1}

In this section we perform a full Monte Carlo analysis of the stau LSP
benchmark scenario BC1 (with $\lambda_{121} = 0.032$ at $M_{GUT}$; 
{\it cf.} Tab.~\ref{BCs}) and the dominant SM backgrounds. 
Our signal process is pair production of all SUSY particles. We
also employ fast detector simulations.  The mass spectrum and
branching ratios (BRs) of BC1 are given in
App.~\ref{App:properties-BC12}.

\subsection{Major Backgrounds}
\label{bkg_BC1}

In what follows, we only consider SM backgrounds
that can lead to at least one (parton level) electron, muon or 
tau in the final state. Furthermore, we expect
from most of the SUSY (signal) events additional energy in the form of
hard jets, that arise from decays in the upper parts of the decay
chain.

\mymed

We thus consider the following SM processes as the major backgrounds
in our analysis:
\begin{itemize}
\item $\ttbar$ production.
\item $Z+\text{jets}$. The $Z$ can decay into a pair of charged
leptons. Because the SUSY events posses additional large amounts of
energy from jets, we only consider $Z$ production with at least one
hard jet at parton level.
\item $W+\text{jets}$. The $W$ can decay into a charged lepton and a
neutrino. In order for the $W$ production to be competitive with the
SUSY processes, we demand at least two additional hard jets at parton
level \cite{Wjets}.
\item Di-boson ($WW$, $WZ$ and $ZZ$) production \footnote{In
principle, tri-boson production can also contribute to the SM
background.  However, the production cross section is much smaller
than the di-boson cross-section \cite{Binoth:2008kt}. In addition, we
expect that the cut on $HT^\prime  = \sum_\text{jet 1-4} \pT$,
{\it i.e.} on the scalar sum of the transverse momenta of the four hardest jets, 
would veto nearly all tri-boson events
that pass the lepton cuts, like the di-boson backgrounds;
\cf~Tab.~\ref{Tab:CutFlowBC1}.}.
\end{itemize} 

\begin{table*}[tb!]
\begin{ruledtabular}
\begin{tabular}{c | l | r | l | l} 
 Sample     & sub-sample            & simulated events & Generator                                 & comments \\
\hline 
 $\ttbar$   &                       & 240 000 & {\tt MC$@$NLO + Jimmy/Herwig}             & \\
\hline
 $Z$ + jets & $\Zee + \geq 1$ jet   & 222 000 & {\tt Alpgen + Jimmy/Herwig}   & for each process split in the\\
            & $\Zmumu + \geq 1$ jet & 231 000 &                                           & exclusive samples $1, \ldots, 4$ extra partons \\
            & $\Zzero\ra\tau\tau + \geq 1$ jet & 232 000 &                             & and the inclusive sample $\geq 5$ extra partons\\
\hline
 $W$ + jets & $W\ra e\nu \,+ \geq 2$ jets & 518 000 & {\tt Alpgen + Jimmy/Herwig} & for each process split in the\\
            & $W\ra \mu\nu + \geq 2$ jets & 642 000 &                                     & exclusive samples $2, \ldots, 4$ extra partons\\
            & $W\ra \tau\nu + \geq 2$ jets & 659 000 &                                    & and the inclusive sample $\geq 5$ extra partons\\
\hline
 di-boson   & $WW$                  & 30 000 & {\tt Jimmy/Herwig}                   &  \\
            & $WZ$                  & 20 000 &                                           & \\
            & $ZZ$                  & 9 990 &                                           & \\
\hline 
signal & & $\approx$10 000 each & {\tt Jimmy/Herwig} & \\
\end{tabular}
\caption{\label{Tab:MC_samples} Monte Carlo samples of the SM
background and signal events used for our analysis.  The third column
shows the number of simulated events (for the {\tt Alpgen} samples
after MLM matching
\cite{Caravaglios:1998yr,Mangano:2001xp,Mangano:2002ea}) and the
fourth column shows the employed generators.  For the simulation of
the signal we employed a special version of {\tt Herwig} which also
incorporates the 4-body decays of the stau LSP \cite{Peter}.  Note,
that we simulated approximately $10\,000$ signal events for each
mSUGRA parameter point in Sec.~\ref{discovery_potential} including
BC1.  }
\end{ruledtabular}
\end{table*}

Table~\ref{Tab:MC_samples} gives an overview of the background samples
used in our analysis.  QCD di- and multi-jet events have been
neglected.  It has been shown \eg~in Ref.~\cite{ATLAS:CSCbook} that QCD
background can efficiently be suppressed in multi-lepton final states.

\subsection{Simulation and Selection Cuts}
\label{sim_and_select}

The SUSY mass spectra were calculated with {\tt SOFTSUSY3.0}
\cite{Allanach:2001kg,Allanach:2009bv}.  The {\tt SOFTSUSY} output was
fed into {\tt ISAWIG1.200} and {\tt ISAJET7.75} \cite{Paige:2003mg} in
order to calculate the decay widths of the SUSY particles (beside the
stau LSP).  The signal processes, \ie~pair production
of SUSY particles, was simulated with a modified version of {\tt
Herwig6.510} \cite{Corcella:2000bw,Corcella:2002jc,Moretti:2002eu}
which also simulates the 4-body decays of the stau LSP
\cite{Peter}. We employed {\tt Jimmy4.31} \cite{Butterworth:1996zw} to
simulate the underlying event.

\mymed

The $\ttbar$ background was simulated with {\tt MC$@$NLO3.41}
 \cite{Frixione:2002ik,Frixione:2003ei}, the di-boson background in
 {\tt Herwig6.510}. We used {\tt Alpgen2.13}
 \cite{Caravaglios:1998yr,Mangano:2001xp,Mangano:2002ea} interfaced
 with {\tt Herwig6.510} to simulate the $Z+\text{jets}$ and
 $W+\text{jets}$ backgrounds. We also employed {\tt Jimmy4.31} in all
 cases for modeling of the underlying event. All Monte Carlo samples
 are based on the parton distribution functions given by
 \texttt{CTEQ6L1} \cite{Pumplin:2002}.  The employed Monte Carlo
 generators are also summarized in Table~\ref{Tab:MC_samples}.

\mymed

Detector effects on signal and background were accounted for using the
generic detector simulation {\tt Delphes1.8} \cite{Delphes}. Its
detector settings were tuned to an ATLAS-like detector at the LHC
\footnote{We changed the default ATLAS settings in Delphes, as
follows. The preselection cuts were adjusted according to
Tab.~\ref{selection_cuts}. Furthermore we chose as the jet algorithm
{\tt SIScone} \cite{Salam:2007xv} with a cone radius of 0.4. 
The calculation of the missing transverse energy $\met$ was extended
with respect to the default to take muons into account.}.  The
results of the detector simulation were cross checked with the {\tt
PGS4} simulation \cite{PGS} using its generic LHC tune. A sufficient
agreement of the two codes was observed for most observables.
However, the identification of electrons and especially tau leptons
showed some discrepancies, \cf~Sec.~\ref{tauID}.  Detector simulations
with full detail of the calorimetry and tau identification algorithms
need to be done to get more reliable results. In the comparison
between {\tt PGS} and {\tt Delphes} we observe some shifts in the energy
scale especially for electrons in our signal events. Using the differences
between the two fast simulations as an estimate for the energy scale uncertainty
would lead to unreasonably large systematic uncertainties in the estimate
of the expected discovery significance. Instead, we take an estimate of the
ATLAS collaboration \cite{ATLAS:CSCbook} for the expected background uncertainties to calculate significances
in the following.

\mymed

\begin{table}[htb!]
\begin{ruledtabular}
\begin{tabular}{l|lll}
particle & transverse momentum & pseudorapidity\\
\hline
electron & $\pT > 7\GeV$ & $|\eta| < 2.5$ \\
muon & $\pT > 6\GeV$ & $|\eta| < 2.7$  \\
tau & $\pT > 10\GeV$ & $|\eta| < 2.5$  \\
jet & $\pT > 20\GeV$ & $|\eta| < 5.0$  \\
\end{tabular}
\caption{\label{selection_cuts} Cuts for the particle selection for
the signal and background.}
\end{ruledtabular}
\end{table}

The particle selection was guided by the definitions used by the ATLAS
collaboration for SUSY studies, {\it cf.} 
Ref.~\cite[pp. 1518]{ATLAS:CSCbook}. The selection cuts are given in
Tab.~\ref{selection_cuts}. We similarly followed these
guidelines for the overlap removal of reconstructed objects. This is
needed, because a single particle may be reconstructed as several
different objects.

\begin{figure*}[t!]
  \subfigure[\label{Fig:BC1-n-particles-a}Number of jets.]{
    \includegraphics[width=0.48\textwidth,trim=0cm 0cm 0cm 0cm]{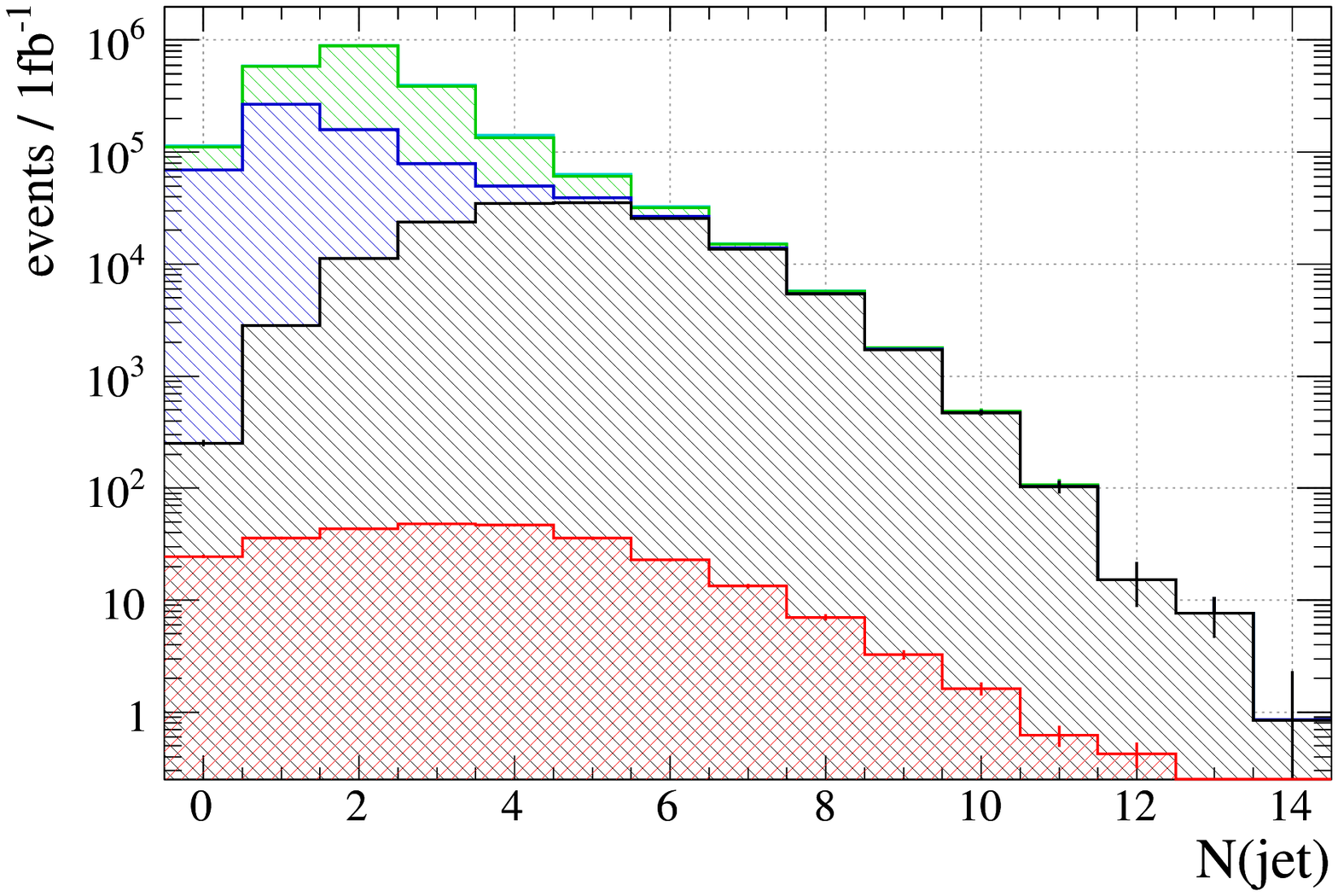}
  }
  \subfigure[\label{Fig:BC1-n-particles-b}Number of electrons.]{
    \includegraphics[width=0.48\textwidth,trim=0cm 0cm 0cm 0cm]{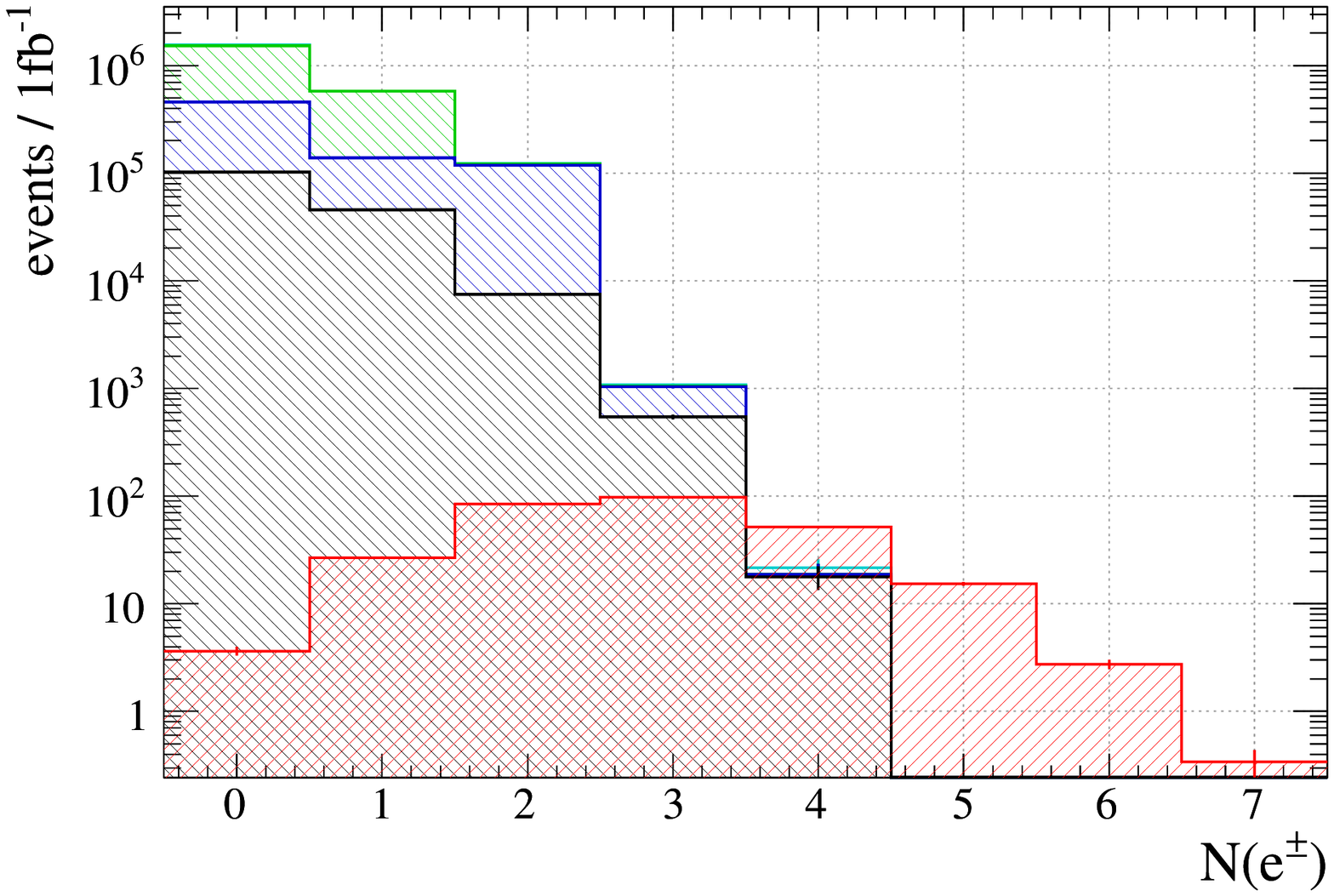}
  }
    \hfill
  \subfigure[\label{Fig:BC1-n-particles-c}Number of muons.]{
    \includegraphics[width=0.48\textwidth,trim=0cm 0cm 0cm 0cm]{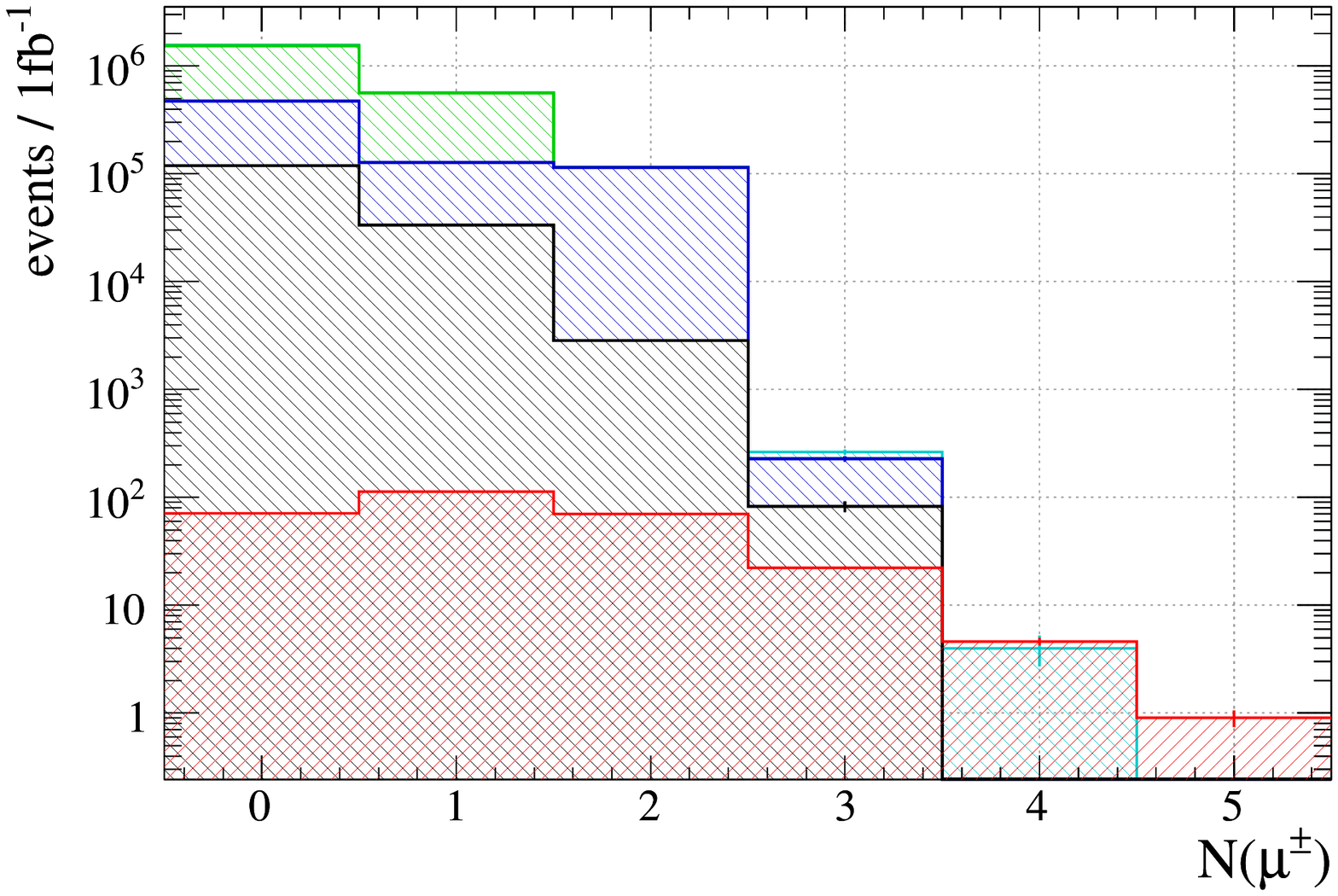}
  }
  \subfigure[\label{Fig:BC1-n-particles-d}Number of taus.]{
    \includegraphics[width=0.48\textwidth,trim=0cm 0cm 0cm 0cm]{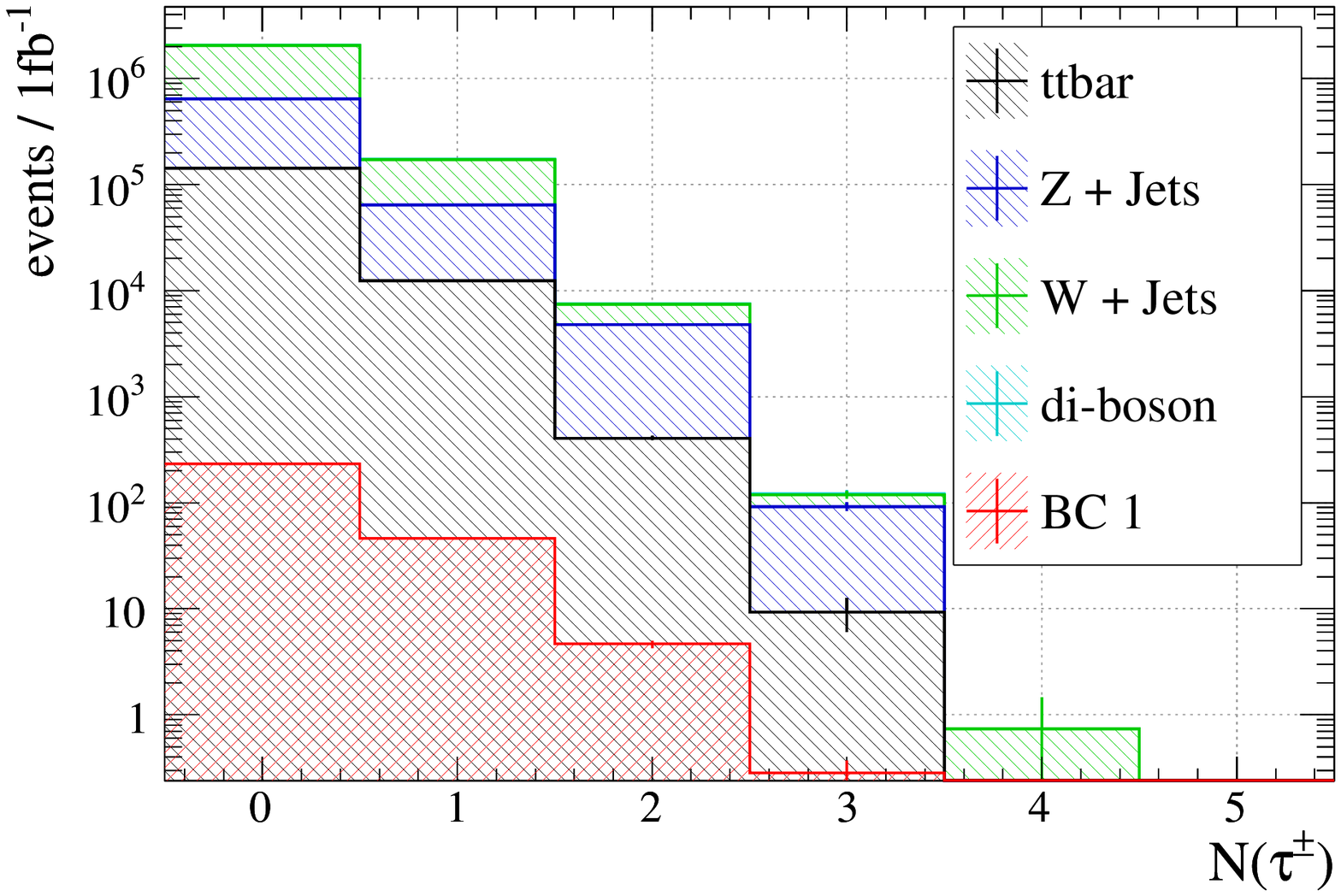}
  }
\caption{\label{Fig:BC1-n-particles} The number of reconstructed
particles per event after selection cuts and overlap
removal as described in Sec.~\ref{sim_and_select}. The color code for
the curves in all four plots is given in (d).  The different
background contributions are stacked on top of each other, while the
expected signal is shown in front of the histograms in red. The signal
corresponds to the benchmark scenario BC1
\cite{Allanach:2006st}. The number of events are scaled to
1~$\ifb$ at $\sqrt{s}=7\TeV$.}
\end{figure*}

\mymed

For the overlap removal we simply use the distance in pseudorapidity
$\eta$ and azimuthal angle $\phi$, defined as $\Delta R =
\sqrt{(\Delta\phi)^2 + (\Delta\eta)^2}$. Objects are selected in the
following order:
\begin{enumerate}
    \item Muons, if no jet and no previously selected muon is present
    within $\Delta R < 0.4$.
    \item Electrons, if no jet is present within $0.2 < \Delta R <
            0.4$ and no previously selected electron within $\Delta R
            < 0.4$.
    \item Hadronically decayed taus, if no electron or tau has already
    been selected within $\Delta R < 0.4$.
    \item Jets, if no electron, tau or another jet has already been
    selected within $\Delta R < 0.4$.
\end{enumerate}
These cuts take care of the fact that electrons and taus are usually
also reconstructed as jets.  Furthermore, electrons are more reliably
identifiable than taus. In addition, one does not want to select
electrons or muons, that stem from heavy flavor decays within jets.

\subsection{Particle Multiplicities and Kinematic Properties}
\label{Sec:kinematics-BC1}

In this section we show the basic kinematic properties of the BC1
scenario compared to the most important Standard Model backgrounds. We
also motivate the cuts to obtain a good significance and a
good signal over background ratio. No further event selection cuts
beyond the object selection cuts given in Sec.~\ref{sim_and_select}
are applied here. All samples are scaled to an integrated luminosity
of $1\,\ifb$ at a center-of-mass energy of $\sqrt{s}=7$ TeV. 

\mymed

We show in Fig.~\ref{Fig:BC1-n-particles} the number of reconstructed
jets, electrons, muons and taus, respectively. The
different background contributions are stacked on top of each other,
while the expected BC1 signal is shown in front of the background
histograms. Error bars correspond to statistical uncertainties of the
generated samples. From the discussion in
Sec.~\ref{Sec:stau-lsp-signatures} we expect 2-4 jets at parton level
for the scenario BC1.  This behavior can also be seen in
Fig.~\ref{Fig:BC1-n-particles-a} (red histogram), where the
signal distribution is maximal at 3 jets. In general,
we can get in addition to the parton-level jets also jets from parton
shower radiation and possibly from non-identified hadronically
decaying taus.

\mymed

\begin{figure*}[t!]
  \subfigure[\label{Fig:BC1-jet-pt-a}$\pT$ distribution of the hardest
    jet.]{ \includegraphics[width=0.48\textwidth,trim=0cm 0cm 0cm
    0cm]{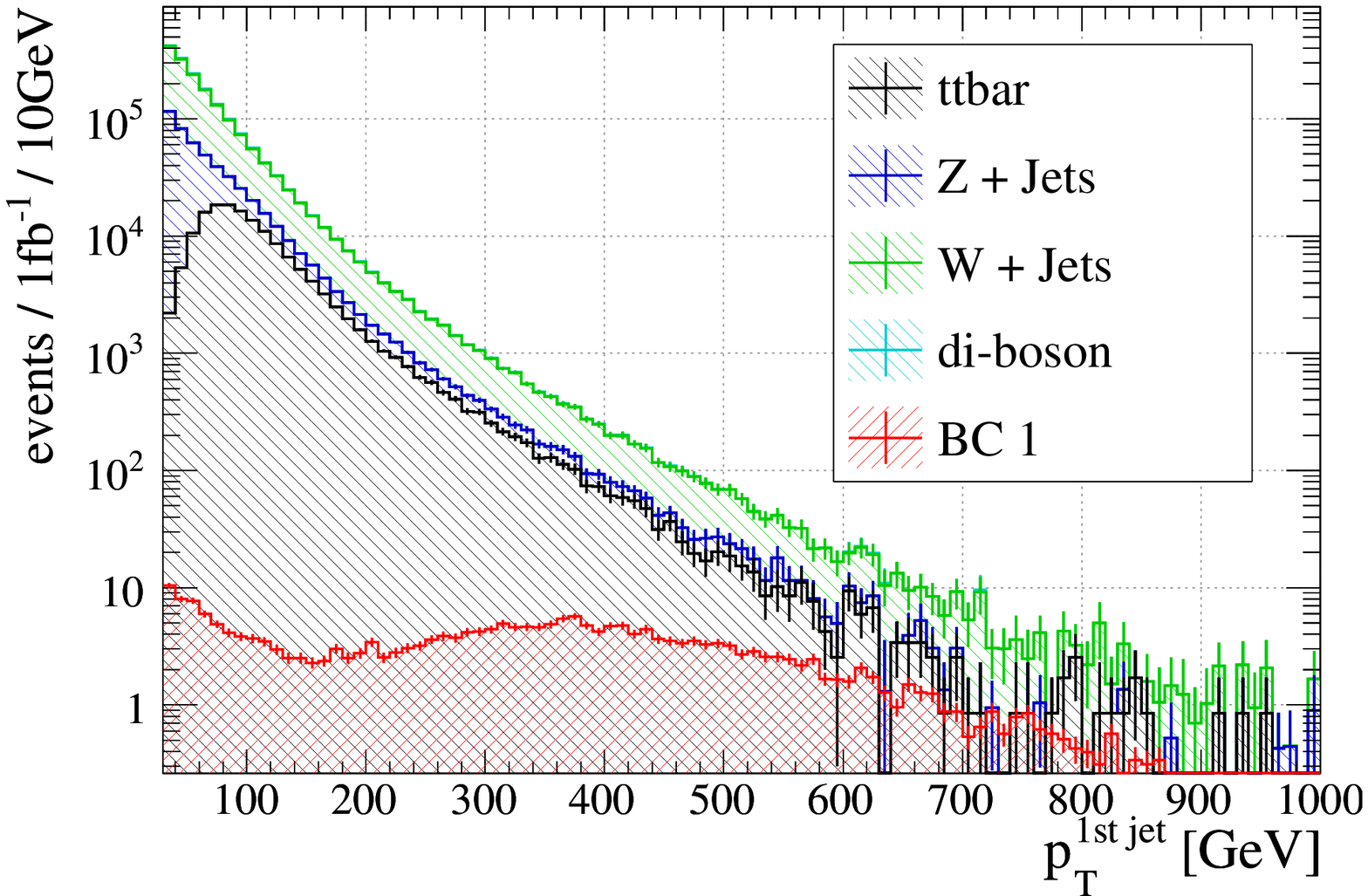} }
    \subfigure[\label{Fig:BC1-jet-pt-b}$\pT$ distribution of the 2nd
    hardest jet.]{ \includegraphics[width=0.48\textwidth,trim=0cm 0cm
    0cm 0cm]{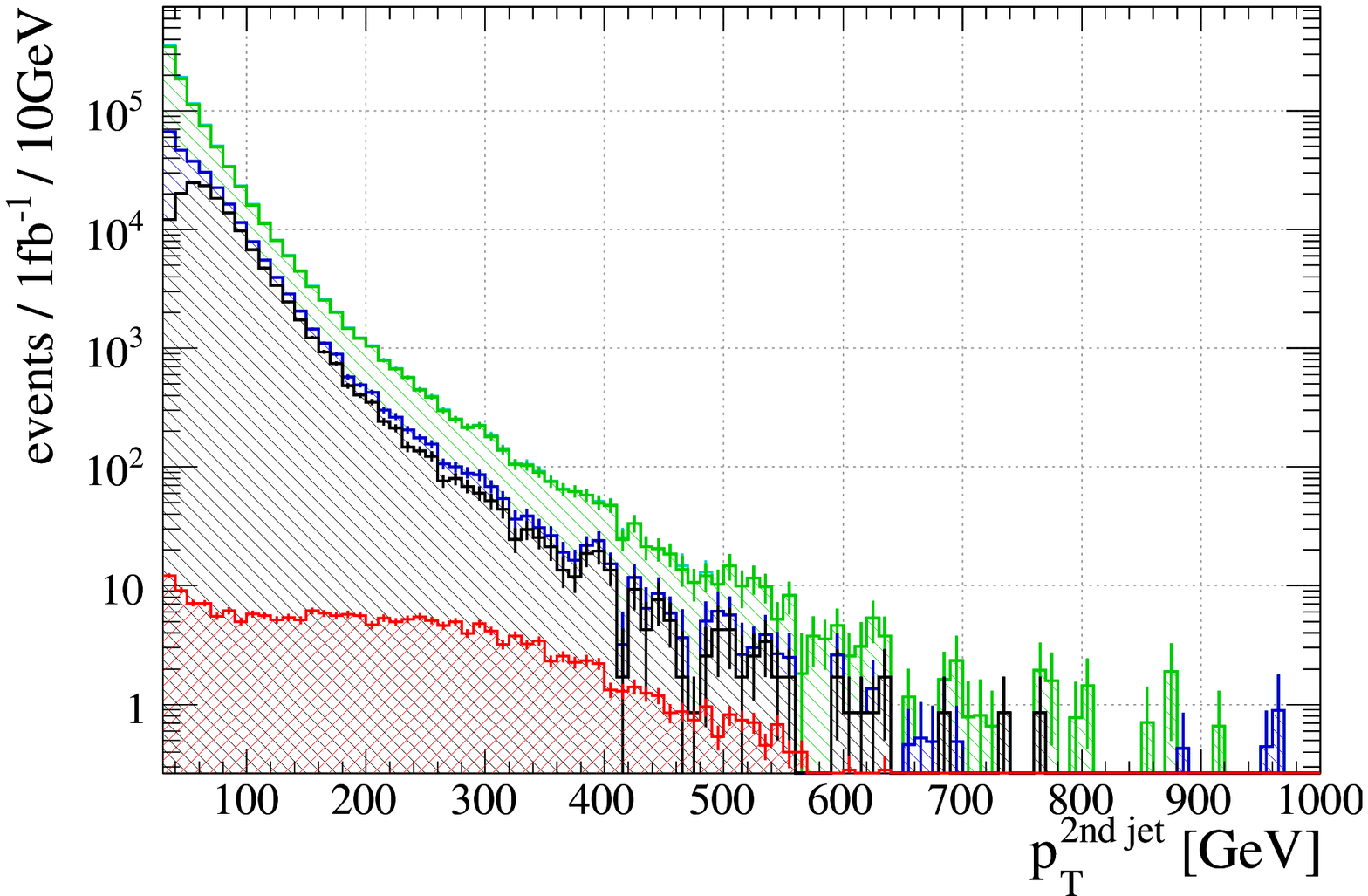} } \hfill
    \subfigure[\label{Fig:BC1-jet-pt-c}$\pT$ distribution of the 3rd
    hardest jet.]{ \includegraphics[width=0.48\textwidth,trim=0cm 0cm
    0cm 0cm]{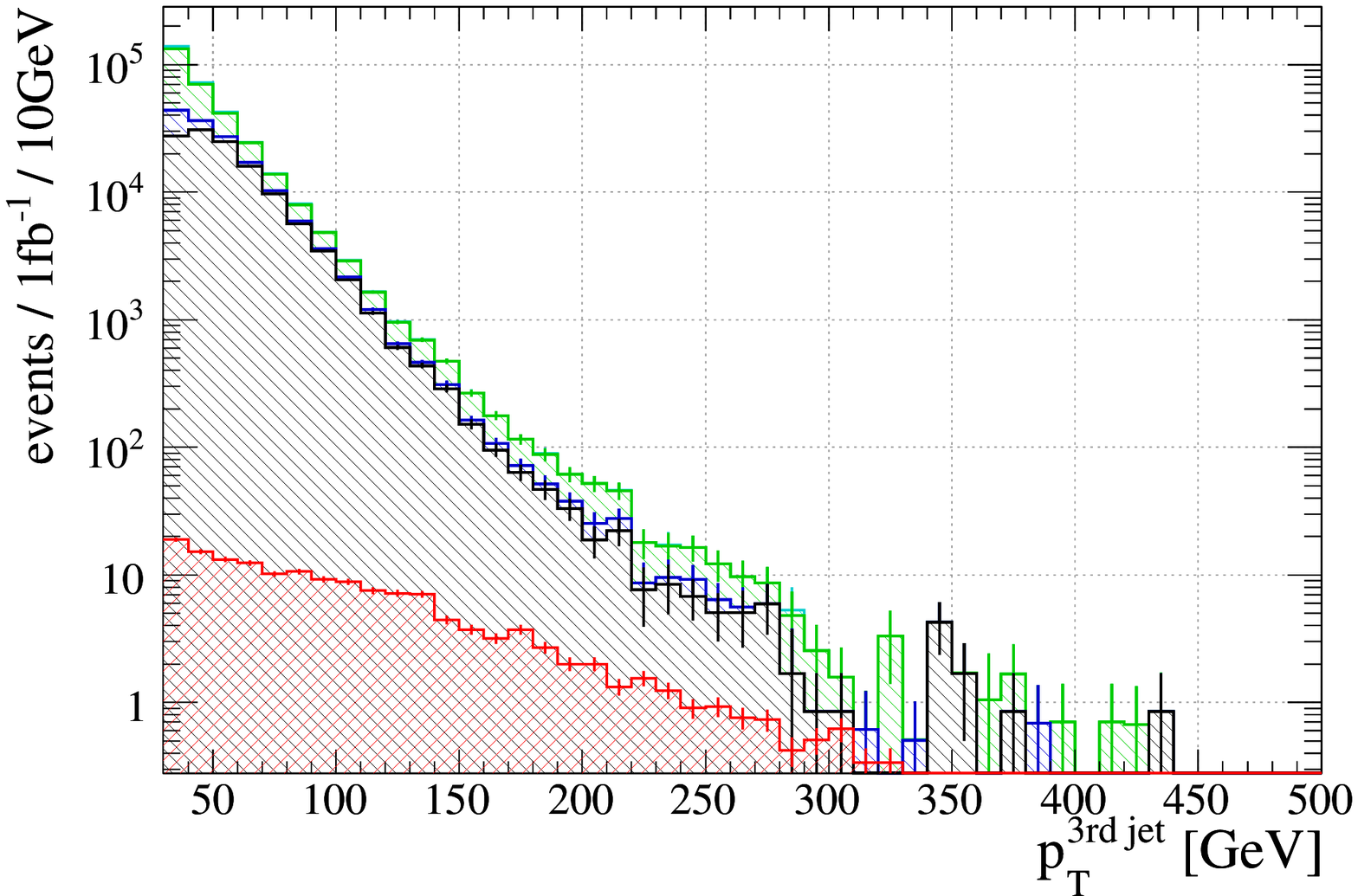} }
    \subfigure[\label{Fig:BC1-jet-pt-d}$\pT$ distribution of the 4th
    hardest jet.]{ \includegraphics[width=0.48\textwidth,trim=0cm 0cm
    0cm 0cm]{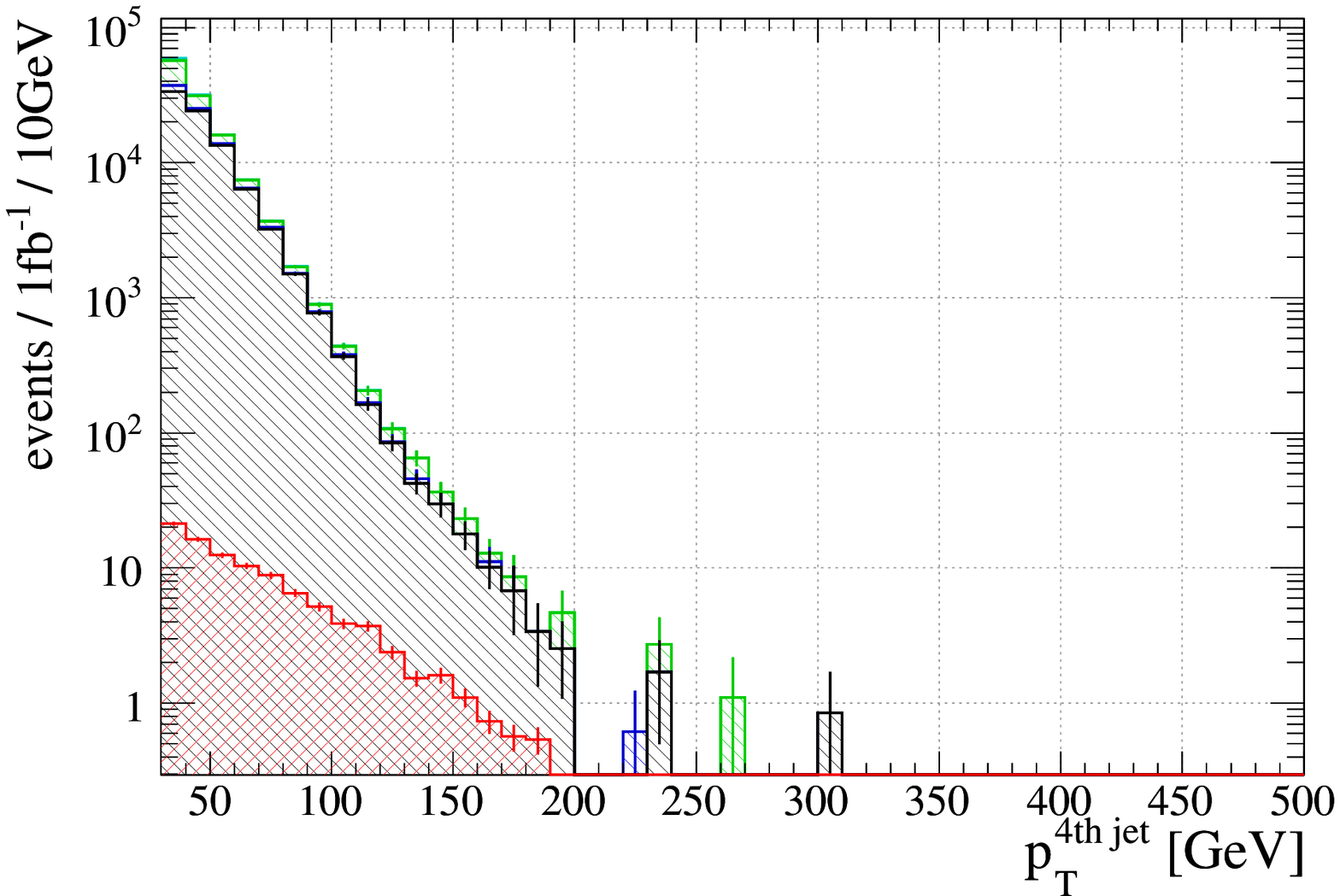}
    }\caption{\label{Fig:BC1-jet-pt} $\pT$ distributions of the four
    hardest jets after object selection cuts and overlap removal.
}
\end{figure*}

According to Fig.~\ref{Fig:BC1-n-particles-a} the SM background is
predominant even for high jet multiplicities.  The $t \bar t$
background contributes via the two (parton-level) $b$-jets from top
decays as well as jets from a hadronically decaying $W$. Therefore, a
$b$-jet veto may in principle suppress this background. However, the
signal can contain $b$-jets as well, and we will not consider
$b$-tagging in our analysis.

\mymed

The bins in Fig.~\ref{Fig:BC1-n-particles-a} with $\leq$ 1 jet are
dominated by the $Z+\geq 1$~jet background (blue histogram). This is
expected due to the large $Z+\text{jets}$ cross section. Note that
Fig.~\ref{Fig:BC1-n-particles-a} displays the event
numbers on a logarithmic scale. In contrast, $W+\text{jets}$ (green
histogram) mostly dominates for $\geq$ 2 jets, because we only show
$W$ production in association with two parton-level jets in
Fig.~\ref{Fig:BC1-n-particles-a}.

\mymed

The transverse momentum, $\pT$, distributions of the four hardest jets
are shown in Fig.~\ref{Fig:BC1-jet-pt}.  We observe in
Fig.~\ref{Fig:BC1-jet-pt-a} that the $\pT$ distribution of the hardest
signal jet is relatively flat over several hundreds of GeV whereas the
distribution of the background falls off steeply.  One can also see a
peak in the $\pT$ distribution of the signal around 360 GeV.
This peak is due to the mainly hard jets from the decay of
a squark into the $\ninoone$. Note that the mass difference
between most of the squarks and the $\ninoone$ in BC1 is
400-500 GeV, \cf~App.~\ref{App:properties-BC12}.  The invariant
mass of the hardest jet with the decay products of the
$\ninoone$ might thus allow a reconstruction of the squark
masses.

\mymed

Gluinos decaying into a jet and a squark will in general produce
softer jets, because the respective mass difference is only $\approx
100$ GeV. Therefore, these additional jets are most probably found in
Fig.~\ref{Fig:BC1-jet-pt-b}--\ref{Fig:BC1-jet-pt-d}.

\mymed

\begin{figure}[t!]
    \includegraphics[width=0.48\textwidth,trim=0cm 0cm 0cm 0cm]{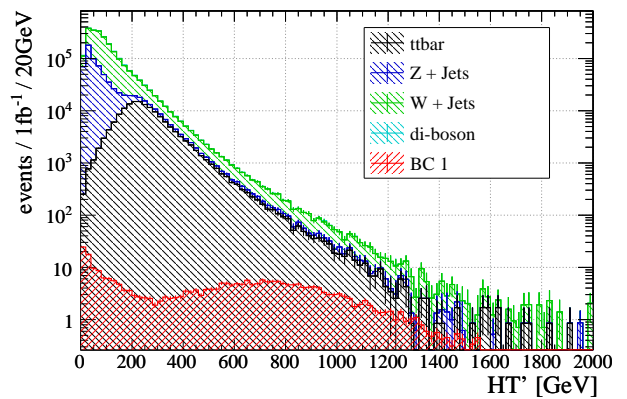}
    \caption{\label{Fig:BC1-HTprime} The number of events per
$1\,\mathrm{fb}^{-1}$ as a func\-tion of the scalar sum of the
transverse momenta of the four hardest jets, $HT^{\prime}$, after
object selection cuts and overlap removal.  }
\end{figure}

\begin{figure*}[t!]
  \subfigure[\label{Fig:BC1-electron-pt-a} $\pT$ distribution of the hardest electron.]{
    \includegraphics[width=0.48\textwidth,trim=0cm 0cm 0cm 0cm]{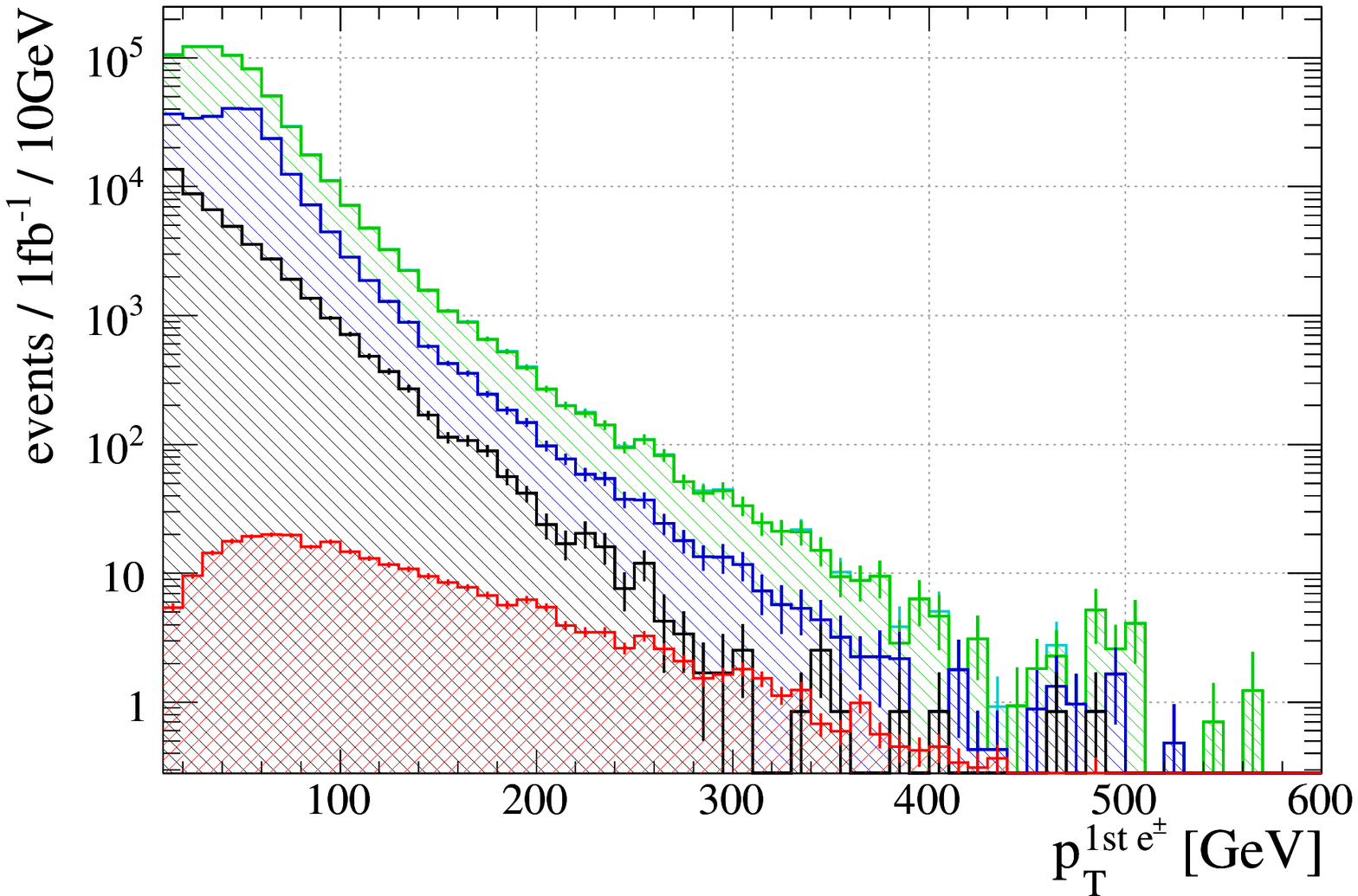}
  }
  \subfigure[\label{Fig:BC1-electron-pt-b}$\pT$ distribution of the 2nd hardest electron.]{
    \includegraphics[width=0.48\textwidth,trim=0cm 0cm 0cm 0cm]{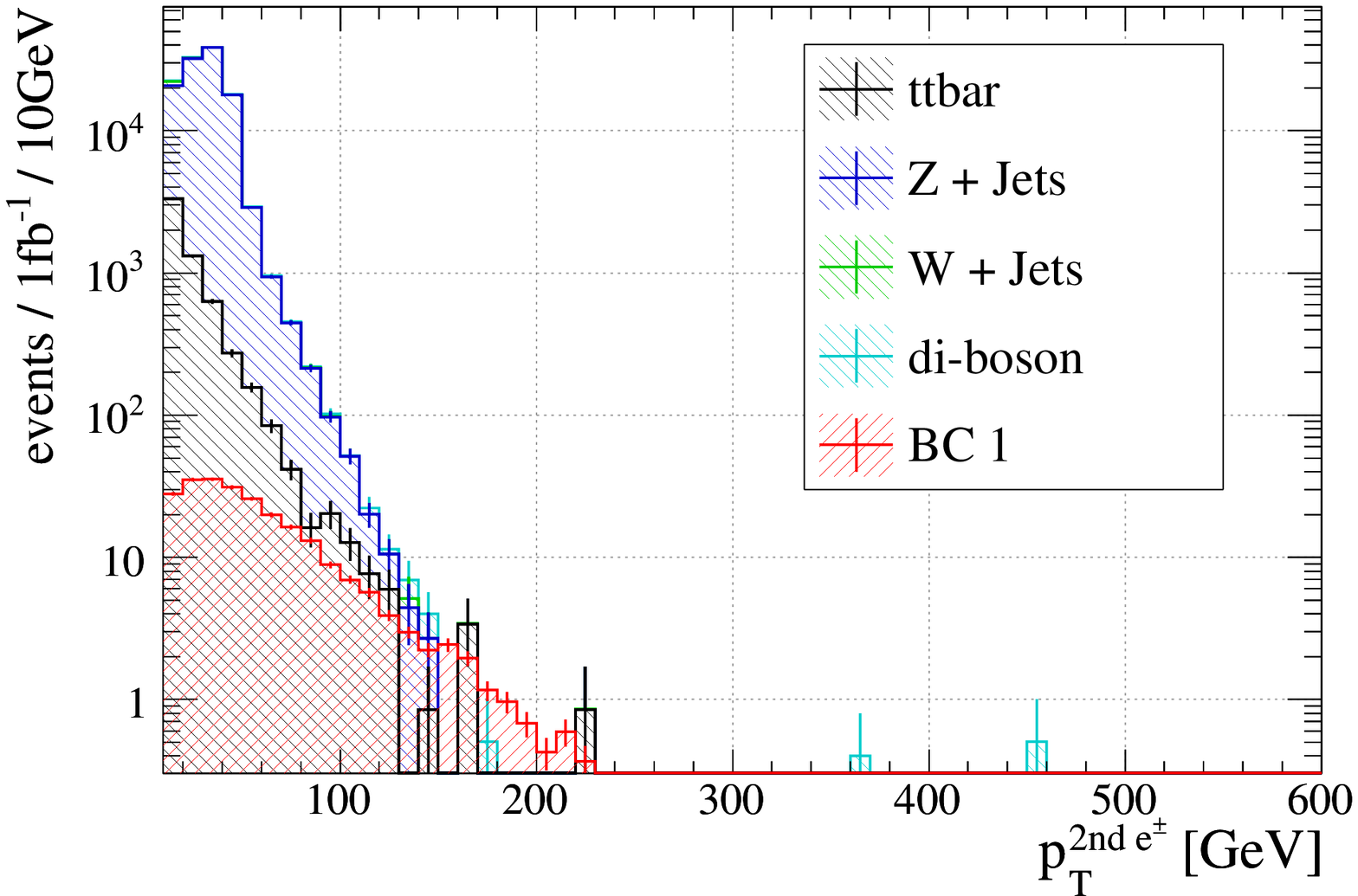}
  }
    \caption{\label{Fig:BC1-electron-pt}$\pT$ distributions of the two hardest electrons  after object selection cuts and overlap removal. 
}
\end{figure*}

\begin{figure*}[t!]
  \subfigure[$\pT$ distribution of the hardest muon]{
    \includegraphics[width=0.48\textwidth,trim=0cm 0cm 0cm 0cm]{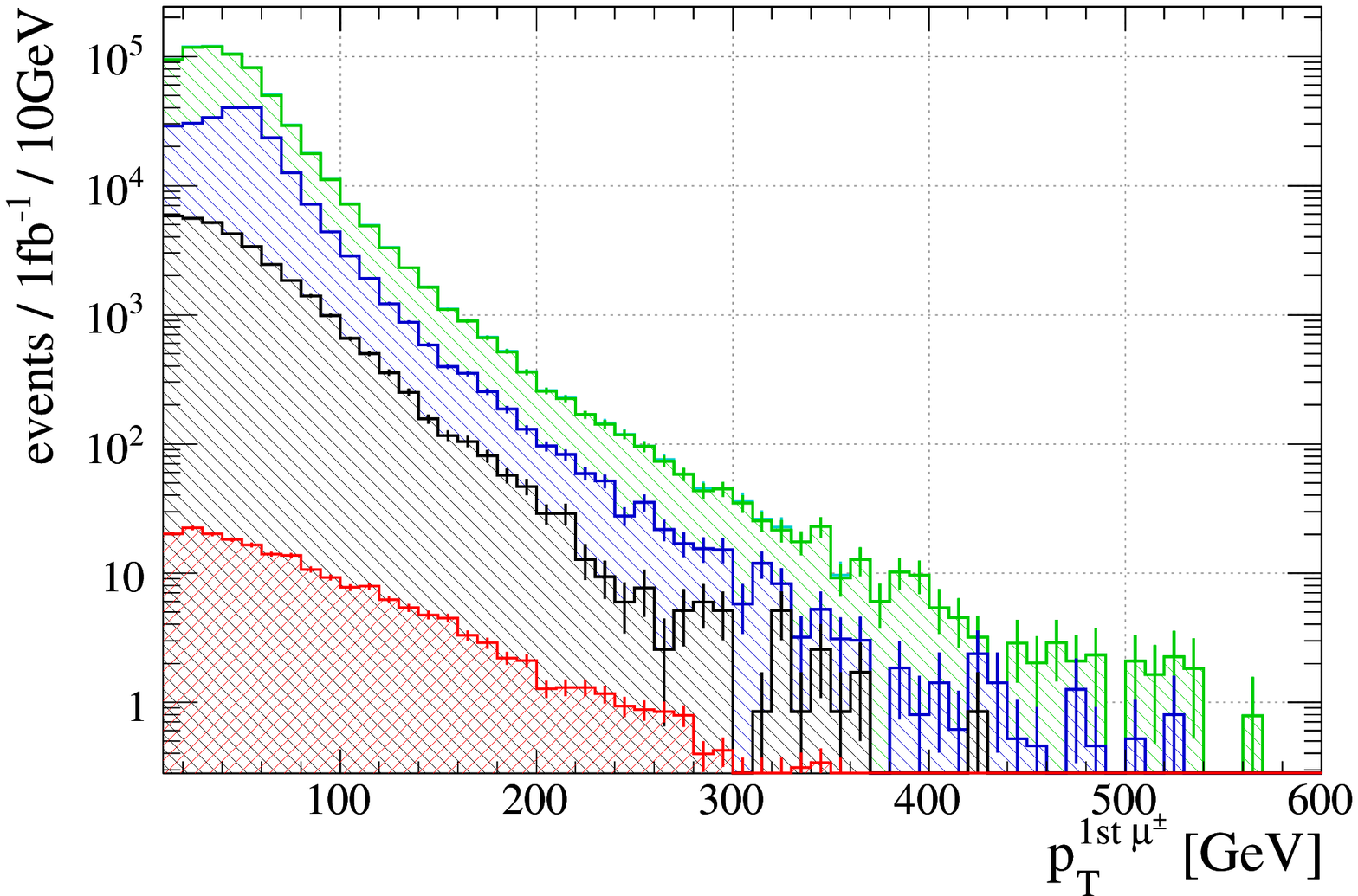}
  }
  \subfigure[$\pT$ distribution of the 2nd hardest muon]{
    \includegraphics[width=0.48\textwidth,trim=0cm 0cm 0cm 0cm]{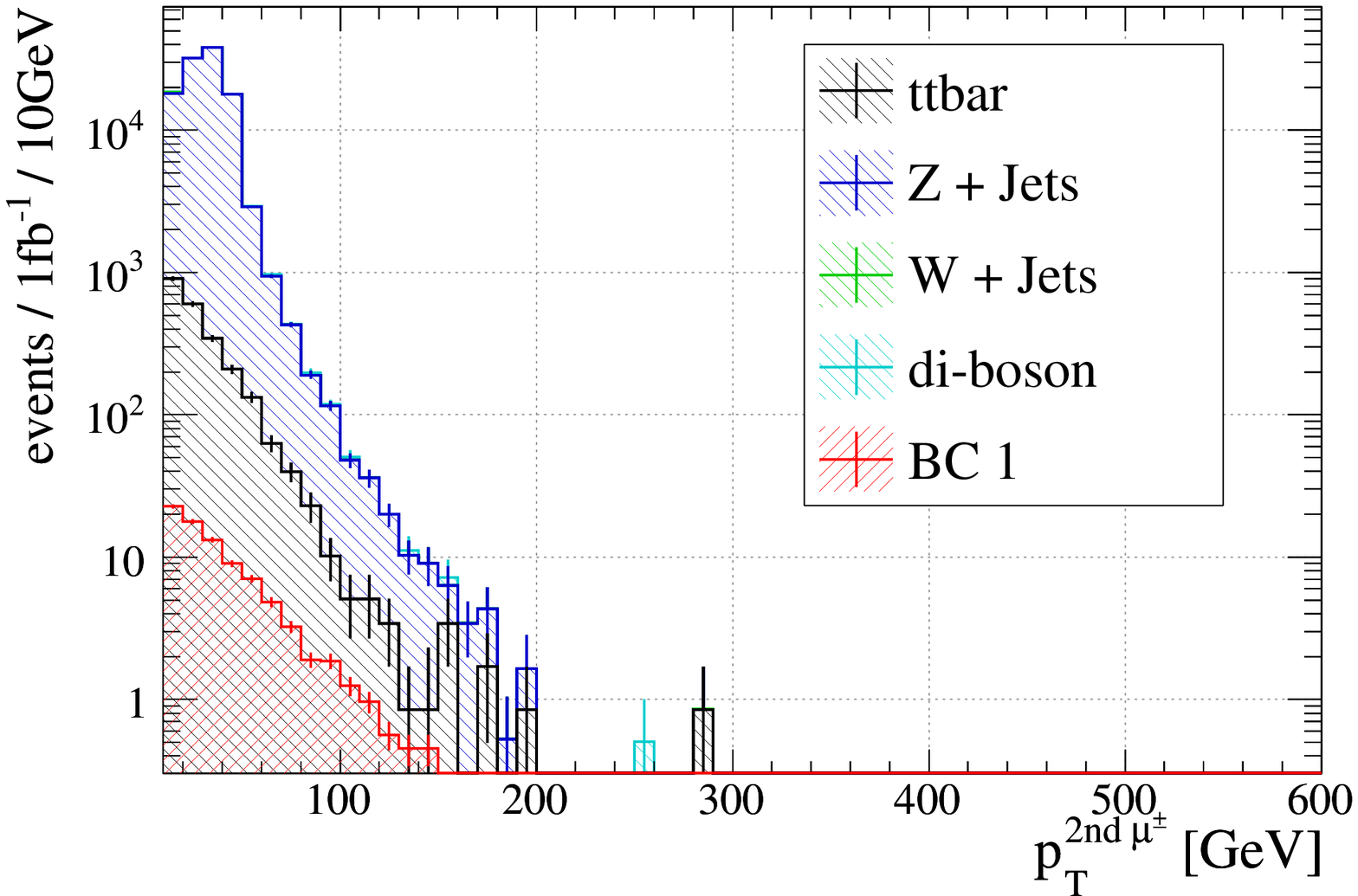}
  }
    \caption{\label{Fig:BC1-muon-pt}$\pT$ distributions of the two hardest muons  after object selection cuts and overlap removal. 
}
\end{figure*}

We see in Fig.~\ref{Fig:BC1-jet-pt} that the signal over
background ratio improves if one only allows for very hard jets.
This is also reflected by the visible hadronic mass, $HT^{\prime} = \sum_\text{jet 1-4} \pT$,
which is the scalar sum of the transverse momenta of the four hardest
jets. The respective distribution is given in
Fig.~\ref{Fig:BC1-HTprime}.  Note, that this cut alone (no
actual cut is defined at this point) would still give an overwhelming
amount of QCD background events.  We thus need to also make use of the
charged leptons in the final state.

\mymed

The most striking signature of the BC1 scenario is the large number of
electrons and muons in the final state. We show the respective
multiplicities in Fig.~\ref{Fig:BC1-n-particles-b} and
Fig.~\ref{Fig:BC1-n-particles-c}.  Even with an integrated luminosity
of only 1~$\ifb$ at $\sqrt{s}=7$ TeV, we expect several dozen events
with four or more electrons in the final state!

\mymed
 
The signal distribution of the number of reconstructed electrons peaks
at 3 whereas for the muons it peaks at 1. This is
exactly the expected behavior from the parton-level signatures
reviewed in Sec.~\ref{Sec:stau-lsp-signatures}, \cf~especially
the row for $\lambda_{121}$ in Tab.~\ref{tab_stau_LSP_sig_LLE}. The
decay of two stau LSPs in a typical BC1 event leads at parton level to
2--4 electrons and 0--2 muons. Furthermore, additional electrons or
muons can arise from other particle decays; \eg~in BC1 from
leptonic tau decays.

\mymed

It is interesting to note that the ratio of the number of
reconstructed electrons and muons carries information about
the involved $\text{B}_3$ coupling. Comparing
Fig.~\ref{Fig:BC1-n-particles-b} with \ref{Fig:BC1-n-particles-c}, we
see that the ratio of the average number of electrons to muons is
roughly three. This is because $\lambda_{121}$ couples two lepton
superfields of the first generation to one of the second
generation. Thus, stau LSP decays produce more electrons than
muons, \cf~Tab.~\ref{tab_stau_LSP_sig_LLE}. If we
instead had $\lambda_{122} \not = 0$, the situation would be reversed
(modulo differences due to the reconstruction efficiencies).

\mymed

According to Fig.~\ref{Fig:BC1-n-particles-b}, it is possible to
obtain a nearly background free sample by requiring more than four
electrons in the final state! However, such a cut would also veto most
of the signal events and is therefore not well suited for a study of
early data.  Furthermore, electrons can easily be faked by early
photon conversions or low multiplicity jets.  Therefore, we demand (as
a cut in the next section) ``only" the presence of at least one
electron with $\pT>32$ GeV, another electron with $\pT>7$ GeV and at
least one muon with $\pT>40$ GeV in the final state.  It is
hard for the QCD background to fake an electron and a muon at
the same time. In addition only the leptonically decaying $W$s from
$t\bar t$ production, leptonically decaying taus from $Z \rightarrow
\tau \tau$ production, and di-boson production can lead to an electron
{\it and} a muon at parton level. We thus expect a strong
suppression of the background from this cut, \cf~Tab.~\ref{Tab:CutFlowBC1}.

\mymed
 
We display the $\pT$ distributions of the two hardest electrons in
Fig.~\ref{Fig:BC1-electron-pt} and those of the two hardest muons in
Fig.~\ref{Fig:BC1-muon-pt}.  The $\pT$ distribution of the background
falls off more rapidly than the signal. In
Fig.~\ref{Fig:BC1-electron-pt-b}, we see that for $\pT>150\,$ GeV the
signal even dominates over the background.

\mymed
\begin{figure}[t!]
    \includegraphics[width=0.48\textwidth,trim=0cm 0cm 0cm 0cm]{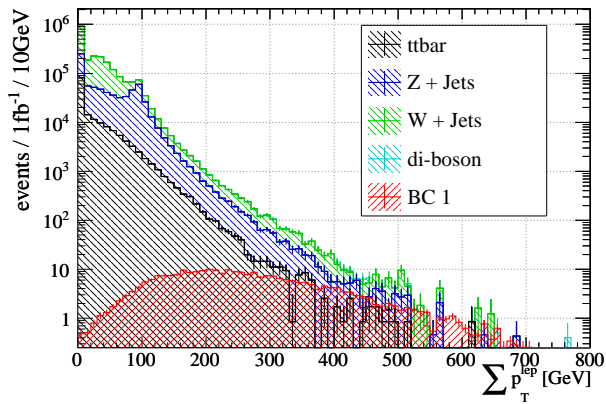}
    \caption{\label{Fig:BC1-pTSumLep} The number of events per
$1\,\mathrm{fb}^{-1}$ as a function of the scalar sum of the
transverse momenta of electrons and muons, $\sum \pT^{\ell}$, after
object selection cuts and overlap removal.  }
\end{figure}
Due to their large multiplicity and large transverse momenta,
leptons beyond the sub-leading electron and leading muon can
contribute significantly to the energy deposition of all leptons. The
cut $\sum \pT^{\ell}>230\GeV$, which we employ in the next
section, accounts for the fact that the signal lepton $\pT$s are on
average larger than the background. The lepton momenta are large
because they lie at the end of a decay chain of heavy SUSY particles,
\cf~ Tab.~\ref{App:BRs-BC1}.  Note, that $\sum \pT^{\ell}$ is
the $\pT$ sum of all electrons and muons.  The distribution is shown
in Fig.~\ref{Fig:BC1-pTSumLep}.

\mymed

\begin{figure*}[t!]
  \subfigure[$\pT$ distribution of the hardest tau]{
    \includegraphics[width=0.48\textwidth,trim=0cm 0cm 0cm 0cm]{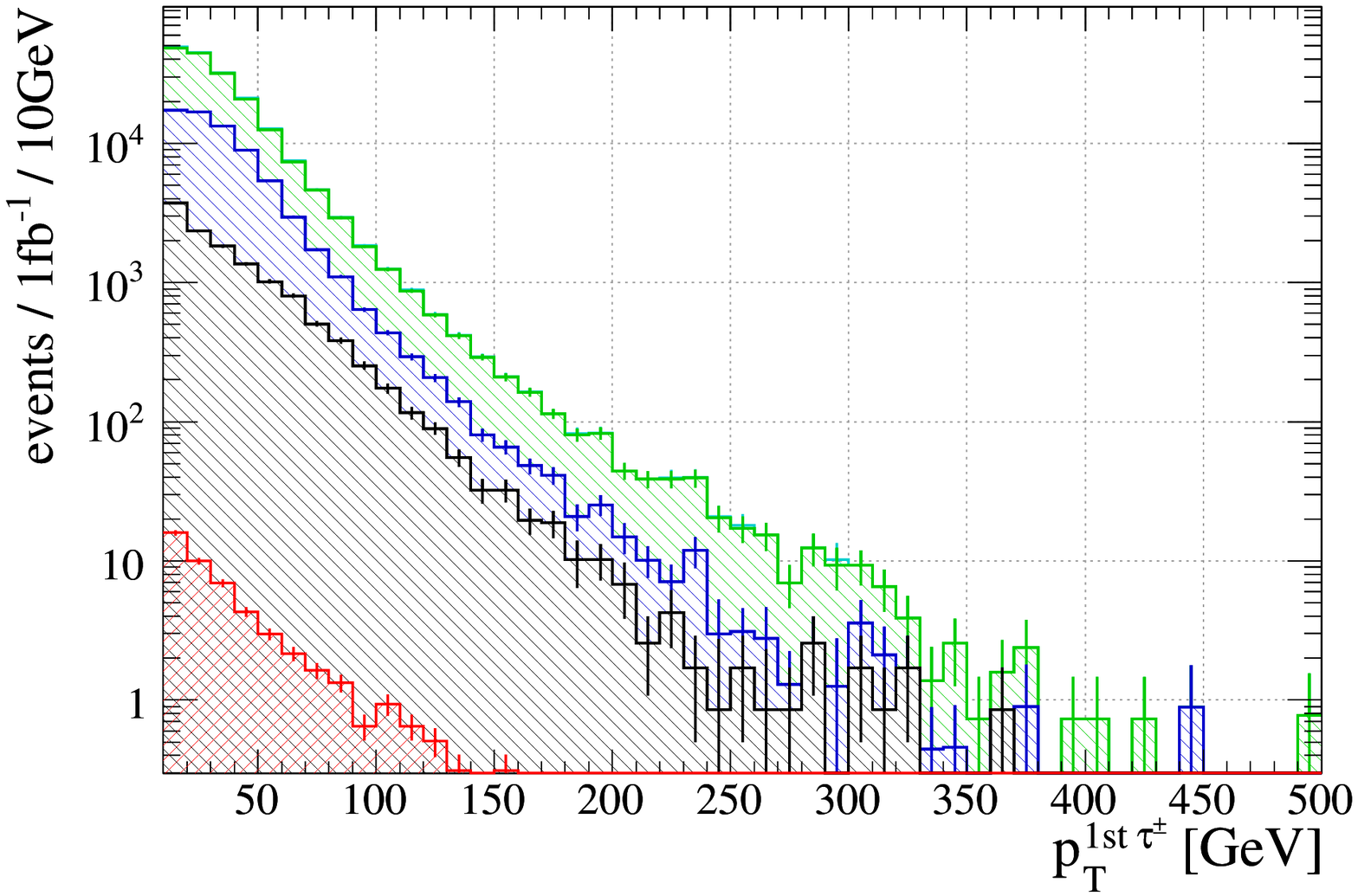}
  }
  \subfigure[$\pT$ distribution of the 2nd hardest tau]{
    \includegraphics[width=0.48\textwidth,trim=0cm 0cm 0cm 0cm]{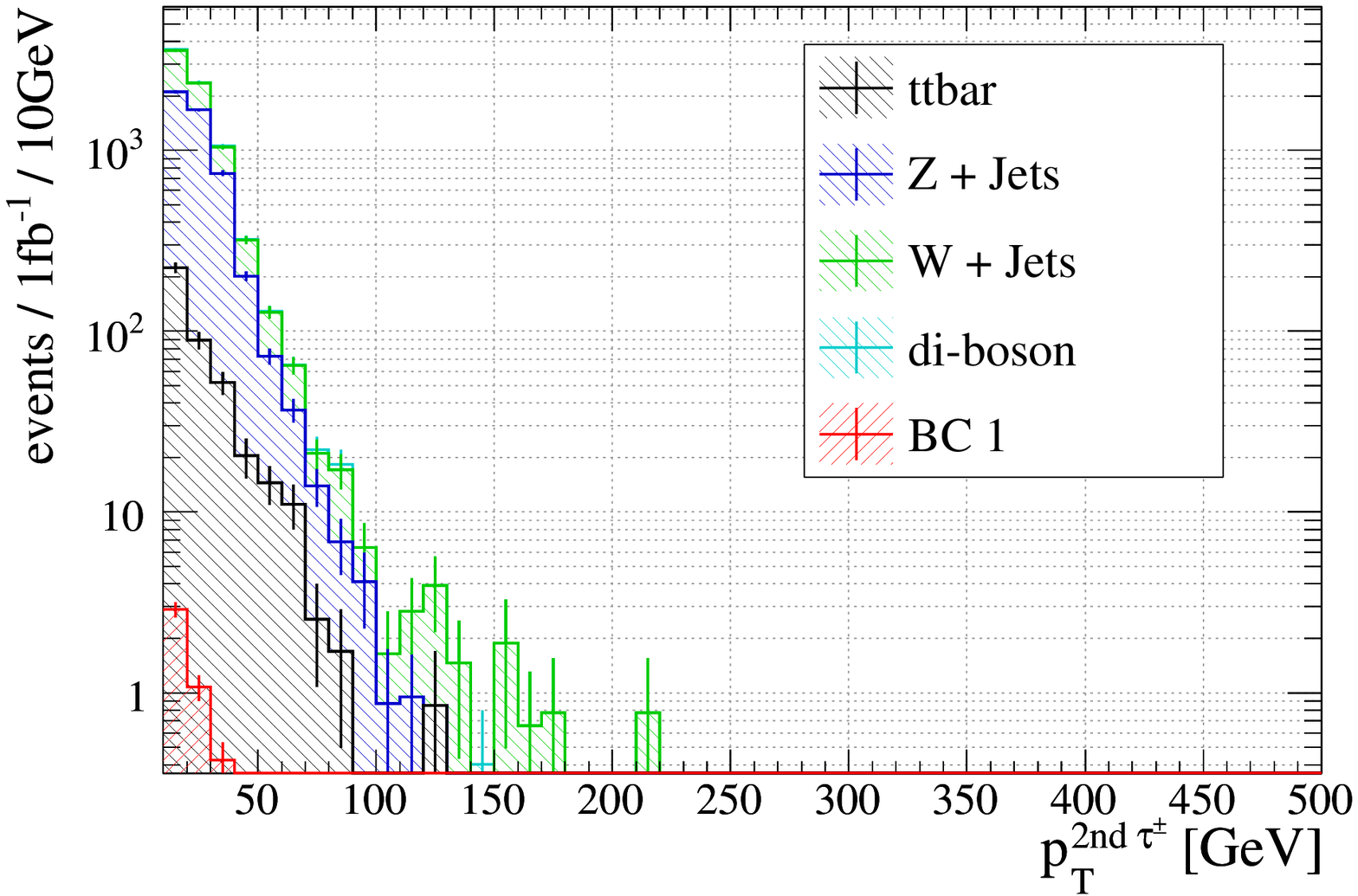}
  }
    \caption{\label{Fig:BC1-tau-pt} 
(Visible) $\pT$ distributions of the two hardest identified taus after object selection cuts and overlap removal. 
}
\end{figure*}
We show in Fig.~\ref{Fig:BC1-n-particles-d} the multiplicity of
reconstructed (hadronically decaying) taus. The signal distribution
peaks at 0. At first glance this is surprising, because from
Tab.~\ref{tab_stau_LSP_sig_LLE} we expect 4 taus in most of the BC1
SUSY events.

\mymed

However, with respect to tau identification (ID), the topologies in
BC1 (\cf~Sec.~\ref{Sec:stau-lsp-signatures}) are special. Due
to a large number of taus and jets from the SUSY decay chains,
overlaps between different tau jets and other jets make standard tau
ID via its hadronic decays very difficult. The low visible momentum of
many tau leptons in this scenario complicates the tau lepton
identification further.  As we show in more detail in
Sec.~\ref{tauID}, tau ID in BC1 has an efficiency of at best $20\%$-$
30\%$. The exact number depends on the working point on the
efficiency-vs-rejection curve of the tau ID algorithm, that is used.
Furthermore, we observe a strong dependence on the nature of the fast
detector simulation used, \ie~{\tt Delphes} or {\tt PGS4}
(\cf~Fig.~\ref{Fig:TauEfficiencyAndFakeRate}).  In any case the ID
efficiency is expected to be a factor of 2--3 smaller in BC1, than
\eg~in $Z\to\tau\tau$ events, even for tau leptons of the same
momentum.

\mymed

Although the number of parton-level taus is much larger in BC1
compared to the SM backgrounds, this does no longer hold for the
reconstructed taus. This can also be seen in
Fig.~\ref{Fig:BC1-tau-pt}, where we display the (visible) $\pT$ of the
two hardest identified taus. Even for large momenta, the background
always exceeds the signal. Naively, we would expect the contrary.
Like the electrons, Fig.~\ref{Fig:BC1-electron-pt}, the taus result
from the decay chains of heavy SUSY particles.
In addition, the tau leptons in BC1 are mostly very soft, which
reduces the ID efficiency further. Therefore, we do not employ the
taus to improve the signal to background ratio.

\mymed

\begin{figure}[t!]
    \includegraphics[width=0.48\textwidth,trim=0cm 0cm 0cm 0cm]{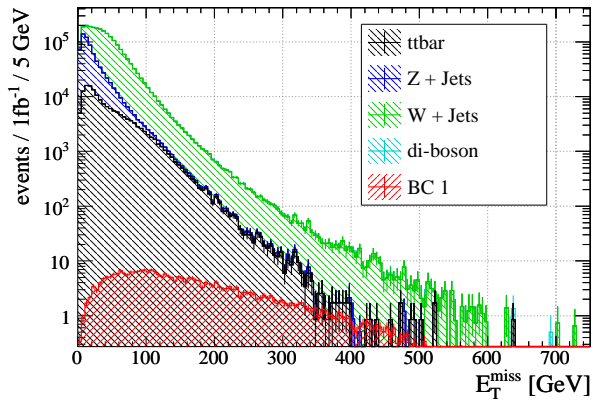}
    \caption{\label{Fig:BC1-ETmiss} Missing transverse energy, $\met$,
distribution after object selection cuts and overlap removal.
}
\end{figure}

In Fig.~\ref{Fig:BC1-ETmiss} we present the missing transverse energy
distribution for the signal and the backgrounds. Even though we
investigate a ${\text{B}_3}$ scenario, where the LSP does not escape
detection, the missing transverse energy, $\met$, can be significant
in BC1. For example, a cut on the missing energy of $\met>400$ GeV
improves the signal over $t \bar t$ background ratio from initially
$\mathcal{O}(0.001)$ to $\mathcal{O}(1)$, \cf~Fig.~\ref{Fig:BC1-ETmiss}.

\mymed

The reason for this are the neutrinos from the stau LSP decays 
\[\eqalign{\ninoone\ra &\stau_1+\tau\cr &\dk &\tau \ell^+ \ell^- \nu}\] 
and the neutrinos from the successive $\tau$ decays.
However, we will not explicitly cut on $\met$ in the event selection in order to keep the analysis 
complementary to searches for $R$-parity conserving SUSY.

\mymed

It should be kept in mind, that the discovery of a $\met$ signal does not necessarily
contradict $\rpv$ models. $\met$ alone is not sufficient to distinguish $R$-parity
violating from $R$-parity conserving SUSY.
Further observables, like kinematic edges, will be needed
to gain insights in the SUSY mass spectrum. Finally, a lepton (linear) collider may be needed
to clarify the nature of the LSP and the source of the missing energy.

\section{Discovery Potential and Mass Reconstruction}
\label{Sec:Dicov-Potential}

\subsection{Cut Selection and Significances}
\label{Sec:BC1-cuts}

We now employ in addition to the pre-selection cuts on reconstructed
objects (see Sec.~\ref{sim_and_select}) further cuts in order to
increase the signal significance and the signal over background
ratio. We show, that with an integrated luminosity of only
200~$\ipb$ at $\sqrt{s}=7$~TeV a discovery of the benchmark scenario
BC1 is possible.

\mymed

From the kinematic properties presented in
Sec.~\ref{Sec:kinematics-BC1}, we implement the following
cuts:
\begin{itemize}
\item $\pT(1\text{st}\; \mu^\pm)>40 \GeV$. We demand at least one muon
with $\pT>40$ GeV in the final state.
\item $\pT(1\text{st}\; e^\pm)>32 \GeV$. We demand an electron with
$\pT>32$ GeV in the final state.
\item $\pT(2\text{nd}\; e^\pm)>7 \GeV$. We demand at least a second
electron in each event. Note, that $\pT>7$ GeV corresponds to the
electron pre-selection cut, \cf~Tab.~\ref{selection_cuts}.
\item $\sum \pT^{\ell}>230 \GeV$. We demand the $\pT$ sum of all
electrons and muons to be larger than 230~GeV.
\item $HT^\prime>200/300/400 \GeV$. We employ also different cuts on
the $\pT$ sum of the four hardest jets, namely 200 GeV, 300 GeV and
400 GeV, respectively.
\end{itemize}     

The cut selection was optimized with the help of the TMVA toolkit
 \cite{TMVA}. The cuts were chosen so as to improve the signal to background
 ratio, based on simulated annealing \cite{metropolis:annealing}. This 
 was iterated with different sets of event selection variables, 
 while some cuts were left fixed. We found that slightly
 different cuts increase the signal significance marginally for our
 Monte Carlo samples, but only at the cost of a higher risk to accept
 QCD events, which we could not simulate.  Several other variables,
 like the $\ell^+\ell^-$ invariant mass as $Z$-veto or the $W$
 transverse mass, were tested as well. However, they turned out to be
 of less relevance in combination with the variables finally used.
 
\mymed

\begin{sidewaystable}
\begin{ruledtabular}
\begin{tabular}{l|*{6}{D{.}{.}{1}@{$\pm$}D{.}{.}{1}|}*{1}{d|}d}
cut                          & \multicolumn{2}{c|}{$\ttbar$} & \multicolumn{2}{c|}{$Z+$jets} & \multicolumn{2}{c|}{$W+$jets} & \multicolumn{2}{c|}{di-boson} &  \multicolumn{2}{c|}{all SM} & \multicolumn{2}{c|}{BC~1} &  \multicolumn{1}{c|}{$S/\sqrt{B}$} &  \multicolumn{1}{c}{$Z_0$} \\
\hline
                             before cuts &   155~500 & 416   &   556~760 & 681 &  1~505~250 & 1127 &    40~720 & 174   &  2~258~230 & 1392 &  
 282.8 & 2.8 &        0.2 & \multicolumn{1}{c}{--}\\
     $\pT(1\text{st}\; \mu^\pm)>40 \GeV$ &    16~745 & 135   &   119~984 & 313 &   180~009 & 375   &     3~236 & 49    &   319~975 & 510   &      141.6 & 2.0 &        0.3 & \multicolumn{1}{c}{--}\\
       $\pT(1\text{st}\; e^\pm)>32 \GeV$ &     1~492 & 40    &       46.9 & 5.9 &      102.4 & 9.0 &      196.2 & 12.7 &     1~837 & 43    &      125.9 & 1.9 &        2.9 &  \multicolumn{1}{c}{--}\\
        $\pT(2\text{nd}\; e^\pm)>7 \GeV$ &      165.6 & 14.4 &        2.2 & 1.3 &        1.9 & 1.3 &       15.2 & 2.7 &      184.9 & 14.8 &      
113.7 & 1.8 &        8.4 &      0.7\\
              $\sum \pT^{\ell}>230 \GeV$ &       13.6 & 4.2 & \multicolumn{2}{c|}{$\lesssim1.0$} &        1.0 & 1.0 &        0.5 & 0.5 &       
15.1 & 4.3 &       85.7 & 1.6 &       22.0 &        4.9 \\
                    $HT^\prime>200 \GeV$ &        5.1 & 2.1 & \multicolumn{2}{c|}{$\lesssim1.0$} &        1.0 & 1.0 &
\multicolumn{2}{c|}{$\lesssim1.0$}&        6.1 & 2.3 &       60.3 & 1.3 &       24.3 &        6.4\\
                    $HT^\prime>300 \GeV$ &        3.4 & 1.7 & \multicolumn{2}{c|}{$\lesssim1.0$} &
\multicolumn{2}{c|}{$\lesssim1.0$}&\multicolumn{2}{c|}{$\lesssim1.0$}&        3.4 & 1.7 &       56.6 & 1.3 &       30.7 &        8.1\\
                    $HT^\prime>400 \GeV$ &\multicolumn{2}{c|}{$\lesssim1.0$}& \multicolumn{2}{c|}{$\lesssim1.0$} &
\multicolumn{2}{c|}{$\lesssim1.0$}&\multicolumn{2}{c|}{$\lesssim1.0$}&\multicolumn{2}{c|}{$\lesssim1.0$}&       52.6 & 1.2 &   \\
\end{tabular}
\end{ruledtabular}
\caption{\label{Tab:CutFlowBC1} Cut flow in BC1 at $\sqrt{s}=7\TeV$,
        scaled to an integrated luminosity of $\int L \text{d}t=
        1\,\ifb$.  The uncertainties include statistical errors
        only. The significance $Z_0$, defined in
        App.~\ref{diff_significances}, assumes a background
        uncertainty of 50\%.}
\end{sidewaystable}

The cut flow is given in Tab.~\ref{Tab:CutFlowBC1}
and visualized in Fig.~\ref{Fig:BC1-cutFlow}. We also quote the
signal to square root of background ratio, $S/\sqrt{B}$ (second last column), 
and the significance $Z_0$ \cite{Cousins:2008} (last column) assuming a relative 
background uncertainty of 50\%. For a definition of
$Z_0$ and related measures of significance see
App.~\ref{diff_significances}.

\mymed

We have chosen such a 
large systematic uncertainty for the SM backgrounds as a 
conservative approach, because uncertainties are expected
from jet and
lepton energy scales as well as the predictions of cross sections.
The ATLAS collaboration for example estimates the systematic uncertainties
for an integrated luminosity of $1\ifb$ to be 50\% for the background from QCD multijet events
and 20\% for the background from $\ttbar$, $W+$jets, $Z+$jets, and $W$ pairs
in the context of SUSY searches \cite{ATLAS:CSCbook}. In our analysis we did not consider QCD multijet
background explicitly. Additionally the systematic uncertainties will be larger
for the very first searches and we therefore use the more conservative estimate of 50\% uncertainty.
The uncertainties in Tab.~\ref{Tab:CutFlowBC1}
stem from the limited statistics of the simulated events 
(see Tab.~\ref{Tab:MC_samples}). Note that for $W$+jets we only show 
events with at least two jets at parton-level \cite{Wjets}.

\begin{figure}[t!]
    \includegraphics[width=0.5\textwidth,trim=1.3cm 0cm 0cm 0cm]{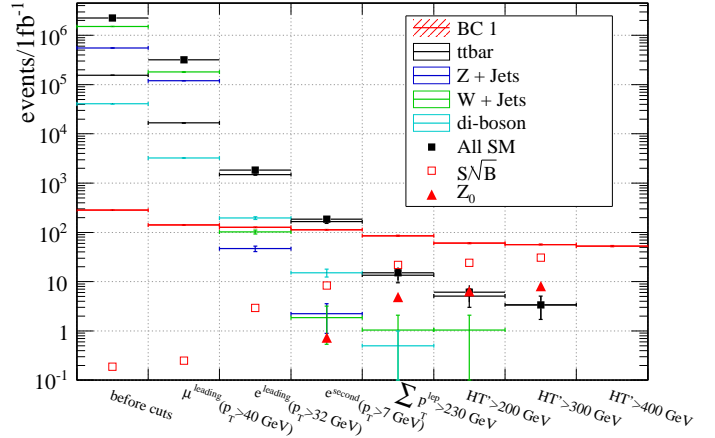}
    \caption{\label{Fig:BC1-cutFlow} Cut flow in BC1 at
     $\sqrt{s}=7\TeV$.  As in Tab.~\ref{Tab:CutFlowBC1}, entries are
     scaled to an integrated luminosity of $\int L \text{d}t =
     1\,\ifb$ and error bars include statistical uncertainties
     only. The significance $Z_0$, defined in
        App.~\ref{diff_significances}, assumes a background uncertainty of
     50\%.}
\end{figure}

\mymed

Employing only selection cuts (second row in Tab.~\ref{Tab:CutFlowBC1}), 
we observe that the number of SM background events (sixth column) 
is several orders of magnitude larger than the number of signal events
(seventh column) \footnote{With only selection cuts, the SM backgrounds 
would be dominated by QCD processes, \cf~Sec.~\ref{bkg_BC1}.}. 
Without any cuts, we expect at $\sqrt{s}=7\TeV$ roughly 300 signal events
for an integrated luminosity of 1~$\ifb$.

\mymed

We obtain a great improvement of the signal to background ratio to
$\mathcal{O}(0.1)$ by demanding one hard muon ($\pT>40$~GeV) and one
hard electron ($\pT>32$~GeV) in the final state.  At this stage the
main backgrounds stem from leptonic decays of $t\bar t$ (second
column) and from leptonic decays of di-bosons (fifth column).  These
processes can produce an electron and muon at parton level. In
addition, we also get some contributions from the $W+$jets and
$Z+$jets backgrounds.  Here, the second charged lepton of the first or
second generation stems mainly from jets that are misidentified as a
charged lepton.  After the first two cuts, we already obtain a
$S/\sqrt{B}$ ratio of roughly three.  However, if one takes into
account systematic uncertainties, that are expected to be large at
the early LHC run, one cannot observe a clear signal,
\cf~the last column in Tab.~\ref{Tab:CutFlowBC1}.

\mymed

Demanding at least a second electron in the final state (fourth row) the 
situation further improves. We now have a signal to background ratio
of roughly one and a $S/\sqrt{B}$ ratio of 8.4. Therefore, by demanding two 
(hard enough) electrons and one muon in the final state, a signal of the scenario
BC1 might already be visible in early LHC data, even though the significance $Z_0$
is still small without further cuts.

\mymed
 
As we have shown in Figs.~\ref{Fig:BC1-electron-pt} and
\ref{Fig:BC1-muon-pt}, the electrons and muons of the signal events
also have on average larger transverse momenta than the charged
leptons from SM processes. We therefore demand in the sixth row in
Tab.~\ref{Tab:CutFlowBC1} the $\pT$ sum of the electrons and muons to
be larger than 230 GeV; {\it cf.} Fig.~\ref{Fig:BC1-pTSumLep}.  With
this cut we obtain a number of signal events that is roughly six times
larger than the number of background events. A clear signal of BC1
should now be visible, because $S/\sqrt{B}=22$ and $Z_0=4.9$.

\mymed

Up to now, we have only employed the electrons and muons that stem
mainly from the stau LSP decays. However, in a typical SUSY event at
the LHC, we will first produce a pair of strongly interacting
sparticles. These sparticles then cascade decay down to the LSP
producing hard jets in the final state. As can be seen in
Fig.~\ref{Fig:BC1-jet-pt}, these jets are in general harder than the
jets from SM backgrounds.  Therefore, we demand in
Tab.~\ref{Tab:CutFlowBC1} as the last cut the $\pt$ sum of the four
hardest jets, Fig.~\ref{Fig:BC1-HTprime}, to be larger than 200~GeV
(third last column), 300~GeV (second last column), or 400~GeV
(last column), respectively.  We show different cuts for $HT^\prime$ in
order to show how the signal over background ratio improves with
harder cuts on $HT^\prime$.

\mymed

For $HT^\prime>200$~GeV, we already obtain a background sample without
di-boson events (fifth row). Because we only have produced a finite
number of background events (see Tab.~\ref{Tab:MC_samples}), we mark a
background free sample by ``$\lesssim 1.0$". If we harden the cut
further to $HT'>300$~GeV, we also veto all $Z+{\rm jets}$ and $W+{\rm
jets}$ events.  Finally for $HT'>400$~GeV, we obtain even a background
free sample!  Note that at the same time, the number of signal events
is only reduced by 30\% ($HT'>200$~GeV), 34\% ($HT'>300$~GeV) and 39\%
($HT'>400$~GeV), respectively, compared to no cut on $HT'$.

\mymed

We thus can further improve the significances $S/\sqrt{B}$ and $Z_0$ 
with a cut on $HT'$. For example, for $HT'>300$~GeV, we obtain $S/\sqrt{B}=24$
and $Z_0=8.1$. A signal is clearly visible. Note, that we cannot give 
meaningful numbers for the significances for $HT'>400$~GeV, because
our background samples are not large enough. In addition, even if we cut away
all the backgrounds in Tab.~\ref{Tab:CutFlowBC1} we still might have
some small (unknown) QCD backgrounds. Therefore, we employ the
$HT'>300$~GeV cut for our parameter scans in the next section in order
to get meaningful results.
It is the main purpose of the $HT'>400$~GeV
cut to show that a (nearly) background free sample is possible.

\mymed

We conclude, that the cuts presented in Tab.~\ref{Tab:CutFlowBC1}
allow a discovery of the benchmark scenario BC1 with early LHC data.
Scaling down the luminosity, we have found that even with an
integrated luminosity of $200\,{\rm pb^{-1}}$ we still get $Z_0>5$! We
also want to point out that we only used electrons, muons and
jets. Our analysis does thus not rely on the reconstruction and
identification of missing energy, $b$--jets and taus, which might be
more difficult with early data. In the next section we show, that
the cuts work also quite well beyond BC1.

\subsection{Discovery Potential with early LHC Data}
\label{discovery_potential}

We now extend our previous analysis to a more extensive
parameter region. We still restrict ourselves to early LHC data.  We
focus on regions of the ${\rm B}_3$ mSUGRA parameter space with
different mass spectra compared to BC1 but with the same non-vanishing
lepton number violating coupling, \ie~$\lambda_{121}=0.032$ at $M_{\rm
GUT}$.  For that purpose we perform a two dimensional parameter scan
in the $M_{1/2}$--$\tan \beta$ plane around BC1 ($M_{1/2}=400$ GeV,
$\tan \beta=13$).

\mymed

We have chosen a scan in $M_{1/2}$ and $\tan \beta$ for the following
reasons.  Due to the RGE running, every sparticle mass
increases with increasing $M_{1/2}$, especially the masses of the
strongly interacting sparticles
\cite{Bernhardt:2008jz,Dreiner:2008ca,Drees:1995hj},
\cf~Fig.~\ref{Fig:gluino_mass} and Fig.~\ref{Fig:squark_mass}. By
varying $M_{1/2}$ we can thus investigate the discovery potential as a
function of the SUSY mass scale.

\mymed

In contrast, changing $\tan\beta$ does not affect most of the
sparticle masses (see Fig.~\ref{Fig:M12_tanb_plane}) and therefore
also leaves the total sparticle production cross section
unchanged. This can be seen in Fig.~\ref{Fig:parameterScan-BC1-xsec},
where we present the total SUSY particle pair production cross section
(at $\sqrt{s}=7 \,{\rm TeV}$) as a function of $M_{1/2}$ and $\tan
\beta$.  For example, increasing $M_{1/2}$ from 320 GeV to 500 GeV
reduces the cross section from $2\,$pb to $0.1\,$pb. Note that
for $M_{1/2}=320$~GeV ($M_{1/2}=500$~GeV) we have squark masses mostly
around 680~GeV (1~TeV) and a gluino mass of roughly 760~GeV (1.1~TeV);
see Fig.~\ref{Fig:squark_mass} and Fig.~\ref{Fig:gluino_mass},
in App.~\ref{App:properties-BC12}. But changing $\tan \beta$
(for fixed $M_{1/2}$) leaves the cross section almost
unchanged.

\mymed

\begin{figure*}[tb!]
  \subfigure[\label{Fig:parameterScan-BC1-xsec}Signal cross section in
     pb at LHC at $\sqrt{s}=7$~TeV.]{
     \includegraphics[width=0.48\textwidth,trim=0cm 0cm 0cm
     0cm]{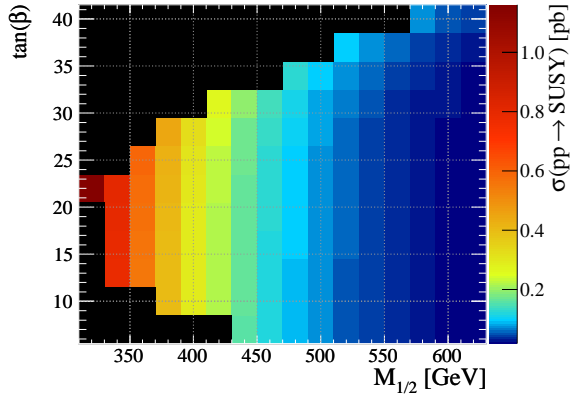} }
     \hfill \subfigure[\label{Fig:parameterScan-BC1-eff}Selection
     efficiency for the signal events at LHC at $\sqrt{s}=7$~TeV.]{
     \includegraphics[width=0.48\textwidth,trim=0cm 0cm 0cm
     0cm]{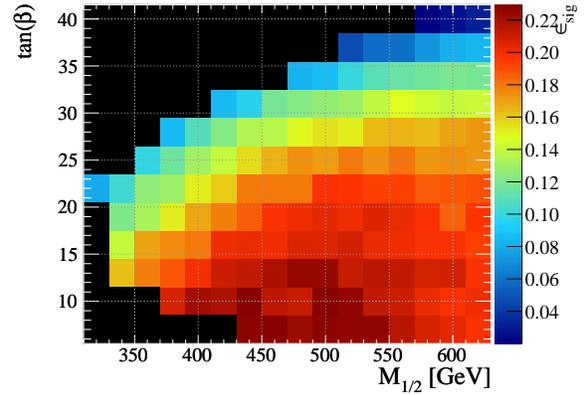} }
     \hfill \subfigure[\label{Fig:parameterScan-BC1-evts}Number of
     selected signal events assuming an integrated luminosity of
     $1\,{\rm fb}^{-1}$ at $\sqrt{s}=7$~TeV.]{
     \includegraphics[width=0.48\textwidth,trim=0cm 0cm 0cm
     0cm]{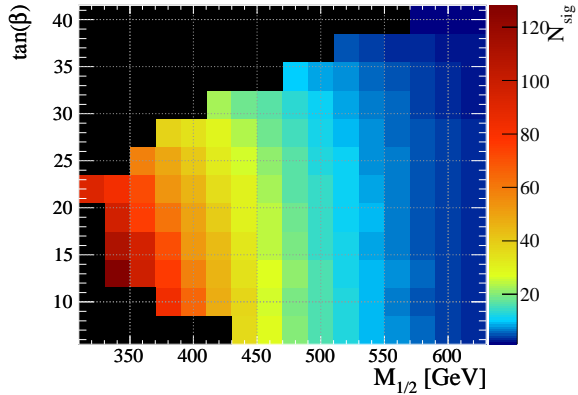} } \hfill
     \subfigure[\label{Fig:parameterScan-BC1-sig} Significance
     $S/\sqrt{B}$ assuming an integrated luminosity of $1\,{\rm
     fb}^{-1}$ at $\sqrt{s}=7$~TeV.]{
     \includegraphics[width=0.48\textwidth,trim=0cm 0cm 0cm
     0cm]{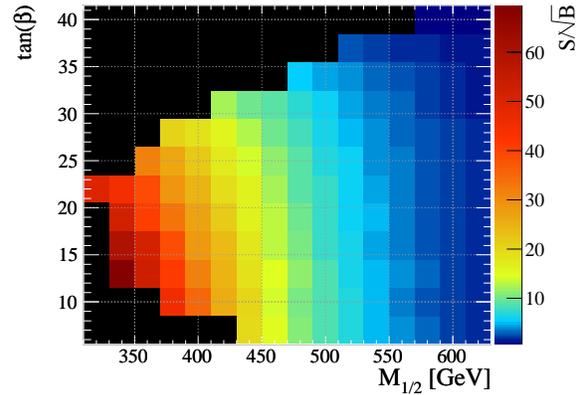} }
     \hfill \subfigure[\label{Fig:parameterScan-BC1-Z0}Significance
     $Z_0$ assuming an integrated luminosity of $1\,{\rm fb}^{-1}$ at
     $\sqrt{s}=7$~TeV.]{
     \includegraphics[width=0.48\textwidth,trim=0cm 0cm 0cm
     0cm]{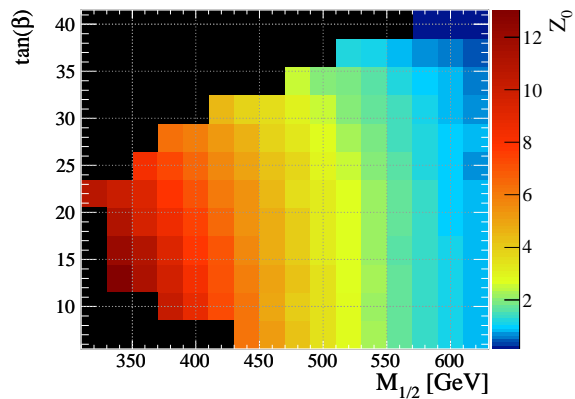} }
    \caption{\label{Fig:parameterScan-BC1} Parameter scans in the
   $M_{1/2}$--$\tan(\beta)$ plane.  The other mSUGRA parameters are
   those of BC1, \ie~$M_0=A_0=0$~GeV and ${\rm
   sgn}(\mu)=+1$. $Z_0$ is defined in
   App. \ref{diff_significances}.}
\end{figure*}

\begin{figure*}[tb!]
  \subfigure[\label{Fig:parameterScan-BC1-lumi}Estimated contours of
     minimum required integrated luminosity ($\ifb$) to reach
     $S/\sqrt{B} > 5$]{ \includegraphics[width=0.48\textwidth,trim=0cm
     0cm 0cm 0cm]{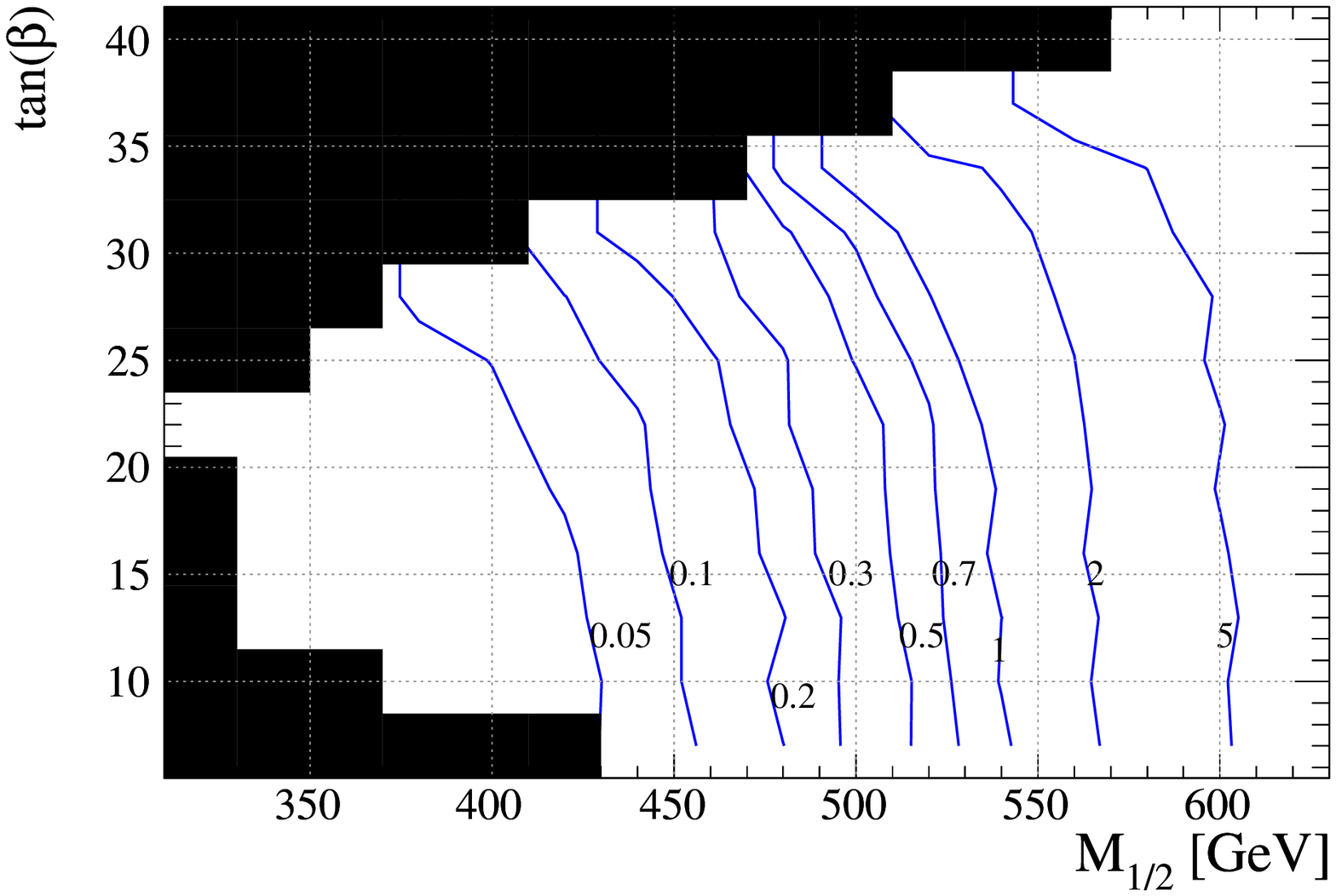}
     } \hfill
     \subfigure[\label{Fig:parameterScan-BC1-lumi-Z0}Estimated
     contours of minimum required integrated luminosity ($\ifb$) to
     reach $Z_{0} > 5$ assuming a relative background uncertainty of
     50\%]{ \includegraphics[width=0.48\textwidth,trim=0cm 0cm 0cm
     0cm]{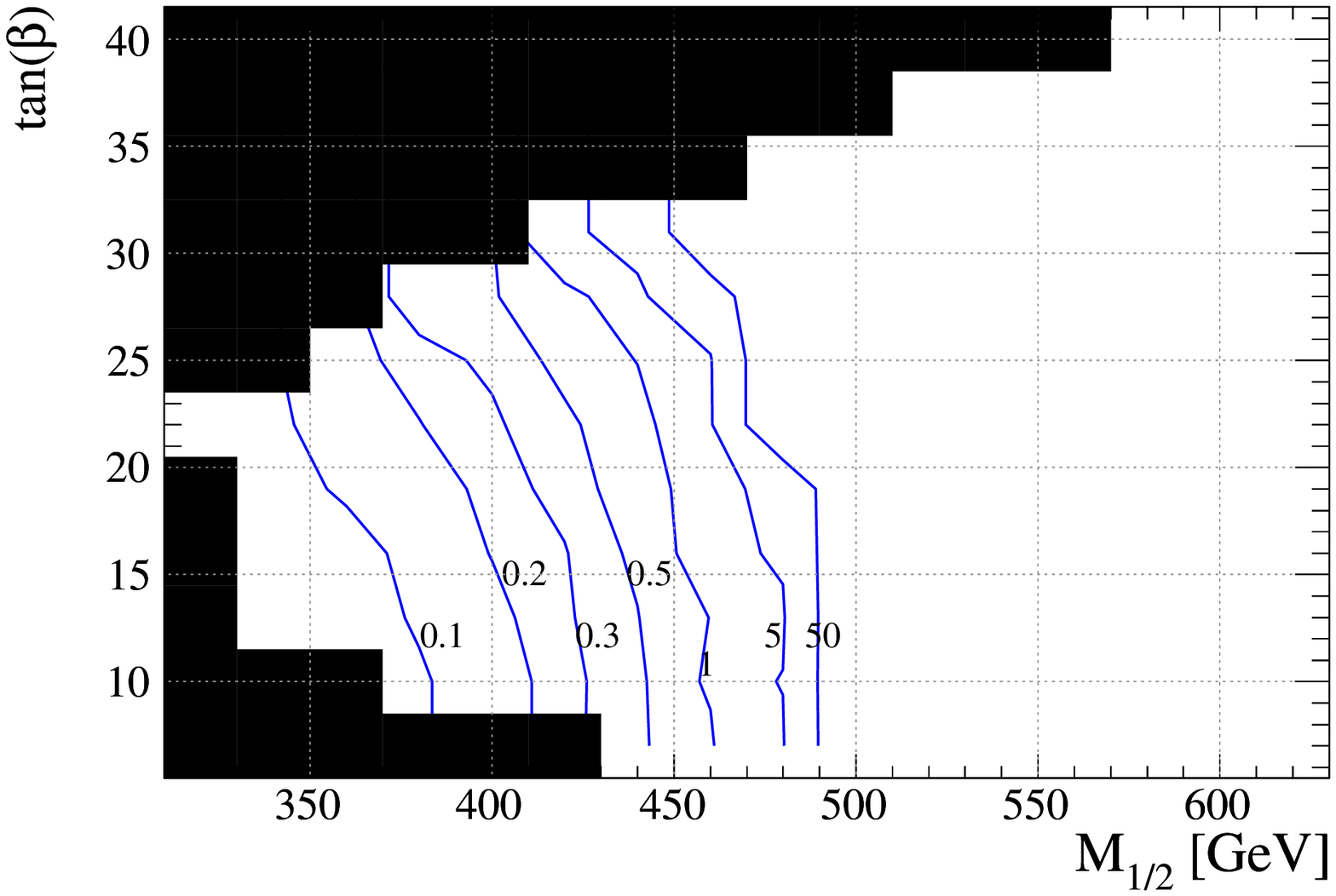} }
     \hfill
    \caption{\label{Fig:parameterScan-BC1-minlumi} Minimum required
    integrated luminosity for a discovery at $\sqrt{s}=7$~TeV
    without [Fig.~\ref{Fig:parameterScan-BC1-lumi}] and with
    [Fig.~\ref{Fig:parameterScan-BC1-lumi-Z0}] systematic
    uncertainties included. The parameters are as in
    Fig.~\ref{Fig:parameterScan-BC1}.}
\end{figure*}

However, increasing $\tan \beta$ decreases the stau LSP mass,
\cf~Fig.~\ref{Fig:stau_mass}. Increasing $\tan \beta$ increases the
tau Yukawa coupling and thus its (negative) contribution to the stau
mass from RGE running
\cite{Bernhardt:2008jz,Dreiner:2008ca,Drees:1995hj}. Additionally, a
larger value of $\tan \beta$ leads normally to a larger mixing between
the left- and right-handed stau \footnote{Beside the term proportional
to $\tan \beta$, the off diagonal element of the stau mass matrix
possesses a term proportional to a softbreaking trilinear
coupling. However, unless $A_0=\mathcal{O}({1 \, {\rm TeV}})$,
this term plays only a subleading role. We have always set
$A_0=0$ in our parameter scans.}.  This further reduces the mass of
the stau LSP. Therefore, with the help of $\tan \beta$ we can change
the kinematics of the LSP decay products and also the momentum of the
tau from the decay of the $\ninoone$ into the stau (and a tau). Note
that the mass of the $\ninoone$ is nearly independent of $\tan \beta$
(see Fig.~\ref{Fig:neut1_mass}), because it is bino-like around BC1.

\mymed

We show the results of the parameter scans in
Fig.~\ref{Fig:parameterScan-BC1}. The black region corresponds to
excluded parameter points, with tachyons
\cite{Allanach:2003eb} or which violate the Higgs mass or stau mass
bounds from LEP \footnote{Following Ref.~\cite{Allanach:2006st} we use
the lower stau mass bound from LEP obtained in the $R$-parity
conserving limit, because LEP did not explore the possibility of a
stau LSP decaying via a four-body decay.}, \ie~$m_{h^0}>90$--$114.4$~GeV 
\cite{Schael:2006cr,Barate:2003sz}, depending on the SUSY parameters 
\footnote{In order to calculate the correct Higgs mass bound we used 
the respective routine implemented in {\tt SOFTSUSY}.},
and $m_{\tilde{\tau}_1}>86$~GeV \cite{Allanach:2006st}. We lowered the
Higgs mass bound by 3~GeV to account for numerical uncertainties of
{\tt SOFTSUSY}
\cite{Allanach:2003eb,Allanach:2003jw,Degrassi:2002fi,Allanach:2004rh}.
Note, that most of the parameter points in
Fig.~\ref{Fig:parameterScan-BC1} are also consistent with the observed
anomalous magnetic moment of the muon, with ${\rm BR}(b\rightarrow s
\gamma)$ and with the upper bound on ${\rm BR}(B_s\rightarrow \mu^+
\mu^-)$ \cite{Allanach:2006st}.

\mymed

We present in Fig.~\ref{Fig:parameterScan-BC1-eff} the selection
efficiency for the signal events, \ie~the fraction of signal events
that pass all the cuts in Tab.~\ref{Tab:CutFlowBC1} with
$HT^{\prime}>300$~GeV. For the scenario BC1 we obtain an efficiency of
roughly $20\%$. Going beyond BC1, we can see in
Fig.~\ref{Fig:parameterScan-BC1-eff} that the fraction of signal
events that pass the cuts lies mostly between $10\%$ and $23\%$. We
conclude that the cuts work also quite well in other regions
of the stau LSP parameter space than BC1.

\mymed

Fig.~\ref{Fig:parameterScan-BC1-eff} also shows a correlation between
the selection efficiency and $\tan \beta$: For fixed $M_{1/2}$ the
efficiency decreases if $\tan \beta$ increases. This behavior can be
easily understood. As mentioned above, increasing $\tan \beta$
decreases the mass of the stau LSP. In this case, the decay products
of a light stau LSP have on average smaller momenta compared to a
heavy one.  Therefore, less electrons and muons from stau LSP decays
will pass the cuts in Tab.~\ref{Tab:CutFlowBC1}.

\mymed

This also explains why we very often obtain  
a better selection efficiency for larger values of $M_{1/2}$ 
(and fixed $\tan \beta$). Increasing $M_{1/2}$ increases the 
mass of all SUSY particles including the stau LSP. The SM particles 
from cascade decays and LSP decays have then on
average larger momenta and thus pass the cuts more
easily. However, when going to very large $M_{1/2}$
the efficiency can again decrease, because the strongly interacting 
particles are very heavy and thus their decay products are highly
boosted. The final state particles might then fail the isolation cuts.

\mymed

By multiplying the signal cross section
[Fig.~\ref{Fig:parameterScan-BC1-xsec}] with the signal efficiency
[Fig.~\ref{Fig:parameterScan-BC1-eff}], we obtain in
Fig.~\ref{Fig:parameterScan-BC1-evts} the number of signal events that
pass the cuts for an integrated luminosity of $1\,{\rm fb^{-1}}$ at
$\sqrt{s}=7$~TeV.  For a light spectrum, \ie~$M_{1/2}\lesssim
350$~GeV, we observe that 100 or more signal events pass the
cuts. If we go to a heavier spectrum ($M_{1/2} \gsim 500$~GeV), where
the squarks and gluinos masses are $\gsim 1$~TeV, we see that the
number of signal events reduces to 20 or less.  Although
$\mathcal{O}(10)$ signal events are enough to claim a discovery of new
physics, we do not have enough events for a reconstruction of
the sparticle masses, especially the stau LSP mass. We will address
this issue in Sec.~\ref{mass_reco}.

\mymed

We finally present the resulting significances: In
Fig.~\ref{Fig:parameterScan-BC1-sig} we give $S/\sqrt{B}$ as a naive
estimator and in Fig.~\ref{Fig:parameterScan-BC1-Z0} $Z_0$ as a more
realistic estimator. Note that $Z_0$ is defined in
App.~\ref{diff_significances}.

\mymed

With an integrated luminosity of $1\,{\rm fb^{-1}}$ at
$\sqrt{s}=7$~TeV, $S/\sqrt{B}$ suggests that stau LSP scenarios up to
$M_{1/2}\lesssim 540$~GeV can be discovered. Note that $M_{1/2} =
540$~GeV corresponds to squark masses [gluino masses] of 1.1~TeV
[1.2~TeV]; see Fig.~\ref{Fig:squark_mass}
[Fig.~\ref{Fig:gluino_mass}], in
App.~\ref{App:properties-BC12}. If we include a $50\%$ systematic
uncertainty on the background estimate the discovery reach is
reduced to $M_{1/2} \lesssim 460$~GeV. In this case we have squark
masses (gluino masses) around 950~GeV (1.1~TeV). We want to point out
that due to the striking multi-lepton signature, the discovery reach
for stau LSP scenarios with early LHC data is larger
than that for $R$-parity conserving mSUGRA models
\cite{Baer:2010tk}.

\mymed

We have also translated the significances of
Fig.~\ref{Fig:parameterScan-BC1-sig} and
Fig.~\ref{Fig:parameterScan-BC1-Z0} to the minimum integrated
luminosity that is required for a five sigma discovery. For
$S/\sqrt{B}$ this can easily be done, because $S/\sqrt{B}$ scales with
the square root of the luminosity.  For $Z_0$ the procedure is more
involved, \cf~App.~\ref{diff_significances}.  The results for
$S/\sqrt{B}$ [$Z_0$] are shown in
Fig.~\ref{Fig:parameterScan-BC1-lumi}
[Fig.~\ref{Fig:parameterScan-BC1-lumi-Z0}]. We can see in
Fig.~\ref{Fig:parameterScan-BC1-lumi-Z0} that the benchmark
scenario BC1 can be discovered with an integrated luminosity of less
than $200\,{\rm pb}^{-1}$.  If the systematic uncertainties are under
good control one might even claim a discovery with roughly $50\,{\rm
pb}^{-1}$, \cf~Fig.~\ref{Fig:parameterScan-BC1-lumi}.

\subsection{Stau Mass Estimate}
\label{mass_reco}

We have shown in the last section that the stau LSP scenario with a
non-vanishing $\lambda_{121}$ coupling can already be tested
quite stringently with early LHC data. If a discovery has
been made, one would try to reconstruct the sparticle mass spectrum;
especially the stau LSP mass. We therefore propose a method to
estimate the latter. Note, that we do not include systematic
uncertainties. This has to be done after a discovery.

\mymed

In and around BC1, the stau LSP decays via the 4-body mode
\begin{equation}
\tilde{\tau}_1^\pm \rightarrow \tau^\pm \ell^+ \ell^- \nu \, ,
\end{equation}
where $\ell$ denotes an electron or muon; see
Tab.~\ref{tab_stau_LSP_sig_LLE}.  By calculating the invariant mass of
the stau LSP decay products one can in principle reconstruct its
mass. But due to the initial neutrino and the neutrino in the
subsequent tau decay, not all decay products are visible. However, we
can still build the invariant mass of the $\ell^+\ell^-$ pair with the
visible part of the (hadronically decaying) tau. We then expect a
kinematic endpoint in the invariant mass distribution, which should
lie at the true stau LSP mass.

\mymed

\begin{figure}[t!]
    \includegraphics[width=0.48\textwidth,trim=0cm 0cm 0cm
    0cm]{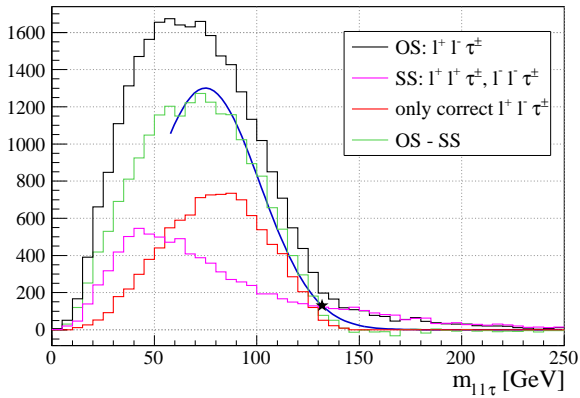}
    \caption{\label{Fig:BC1-inv_mass} Invariant mass distribution of
    the visible part of the hardest tau, $\tau^\pm$, with the two
    nearest (in $\Delta R$) charged leptons, $\ell$, of the first or
    second generation. The black line (purple line) gives the
    distribution for the opposite-sign (same-sign) lepton pair,
    $\ell^+ \ell^-$ ($\ell^\pm \ell^\mp$), plus the tau. The
    distribution is denoted by OS (SS). The green line shows the
    difference of the OS and SS distributions. The red histogram
    corresponds to the correct $\tau^\pm \ell^+ \ell^-$ combination,
    \ie~all three leptons stem from the same stau decay. We also
    fitted a Gaussian (blue line) to the green histogram.}
\end{figure}

The black line in Fig.~\ref{Fig:BC1-inv_mass} shows the invariant mass
distribution of the $\ell^+ \ell^-$ pair plus the visible part of the
tau. The distribution corresponds to $50\,000$ BC1 signal
events \footnote{This corresponds roughly to an integrated luminosity of
$10~\ifb$ at $\sqrt{s}=14$ TeV.}. We do not include the SM backgrounds,
because as shown in the previous section, it is possible to select a
(nearly) background free data sample, \cf~Tab.~\ref{Tab:CutFlowBC1}.
Leptons are combined as follows. In each event we select the hardest
identified tau.  We then look for the two closest opposite sign (OS)
$\ell$s in $\Delta R=\sqrt{\Delta \phi^2 + \Delta \eta^2}$. Finally,
we calculate the invariant mass of these three leptons. If such a
lepton triplet cannot be found in an event, we discard it for mass
reconstruction. Note, that we do not employ a detector simulation
here.  We only employ the following cuts. Electrons and muons
(hadronically decaying taus) are identified if their (visible) $\pt$
is larger than 7~GeV (10~GeV). The electrons and muons (tau decay
products) must also lie within $|\eta|<2.5$. We use the same setup in
what follows if not otherwise mentioned. As an additional cut
only those selected combinations are used, where the distance in
$\Delta R$ between both leptons and the tau is smaller than 1.5.  In
principle this cut may distort the invariant mass spectrum, especially
for very high stau masses.  We checked both options with and without
this cut and observed, that both give nearly identical results in the
precision of the estimated stau mass.
\mymed

\begin{figure*}[t!]
  \subfigure[\label{Fig:estimate-vs-true-mass-a}]{
    \includegraphics[width=0.48\textwidth,trim=0cm 0cm 0cm
    0cm]{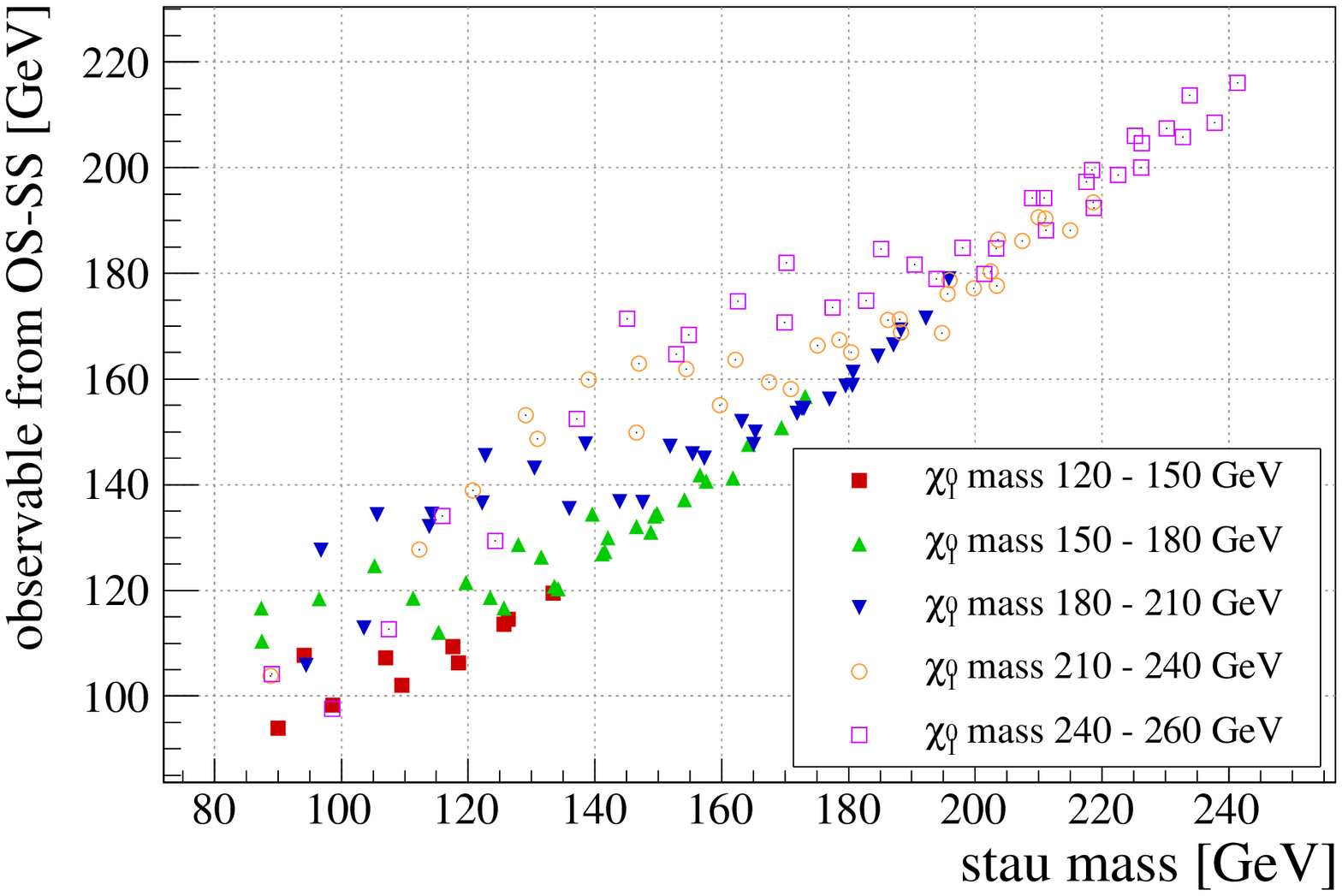} }
    \subfigure[\label{Fig:estimate-vs-true-mass-b}]{
    \includegraphics[width=0.48\textwidth,trim=0cm 0cm 0cm
    0cm]{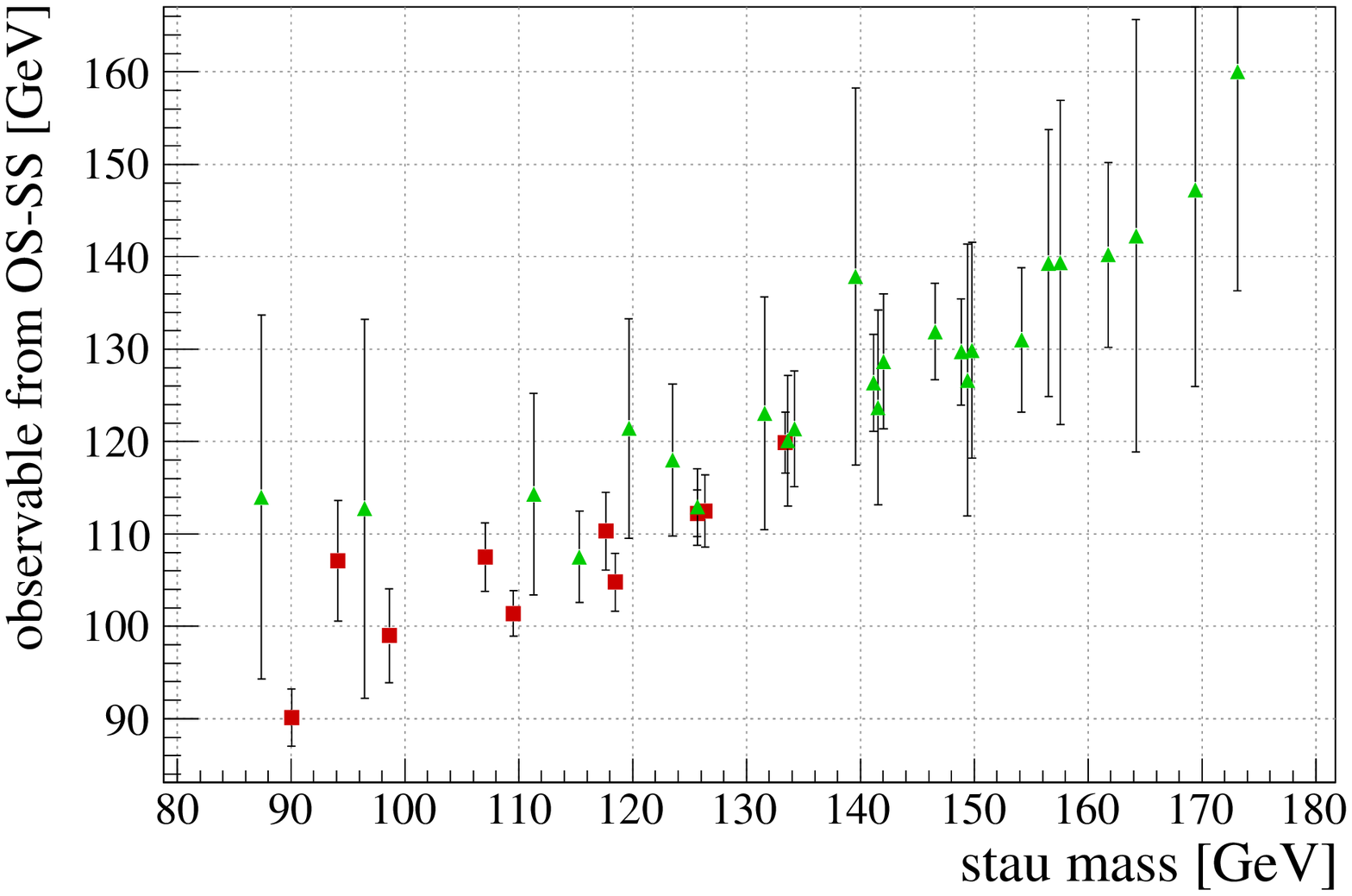}
    } \caption{\label{Fig:estimate-vs-true-mass} Stau mass sensitive
    observable versus true stau mass (see text for definition) for the
    scenarios presented in Sec.~\ref{discovery_potential}.  Different
    colors and shapes of the points correspond to different intervals
    of the $\tilde{\chi}_1^0$ mass.
    Fig.~\ref{Fig:estimate-vs-true-mass-a}: For each parameter point,
    $10\,000$ signal events were simulated.
    Fig.~\ref{Fig:estimate-vs-true-mass-b}: Estimates for an
    integrated luminosity of 5$\,\ifb$ including event selection cuts.
    Only scenarios where at least 150~events pass the cuts are used.
    The error bars show to what precision the estimated stau mass can
    be measured. The errors correspond to statistical fluctuations and
    are estimated as described in the text.}
\end{figure*}

We see in Fig.~\ref{Fig:BC1-inv_mass} that the black line has
a poorly defined endpoint at the true stau mass of 148~GeV.
This is due to combinatorial backgrounds, \ie~we sometimes combine the
wrong tau and electrons or muons with each other. Various other
combination methods have also been studied, for example
starting from the hardest electron or muon or using $\Delta
\phi$ instead of $\Delta R$.  A small improvement is possible by
vetoing the combination $\tau \mu^+ \mu^-$. This does not increase the
fraction of correct combinations but reduces the number of wrong
combinations.  The reason is, that the stau LSP can not decay to $\tau
\mu^+ \mu^- \nu$ via a coupling $\lambda_{121}$;
\cf~Tab.~\ref{tab_stau_LSP_sig_LLE}.  We did not to veto the $\tau
\mu^+ \mu^-$ combination as this keeps the method more model
independent.

\mymed

In 18\% of all events no combination can be found, because no hadronic
tau in the given kinematic range exists. In an independent
18\% of all events there is only a tau, which does not stem from a stau
decay, \ie~every method will choose the wrong tau. With our method,
30\% of the chosen combinations are correct, while 25\% include a
(wrong) tau lepton that stems from the $\ninoone$ decay, which belongs
to the same decay chain as the chosen leptons. In 14\% of the
combinations the tau neither stems from a stau nor a $\ninoone$ decay
and in 15\% at least one lepton stems from the other stau decay in the
event. In 10\% at least one lepton comes from another source, \ie~not
from a stau decay.

\mymed

In order to reduce the combinatorial backgrounds and thus to sharpen
the kinematic endpoint of the invariant mass, we also combine the
hardest tau with the nearest same sign (SS) lepton pair, $\ell^\pm
\ell^\pm$.  The respective invariant mass distribution is given by the
purple line in Fig.~\ref{Fig:BC1-inv_mass}.  We then subtract the
$\tau$+SS distribution (purple line) from the $\tau$+OS distribution
(black line) and obtain the distribution given by the green line
denoted by OS$-$SS.

\mymed

The OS$-$SS invariant mass distribution now follows much closer the
distribution that arises from the correct $\tau \ell^+ \ell^-$ triplet
(red histogram in Fig.~\ref{Fig:BC1-inv_mass}). Without the cut on the
angular distance between the two leptons and the tau the same sign
distribution shows a long tail at high invariant masses, which leads
to an over-subtraction at high masses. This can be explained by the
fact, that one of the same sign leptons mostly stems from the other
stau decay in the event or another source and therefore often has a
larger angle to the first lepton leading to higher invariant masses.
Still one can find stau mass sensitive observables, like the
intersection of the OS$-$SS distribution with the $x$-axis, which show
a good correlation with the true stau mass. 

\mymed

We observe that the
OS$-$SS histogram has an endpoint near the true endpoint.  As the stau
mass sensitive observable, we fit a Gaussian on the OS$-$SS
distribution and take the value, where it drops to 10\% of its maximum
(marked by a star in Fig.~\ref{Fig:BC1-inv_mass})
\footnote{The fit range is crucial and has been determined in an
iterated Gaussian fit, which starts with the maximum bin position and
the RMS of the histogram and uses the range $\mu-\sqrt{2\ln
2}/2\cdot\sigma$ to 200~GeV, where $\mu$ is the mean of the previous
Gaussian fit and $\sigma^2$ its variance.}.  Although the observable
lies below the {\it true} stau mass of 148~GeV, we can estimate it
from the observable as long as there is a clear and known correlation
between the two. This has successfully been demonstrated for example
in Ref.~\cite{Nattermann:2009gh, Aad:2009wy}.

\mymed

This is indeed the case as one can see in
Fig.~\ref{Fig:estimate-vs-true-mass-a}. Here we take the stau LSP
scenarios of our parameter scans in Sec.~\ref{discovery_potential}.
We then simulated $10\,000$ signal events for each scenario and
determined from these the estimated stau mass as described above.
The different colors of the points in
Fig.~\ref{Fig:estimate-vs-true-mass-a} correspond to different
$\tilde{\chi}_1^0$ masses.

\mymed

We can see in Fig.~\ref{Fig:estimate-vs-true-mass-a} a clear
correlation between the true stau LSP mass and the observable from the
OS$-$SS invariant mass distribution. We also see that there is only a
small systematic dependence of the estimated mass on the
$\tilde{\chi}_1^0$ mass. For example, for a stau mass of 120~GeV, the
observable can increase roughly from 100~GeV to 140~GeV, if the
$\tilde{\chi}_1^0$ mass increases from 120~GeV to 240~GeV. This is
because a heavier $\tilde{\chi}_1^0$ leads to a harder tau from the
$\tilde{\chi}_1^0 \rightarrow \tilde{\tau}_1
\tau$ decay. 

\mymed

One can use Fig.~\ref{Fig:estimate-vs-true-mass-a} to translate the
observable to the true stau mass.  Such an analysis is even possible in
a limited way with early LHC data, as can be seen in
Fig.~\ref{Fig:estimate-vs-true-mass-b}, where we again show the
observable versus the true stau mass.  Now, we have only included
scenarios, where at least 150 events in 5~$\ifb$ pass our cuts. Otherwise, we would not have
enough statistics for the mass reconstruction. We applied the event
selection cuts that are given in Tab.~\ref{Tab:CutFlowBC1} with
$HT^\prime>300 \GeV$. The error bars correspond to the
precision with which the observable can be measured assuming an
integrated luminosity of 5~$\ifb$. We do not include systematic
uncertainties.

\mymed
\begin{figure}[t!]
    \includegraphics[width=0.48\textwidth,trim=0cm 0cm 0cm
    0cm]{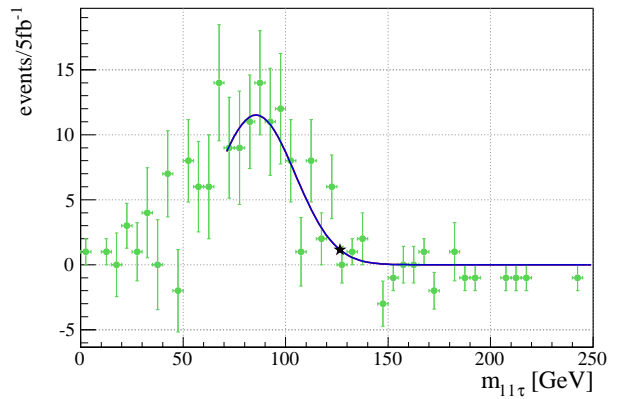}
    \caption{\label{Fig:BC1-inv_mass-5fb} OS$-$SS distribution as in
    Fig.~\ref{Fig:BC1-inv_mass}, but here including event selection
    cuts and randomly selecting events corresponding to an integrated
    luminosity of 5~$\ifb$, \ie~uncertainties and fluctuations
    correspond to those expected for 5~$\ifb$.  }
\end{figure}

The statistical uncertainties in
Fig.~\ref{Fig:estimate-vs-true-mass-b} were estimated in the following
way.  Out of $10\,000$ simulated signal events, events have been
randomly chosen to get a sub-sample corresponding to an integrated
luminosity of 5~$\ifb$. This procedure was repeated 100 times for each
point to obtain different sub-samples. The observable of these
sub-samples follows a Gaussian distribution, where its width
corresponds to the statistical uncertainties in
Fig.~\ref{Fig:estimate-vs-true-mass-b}.
Fig.~\ref{Fig:BC1-inv_mass-5fb} shows an example of one sub-sample for
the BC1 scenario.

\mymed

We can see that with an integrated luminosity of 5~$\ifb$, rough
estimates of the stau LSP mass are possible.

\section{Tau Identification}
\label{tauID}

We have seen in Sec.~\ref{mass_reco} that the identification (ID) of
tau leptons is vital for the mass reconstruction of the stau
LSP. However, as we have pointed out already in the discussion of
Fig.~\ref{Fig:BC1-n-particles-d}, tau ID within the framework of BC1
is difficult.  We now give a short explanation for this observation.

\mymed 

The ID of hadronic decays of tau leptons in the detector makes use of
special properties of the hadronic tau decay. First one expects one or
three charged pion tracks, \ie~{\it single-prong} and {\it
three-prong} events, respectively. Furthermore, jets from tau decays
are usually rather collimated compared to jets from the hadronization
of quarks or gluons. The ratio of energy deposits in the
electromagnetic calorimeter to those in the hadronic calorimeter can
give further hints. Leptonic tau decays ($\tau \to \nu_e \nu_\tau e$
and $\tau \to \nu_\mu \nu_\tau \mu$) are usually not considered,
because the decay length of tau leptons is mostly too small to
distinguish the decay products from prompt electrons or muons.

\mymed

However, the special event topologies in BC1 (and possibly other
beyond-the-SM models) can make it very difficult to reliably identify
tau decays in those events.  In BC1, and also more generally in nearly
all stau LSP scenarios, a very dense environment is expected, \ie~a
high multiplicity of charged particles; see
Tab.~\ref{tab_stau_LSP_sig_LLE}.  Therefore, overlaps between the tau
jets and jets from the SUSY decay chain or leptons from the stau LSP
decay are very likely. In addition, different tau jets might also
overlap. A reduced tau ID efficiency compared to most SM processes is
thus expected.

\mymed

Both {\tt Delphes} \cite{Delphes} and {\tt PGS} \cite{PGS} use
approaches for the detector simulation of tau leptons, which
can be seen as simplified versions of algorithms for tau lepton ID
used by the experiments.  Observables, like electromagnetic
collimation and track isolation, are calculated from the simulated
calorimeter deposits and tracks. Cuts are then applied to
decide, whether a jet is tagged as a tau lepton. In both codes we only
consider 1-prong candidates, \ie~candidates with one assigned track.

\mymed

We show in Fig.~\ref{Fig:TauEfficiencyAndFakeRate} the efficiency for
the ID of hadronic tau decays in BC1 and for (SM) $Z \to \tau\tau +
1\,\text{jet}$ production.  We also give the fraction of wrongly
tagged tau jets per event. We show results for the fast detector
simulation {\tt Delphes}
[Fig.~\ref{Fig:Delphes-ID-Zjet}-Fig.~\ref{Fig:Delphes-fake-BC1}] as
well as for {\tt PGS}
[Fig.~\ref{Fig:PGS-ID-Zjet}-Fig.~\ref{Fig:PGS-fake-BC1}]. In addition,
the standard overlap removal is applied \footnote{The efficiencies are
nearly identical with and without overlap removal, only the number of
fake taus differs significantly as many electrons are also identified
as taus in {\tt Delphes}.}.

\mymed

Both detector simulations show that the ID efficiency is
significantly lower in BC1 than in $Z \to \tau\tau +
1\,\text{jet}$. This has to be kept in mind, as many methods used in
the experiments to estimate the tau ID efficiency rely on $Z \to
\tau\tau$ as a ``standard candle'' \cite{Aad:2009wy}. For example, for
a (visible) tau-$\pT$ of 45 GeV in the inner part of the detector, the
tau ID efficiency is 40\% [60\%] for $Z \to \tau\tau$ but only 10\%
[30\%] in BC1 according to {\tt Delphes} [{\tt PGS}],
\cf~Fig.~\ref{Fig:Delphes-ID-Zjet} and Fig.~\ref{Fig:Delphes-ID-BC1}
[Fig.~\ref{Fig:PGS-ID-Zjet} and Fig.~\ref{Fig:PGS-ID-BC1}] at the
respective working points on the efficiency-vs-rejection curve.

\mymed 

\begin{figure}[t!]
    \includegraphics[width=0.5\textwidth,trim=0cm 0cm 0cm 0cm]{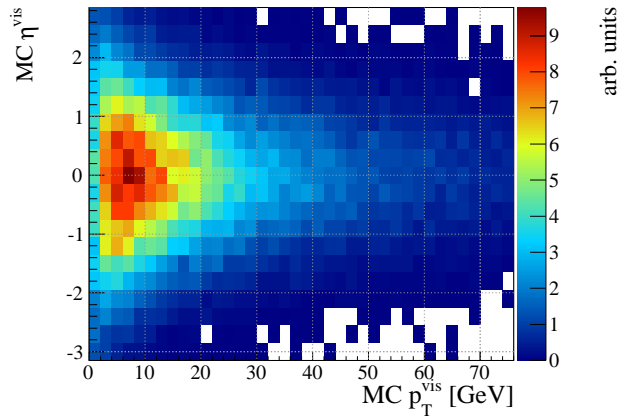}
    \caption{\label{Fig:TauTruthEtaPt} Number of hadronically decaying
        taus in the $\eta$--$\pT$ plane in arbitrary units.  Only the
        visible parts of the taus are considered. No detector
        simulation and acceptance cuts have been applied.
        Bins without entries are white.}
\end{figure}

We also see in Fig.~\ref{Fig:TauEfficiencyAndFakeRate} that in general
the tau ID in {\tt Delphes} has a much lower fake rate than the one in
{\tt PGS}. However this comes at the cost of a slightly worse
identification efficiency giving a different working point of the tau
ID. As a conservative approach, we take the numbers from {\tt
Delphes}. Even in the optimistic case the tau ID efficiency cannot be
expected to be better than about 25\% over a wide $\pT$ range and can
be even lower in the interesting low $\pT$ range,
\ie~(visible) tau-$\pT\lesssim 30$~GeV.  On can see in
Fig.~\ref{Fig:TauTruthEtaPt}, that most tau leptons even have visible
transverse momenta below our ID threshold of 10~GeV.  Given that only
65\% of all tau leptons decay hadronically the efficiency with respect
to all tau leptons is below 10\%.  This means for less than one in
$10\,000$ BC1 events, one can expect all four tau leptons to be
correctly tagged as such. This explains the low tau multiplicity of
Fig.~\ref{Fig:BC1-n-particles-d}.

\begin{figure*}[hp!]
  \subfigure[\label{Fig:Delphes-ID-Zjet} ID efficiency in $Z \to \tau\tau + 1\text{jet}$ for Delphes]{
    \includegraphics[width=0.4\textwidth,trim=0cm 0cm 0cm 0cm]{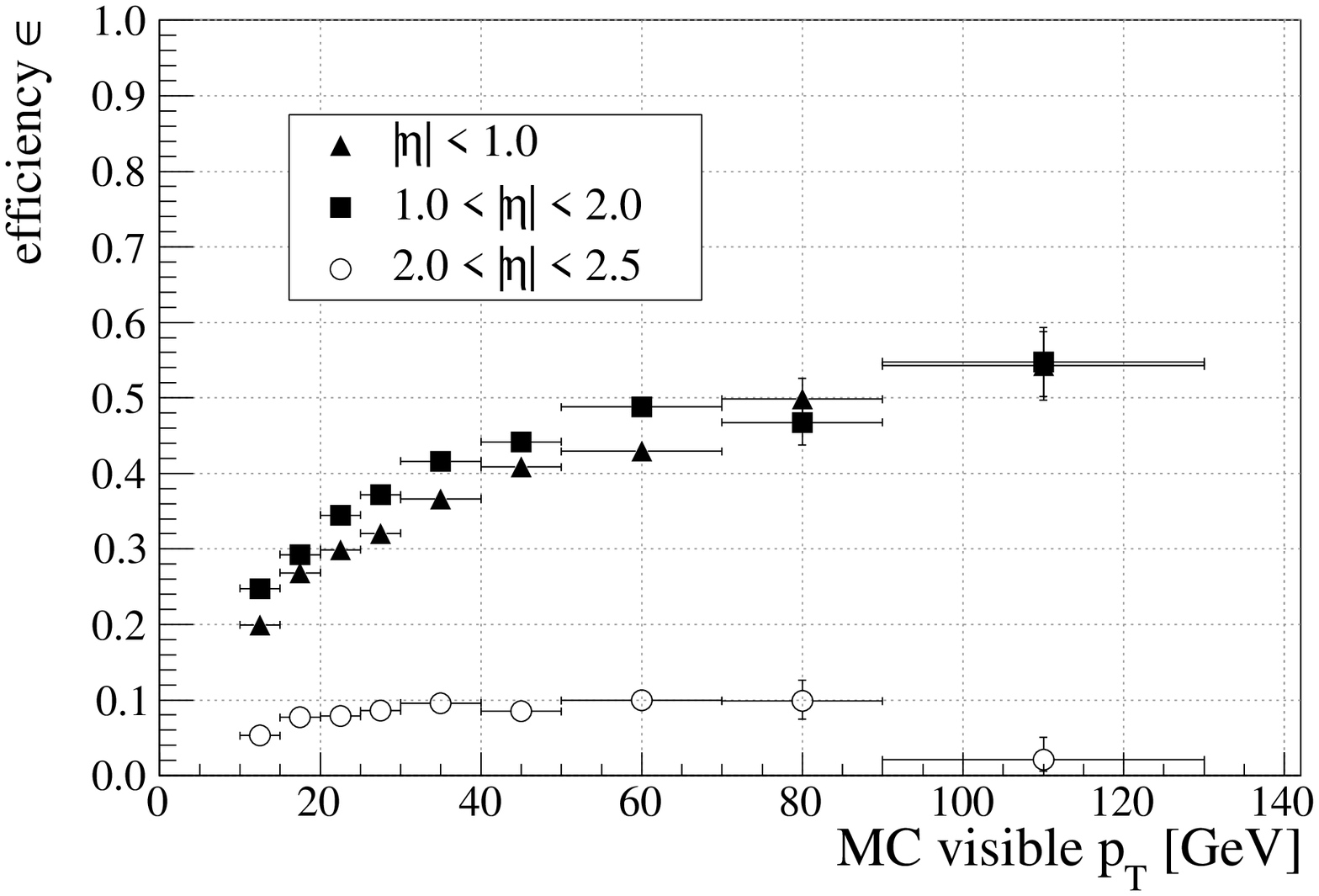}
  }
  \subfigure[\label{Fig:Delphes-fake-Zjet} Fake rate in $Z \to \tau\tau + 1\text{jet}$ for Delphes]{
    \includegraphics[width=0.4\textwidth,trim=0cm 0cm 0cm 0cm]{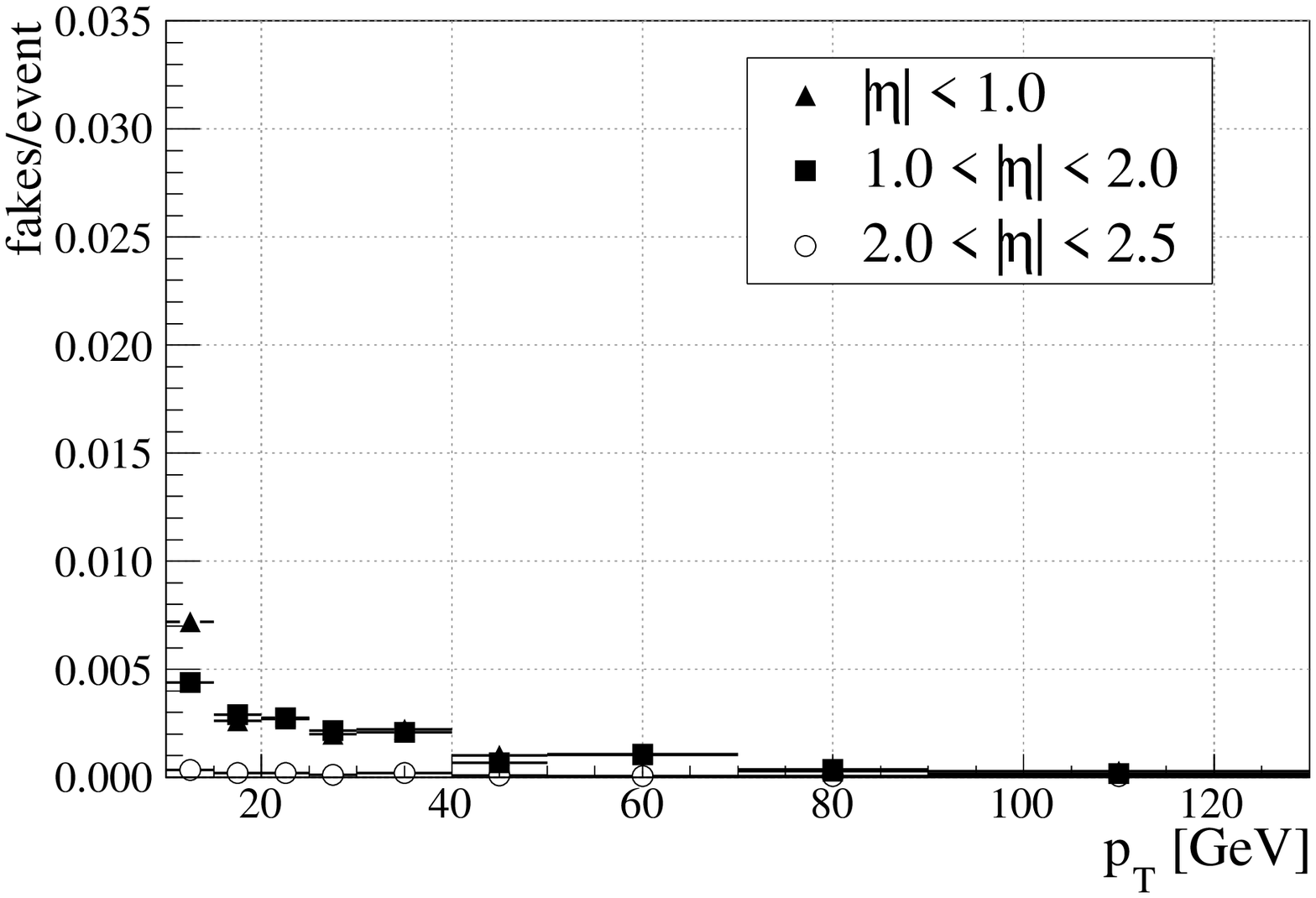}
  }
    \hfill
  \subfigure[\label{Fig:Delphes-ID-BC1} ID efficiency in BC~1 for Delphes]{
    \includegraphics[width=0.4\textwidth,trim=0cm 0cm 0cm 0cm]{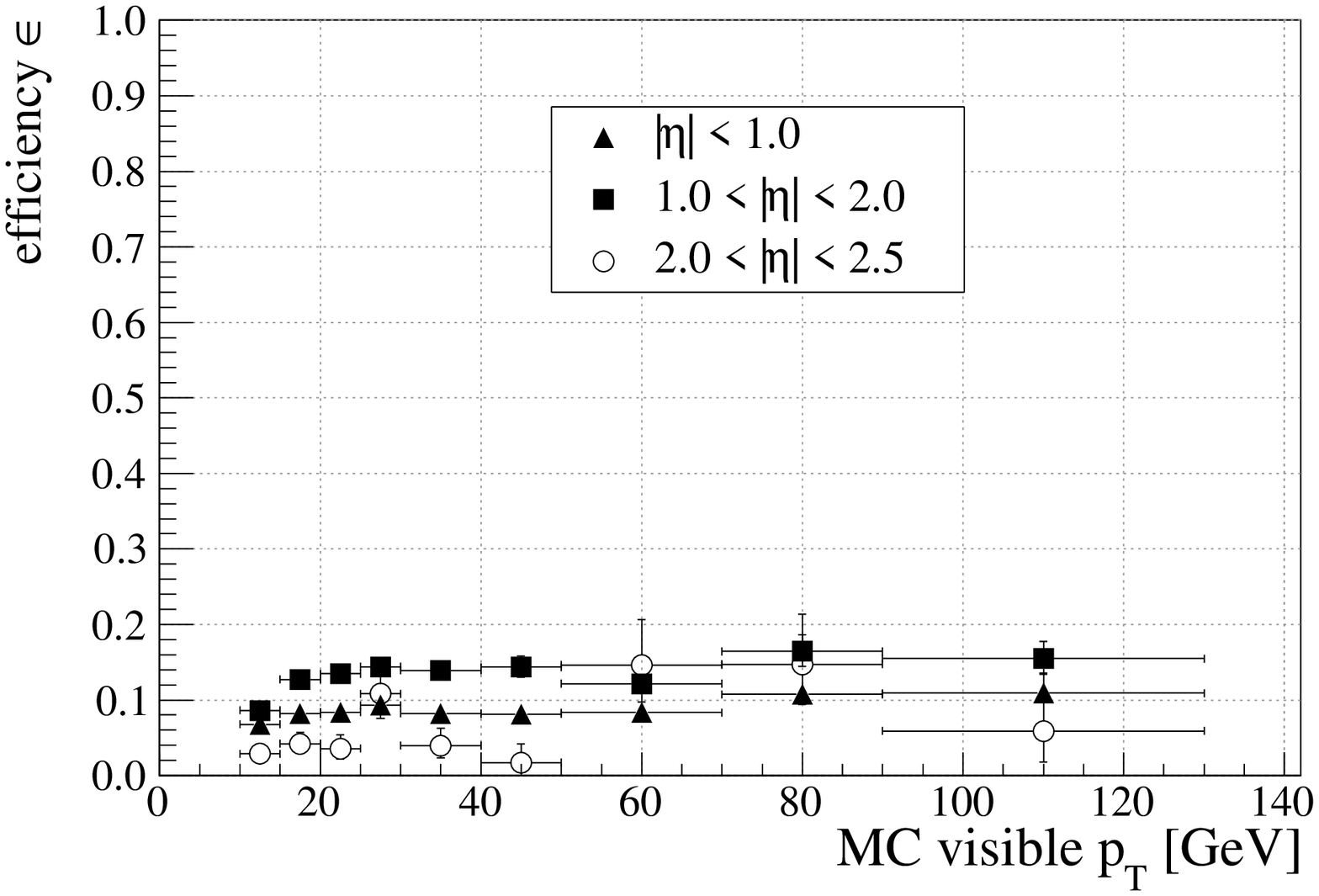}
  }
  \subfigure[\label{Fig:Delphes-fake-BC1} Fake rate in BC~1 for Delphes]{
    \includegraphics[width=0.4\textwidth,trim=0cm 0cm 0cm 0cm]{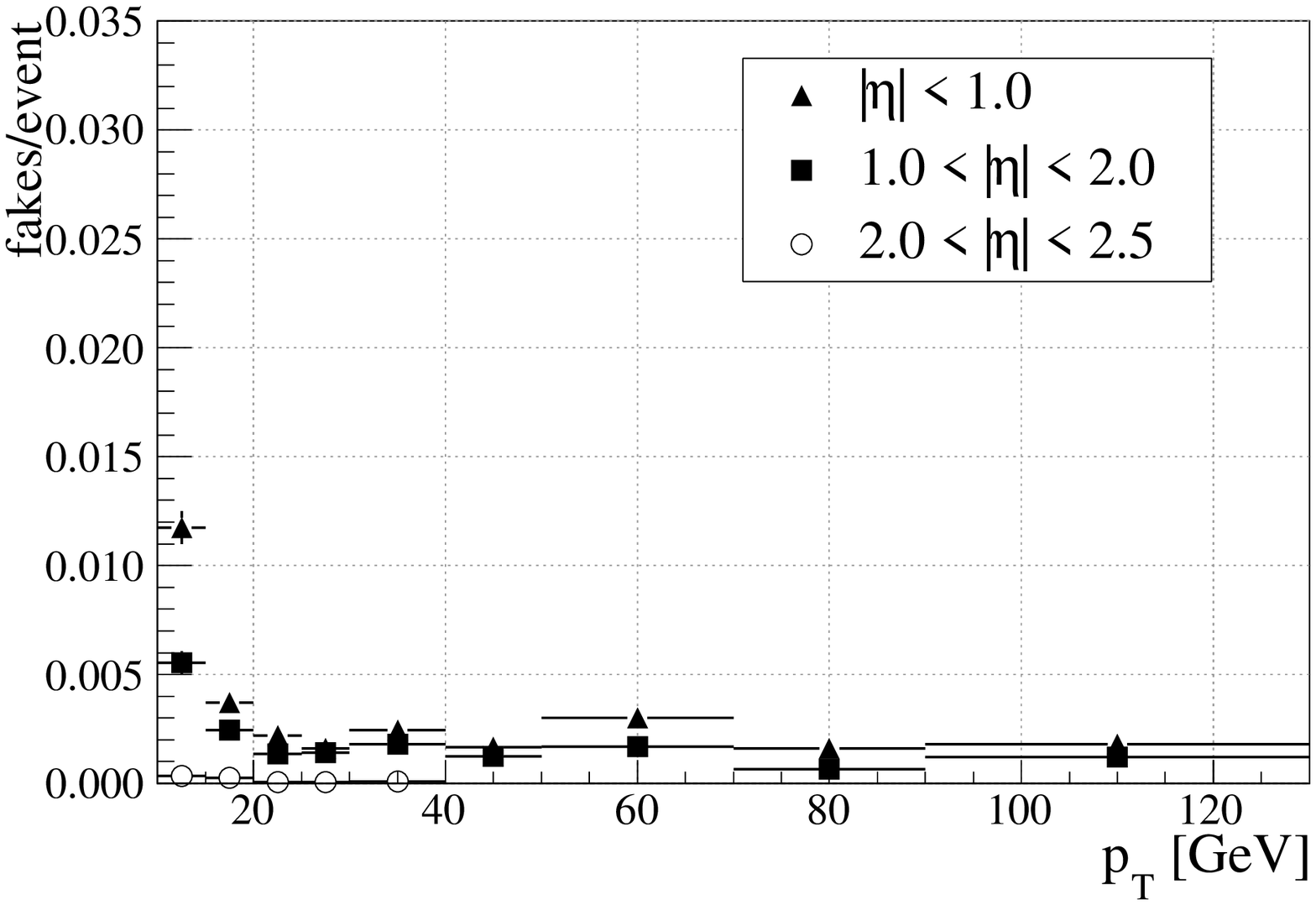}
  }
    \hfill
  \subfigure[\label{Fig:PGS-ID-Zjet} ID efficiency in $Z \to \tau\tau + 1\text{jet}$ for PGS]{
    \includegraphics[width=0.4\textwidth,trim=0cm 0cm 0cm 0cm]{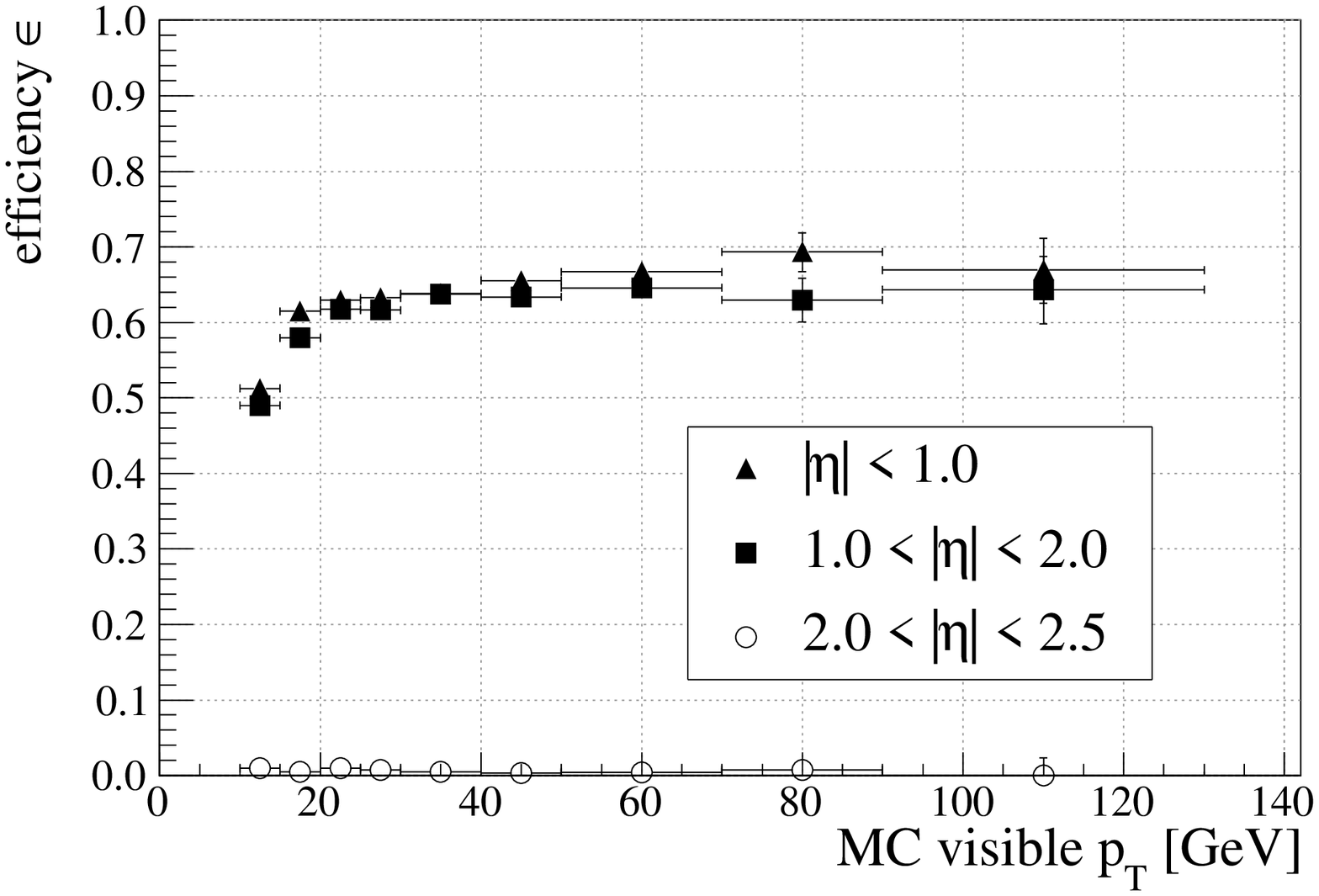}
  }
  \subfigure[\label{Fig:PGS-fake-Zjet} Fake rate in $Z \to \tau\tau + 1\text{jet}$ for PGS]{
    \includegraphics[width=0.4\textwidth,trim=0cm 0cm 0cm 0cm]{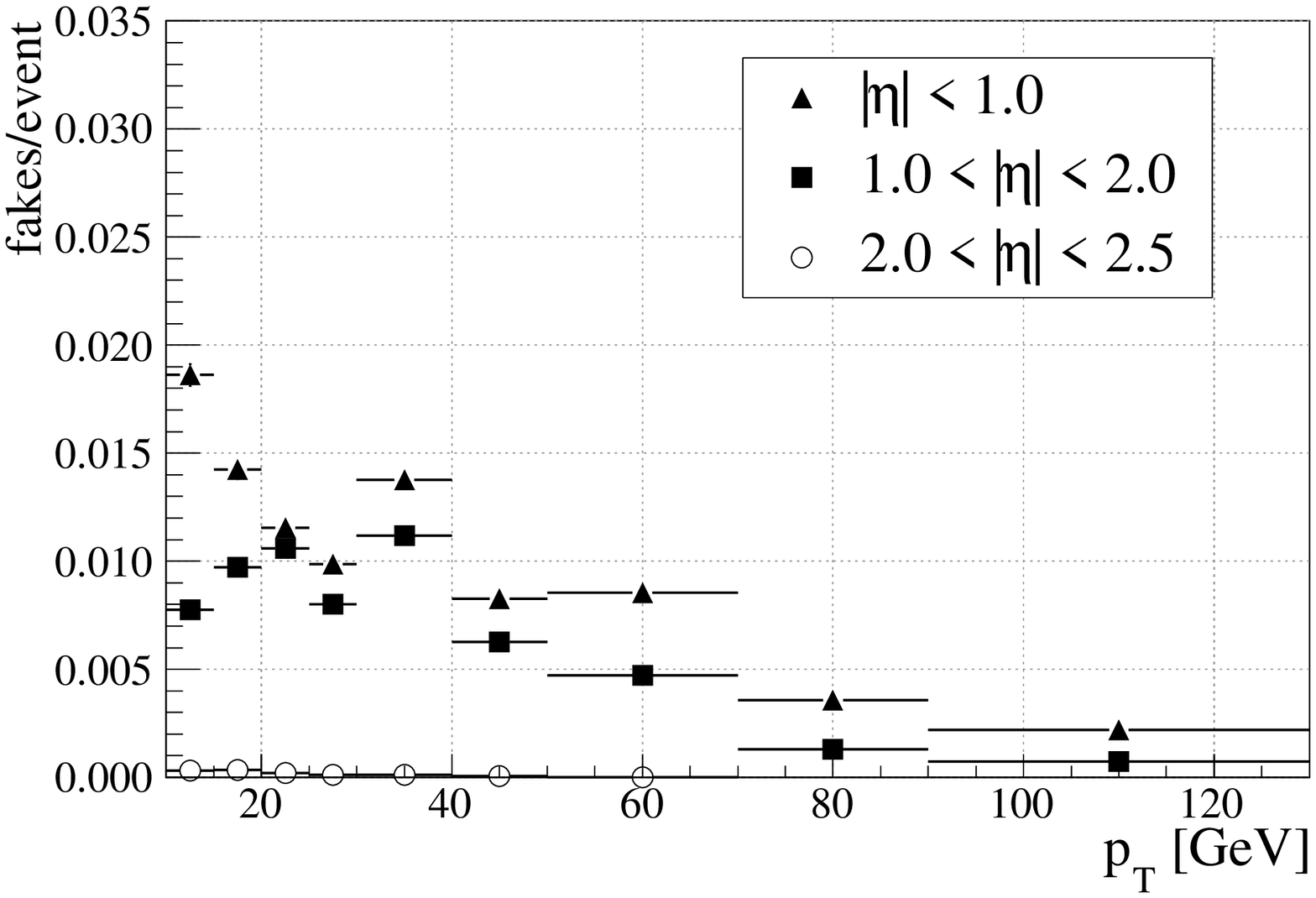}
  }
    \hfill
  \subfigure[\label{Fig:PGS-ID-BC1} ID efficiency in BC~1 for PGS]{
    \includegraphics[width=0.4\textwidth,trim=0cm 0cm 0cm 0cm]{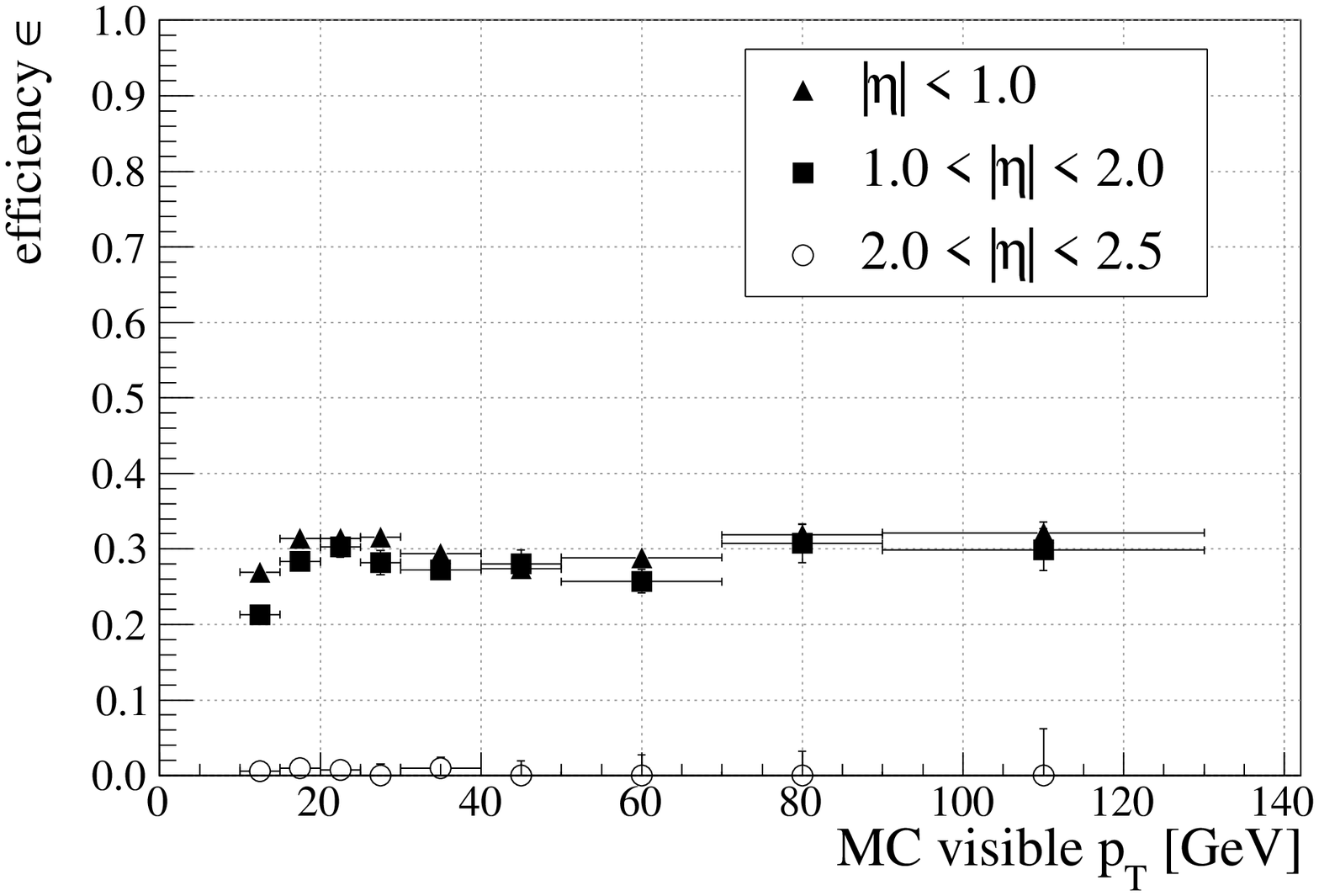}
  }
  \subfigure[\label{Fig:PGS-fake-BC1} Fake rate in BC~1 for PGS]{
    \includegraphics[width=0.4\textwidth,trim=0cm 0cm 0cm 0cm]{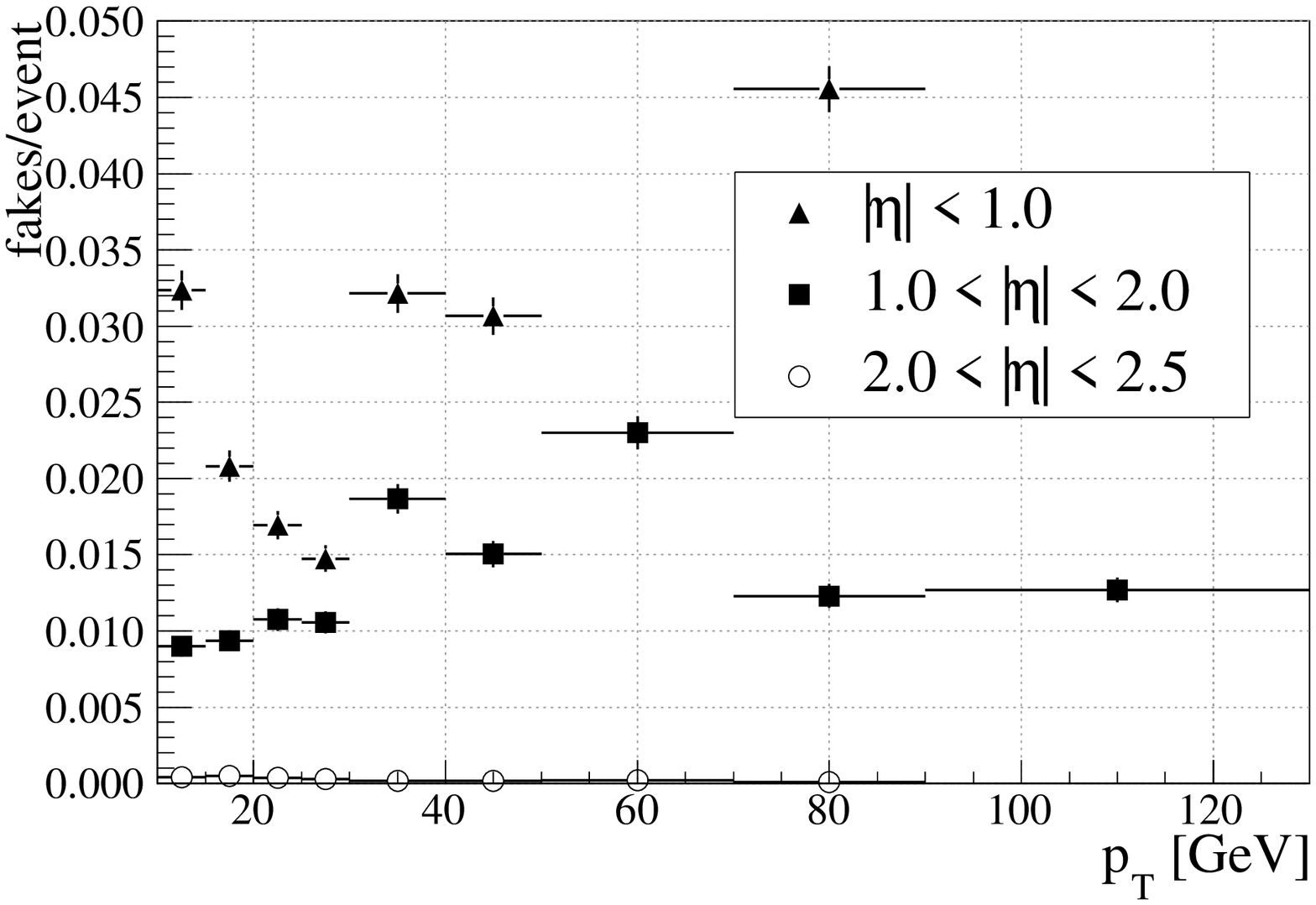}
  }
    \caption{\label{Fig:TauEfficiencyAndFakeRate} Tau ID efficiencies
    and fake rates in BC~1 and $Z \to \tau\tau + 1\text{jet}$ for PGS
    and Delphes.}
\end{figure*}

In conclusion the identification of hadronic tau decays in BC1 and
presumably in similar models as well, is challenging for the
experiments at the LHC.  Precise estimates of the identification
efficiency need studies with full-fledged detector
simulations. The direct estimate of the efficiency from
measured data may be difficult as the numbers are expected to differ
strongly between BC1 events and the usual ``standard candles'', like
$Z \to \tau\tau$.

\section{Summary and Conclusion}
\label{Sec:summary}

If $R$-parity is violated, the lightest supersymmetric particle (LSP)
is unstable. Therefore, it can be charged and any
supersymmetric particle can be the LSP. Within the framework of
minimal supergravity (mSUGRA), this allows for a large region of the
parameter space, where the scalar tau (stau) and not the lightest
neutralino is the LSP.

\mymed

We have investigated the LHC phenomenology of $R$-parity violating
mSUGRA with a stau LSP and with lepton number violation. In
this model, one non-vanishing $R$-parity violating operator is present
at the grand unification scale in addition to the $R$-parity
conserving mSUGRA parameters. Focusing on pair production of strongly
interacting SUSY particles, we classified in
Sec.~\ref{Sec:stau-lsp-signatures} all possible LHC signatures at
parton level;
\cf~Tabs.~\ref{tab_stau_LSP_sig_LLE}-\ref{tab_stau_LSP_sig_UDD} for an
overview. The most promising scenarios for the early LHC are
those where the stau LSP decays via a non-vanishing $\lambda_{121}$
(or $\lambda_{212}$) coupling. This is because each stau LSP
decays purely leptonically via a 4-body decay into two electrons or
muons, a tau and a neutrino; \cf~Tab.~\ref{tab_stau_LSP_sig_LLE}.

\mymed

We have here performed a first comprehensive signal over
background analysis of stau LSP scenarios at the LHC with early
data. We employed pair production of all SUSY particles
as our signal. Our results using fast simulations of the ATLAS detector show,
that the LHC has a good potential to test $R$-parity violating
supersymmetry with a stau LSP and multi-lepton final states
already in the next two years. The benchmark scenario BC1 (see
Tab.~\ref{BCs}) with the 4-body decay $\stauone^\pm \rightarrow
\tau^\pm \ell^+ \ell^- \nu$ has many electrons and muons in its final
states. This allows the selection of a (nearly)
background-free sample of $\mathcal{O}(50)$ BC1 events using an
integrated luminosity of $1\,\ifb$ at $\sqrt{s}=7\,\TeV$.

\mymed 

In our event selection used to derive the discovery potential
we avoid using tau-leptons. Naively one
would expect large numbers of tau leptons in the final states from the
decay $\ninoone \to \stauone + \tau$ and the LSP decay. However, the
reconstruction and identification efficiency for the tau leptons is
expected to be very low, {\it i.e.} not larger than $25\%$ 
(see Sect.~\ref{Sec:summary}), because of their small momenta.
But a precise estimate is only possible with a full detector simulation. Additionally,
overlaps between jets from tau decay products and other jets and
leptons in the event further reduce the efficiency by a factor 2-3
compared to simple event topologies, like $Z\to\tautau$. Our estimates
using fast detector simulations predict no identified tau leptons
in the majority of the BC1 events, even though four tau leptons
are expected at tree level.
Instead, we only
use the transverse momenta of the leading and sub-leading electron and
the leading muon, the scalar sum of the electron and muon
transverse momenta, $\sum \pT^{\ell}$ and the visible hadronic
mass, $HT^\prime = \sum_\text{jet 1-4} \pT$.

\mymed 

In wide ranges of the $M_{1/2}$--$\tan\beta$ parameter space around
the benchmark point one can achieve event selection efficiencies of
about 20\% for nearly background-free samples.  Including a systematic
uncertainty of 50\% on the background estimate the discovery reach is
up to $M_{1/2} = 460\GeV$ for $1~\ifb$. The benchmark scenario
BC1 itself can even be discovered with an integrated luminosity of
less than 200~${\rm pb}^{-1}$.

\mymed 

Despite being potentially easy to discover in the first LHC
data, there are some difficulties in measuring the mass of the
stau-LSP in BC1. Due to large combinatorial backgrounds, the
selection of correct combinations of stau-decay products is difficult
and sometimes even impossible, because the visible momenta of
the tau leptons are too small to be reconstructed. Neutrinos
from the stau decay and the successive tau decay only allow for the
reconstruction of a smooth endpoint in the invariant mass
distribution. However, we could show that observables can be
reconstructed, which show a good correlation with the true stau mass,
even though they will need much more data than 5~$\ifb$ to provide
a precise measurement.


\begin{acknowledgments}
We thank John Conway and Kyoungchul Kong for helpful discussions. S.G. thanks the
Alexander von Humboldt Foundation for financial support. The work of
S.G. was also partly financed by the DOE grant
DE-FG02-04ER41286. S.F. thanks the Bonn-Cologne Graduate School of
Physics and Astronomy for financial support. The work of
H.D. was supported by the BMBF ``Verbundprojekt HEP--Theorie''
under the contract 05H09PDE and the Helmholtz Alliance ``Physics at
the Terascale''. The work of K.D., S.F.\ and P.W.\ was supported by the
German Federal Ministry of Education and Research (BMBF) under the contract 05H09PDA.

\end{acknowledgments}

\appendix

\section{Properties of the benchmark scenario BC1}
\label{App:properties-BC12}

We review in this appendix some properties of the benchmark scenarios
BC1 as described in Ref.~\cite{Allanach:2006st}.

\subsection{Branching Ratios and Mass Spectrum}

\begin{table*}[htp!]
  \centering
\begin{tabular}{cc}
  \begin{tabular}{|lc|ll|ll|}
    \hline
     & mass [GeV] &channel &BR &channel &BR \\
    \hline
$\tilde{\tau}_1^-$ & 148 & $\mu^+ e^- \tau^- \bar \nu_e$ & {\bf 32.2\%} & $e^+ e^- \tau^- \bar \nu_\mu$ & {\bf 32.1\%}\\ 
    & & $\mu^- e^+ \tau^- \nu_e$ & {\bf 17.9\%} & $e^- e^+ \tau^- \nu_e$ & {\bf 17.8\%}\\
    \hline
$\tilde{e}^-_R$ & 161 & $e^- \nu_\mu$ & {\bf 50\%} & $\mu^- \nu_e$ & {\bf 50\%}\\
    \hline
$\tilde{\mu}^-_R$ & 161 & $\tilde{\tau}_1^+ \mu^- \tau^-$ & 51.2\% & $\tilde{\tau}_1^- \mu^- \tau^+$ & 48.7\%\\
    \hline
$\tilde{\chi}_1^0$ & 162 & $\tilde{\tau}_1^+ \tau^-$ & 49.8\% & $\tilde{\tau}_1^- \tau^+$ & 49.8\%\\
    \hline
$\tilde{\nu}_\tau$ & 261 & $\tilde{\chi}_1^0 \nu_\tau$ & 67.2\% & $W^+ \tilde{\tau}_1^-$ & 32.8\%\\
    \hline
$\tilde{\nu}_e \,(\tilde{\nu}_\mu)$ & 262 & $\tilde{\chi}_1^0 \nu_e (\nu_\mu)$ & 92.4\% & $e^- \mu^+ (e^+)$ & {\bf 7.5\%}\\
    \hline
$\tilde{e}_L^- \, (\tilde{\mu}_L^-)$ & 274 & $\tilde{\chi}_1^0 e^- (\mu^-)$ & 91.9\% & $e^- \bar \nu_e (\bar \nu_\mu)$ & {\bf 8.1\%} \\
    \hline
$\tilde{\tau}_2^-$ & 278 & $\tilde{\chi}_1^0 \tau^-$ & 63.0\% & $\tilde{\tau}_1^- Z$ & 17.6\%\\
        & & $h \tilde{\tau}_1^-$ & 19.4\% & & \\
    \hline
$\tilde{\chi}_2^0$ & 303 & $\tilde{\nu}_\tau \bar \nu_\tau$ & 9.1\% & $\tilde{\nu}^*_\tau \nu_\tau$ & 9.1\%\\
        & & $\tilde{\tau}_1^- \tau^+$ & 9.1\% & $\tilde{\tau}_1^+ \tau^-$ & 9.1\% \\
        & & $\tilde{\nu}_e \bar \nu_e$ & 8.5\% & $\tilde{\nu}^*_e \nu_e$ & 8.5\%\\
        & & $\tilde{\nu}_\mu \bar \nu_\mu$ & 8.5\% & $\tilde{\nu}^*_\mu \nu_\mu$ & 8.5\%\\
        & & $\tilde{e}_L^- e^+$ & 4.5\% & $\tilde{e}_L^+ e^-$ & 4.5\% \\
        & & $\tilde{\mu}_L^- \mu^+$ & 4.5\% & $\tilde{\mu}_L^+ \mu^-$ & 4.5\% \\
        & & $\tilde{\tau}_2^- \tau^+$ & 3.1\% & $\tilde{\tau}_2^+ \tau^-$ & 3.1\% \\
        & & $\tilde{\chi}_1^0 h$ & 3.5\% & &\\
    \hline
$\tilde{\chi}_1^-$ & 303 & $\tilde{\nu}_\tau \tau^-$ & 20.2\% & $\tilde{\nu}_\mu \mu^-$ & 18.6\%\\
        & & $\tilde{\nu}_e e^-$ & 18.6\% & $\tilde{\tau}_1^- \bar \nu_\tau$ & 16.7 \%\\
        & & $\tilde{e}_L^- \bar \nu_e$ & 8.1\% & $\tilde{\mu}_L^- \bar \nu_\mu$ & 8.1\% \\
        & & $\tilde{\tau}_2^- \bar \nu_\tau$ & 5.5\% & $\tilde{\chi}_1^0 W^-$ & 4.0\%\\
    \hline
$\tilde{\chi}_3^0$ & 514 & $\tilde{\chi}_1^- W^+$ & 28.9\% & $\tilde{\chi}_1^+ W^-$ & 28.9\%\\
        & & $\tilde{\chi}_2^0 Z$ & 24.1\% & $\tilde{\chi}_1^0 Z$ & 10.2\%\\
        & & $\tilde{\chi}_1^0 h$ & 1.8\% & $\tilde{\tau}_1^- \tau^+$ & 1.0\%\\
        & & $\tilde{\tau}_1^+ \tau^-$ & 1.0\% & &\\
    \hline
$\tilde{\chi}_4^0$ & 529 &$\tilde{\chi}_1^- W^+$ & 26.5\% & $\tilde{\chi}_1^+ W^-$ & 26.5\%\\
        & & $\tilde{\chi}_2^0 h$ & 17.5\% & $\tilde{\chi}_1^0 h$ & 7.1\%\\
        & & $\tilde{\nu}_\tau \bar \nu_\tau$ & 1.8\% & $\tilde{\nu}^*_\tau \nu_\tau$ & 1.8\%\\
        & & $\tilde{\nu}_e \bar \nu_e$ & 1.8\% & $\tilde{\nu}^*_e \nu_e$ & 1.8\%\\
        & & $\tilde{\nu}_\mu \bar \nu_\mu$ & 1.8\% & $\tilde{\nu}^*_\mu \nu_\mu$ & 1.8\%\\
        & & $\tilde{\tau}_2^- \tau^+$ & 1.7\% &  $\tilde{\tau}_2^+ \tau^-$ & 1.7\% \\
        & & $\tilde{\chi}_1^0 Z$ & 1.8\% & $\tilde{\chi}_2^0 Z$ & 1.4\% \\
    \hline
  \end{tabular} &

\hspace{1cm}
  \begin{tabular}{|lc|ll|ll|} 
    \hline
     &mass [GeV] &channel &BR &channel &BR \\
    \hline
$\tilde{\chi}_2^-$ & 532 & $\tilde{\chi}_2^0 W^-$ & 28.3\% & $\tilde{\chi}_1^- Z$ & 25.3\%\\
            & & $\tilde{\chi}_1^- h$ & 19.8\% & $\tilde{\chi}_1^0 W^-$ & 8.1\%\\
            & & $\tilde{\tau}_2 \bar \nu_\tau$ & 4.4\% & $\tilde{e}_L \bar \nu_e$ &  3.7\%\\
            & & $\tilde{\mu}_L \bar \nu_\mu$ &  3.7\% & $\tilde{\nu}^*_\tau \tau^-$ & 2.8\%\\
            & & $\tilde{\nu}^*_e e^-$ & 1.6\% & $\tilde{\nu}^*_\mu \mu^-$ & 1.6\%\\
     \hline
$\tilde{t}_1$ & 647 & $\tilde{\chi}_1^+ b$ & 44.0\% & $\tilde{\chi}_1^0 t$ & 23.7\%\\
            & & $\tilde{\chi}_2^+ b$ & 17.0\% & $\tilde{\chi}_2^0 t$ & 15.4\% \\
    \hline
$\tilde{b}_1$ & 780 & $\tilde{\chi}_1^- t$ & 36.0\% & $\tilde{\chi}_2^- t$ & 25.2\%\\
            & & $\tilde{\chi}_2^0 b$ & 22.0\% & $W^- \tilde{t}_1$ & 12.0\% \\
            & & $\tilde{\chi}_1^0 b$ & 2.4\% & $\tilde{\chi}_3^0 b$ & 1.2\% \\
    \hline
$\tilde{b}_2$ & 816 & $\tilde{\chi}_2^- t$ & 40.8\% & $\tilde{t}_1 W^-$ & 15.2 \% \\
            & & $\tilde{\chi}_1^0 b$ & 12.7\% & $\tilde{\chi}_1^- t$ & 10.0\% \\
            & & $\tilde{\chi}_4^0 b$ & 8.6\% & $\tilde{\chi}_3^0 b$ & 6.7\% \\
            & & $\tilde{\chi}_2^0 b$ & 6.0\% &  & \\
    \hline
$\tilde{d}_R$ $(\tilde{s}_R)$ & 820 & $\tilde{\chi}_1^0 d(s)$ & 99.4\% & &  \\
    \hline
$\tilde{u}_R$ $(\tilde{c}_R)$ & 822 & $\tilde{\chi}_1^0 u(c)$ & 99.4\% & & \\  
    \hline
$\tilde{t}_2$ & 835 & $\tilde{\chi}_4^0 t$ & 23.5\% & $\tilde{\chi}_1^+ b$ & 23.0\%\\
            & & $\tilde{\chi}_2^+ b$ & 15.0 \% & $\tilde{t}_1 Z$ & 12.3\% \\
            & & $\tilde{\chi}_3^0 t$ & 9.6 \% & $\tilde{\chi}_2^0 t$ & 9.6 \% \\
            & & $\tilde{t}_1 h$ & 5.7\% & $\tilde{\chi}_1^0 t$ & 2.3 \% \\
    \hline
$\tilde{u}_L$ $(\tilde{c}_L)$ & 852 & $\tilde{\chi}_1^+ d(s)$ & 64.6\% & $\tilde{\chi}_2^0 u(c)$ & 31.8\%\\
            & & $\tilde{\chi}_2^+ d(s)$ & 1.5\% & $\tilde{\chi}_4^0 u(c)$ & 1.1\%\\
            & & $\tilde{\chi}_1^0 u(c)$ & 1.0\% & & \\
    \hline
$\tilde{d}_L$ $(\tilde{s}_L)$ & 855 & $\tilde{\chi}_1^- u(c)$ & 61.6\% & $\tilde{\chi}_2^0 d(s)$ & 31.8\%\\
            & & $\tilde{\chi}_2^- u(c)$ & 3.8\% & $\tilde{\chi}_1^0 d(s)$ & 1.8\%\\
            & & $\tilde{\chi}_4^0 d(s)$ & 1.4\% & & \\
    \hline
$\tilde{g}$ & 932 & $\tilde{q} \bar q$ & 25.0\% & $\tilde{q}^* q$ & 25.0\%\\
            & & $\tilde{t}_1 \bar t$ & 9.5\% & $\tilde{t}^*_1 t$ & 9.5\%\\
            & & $\tilde{b}_1 \bar b$ & 7.7\% & $\tilde{b}^*_1 b$ & 7.7\%\\
            & & $\tilde{b}_2 \bar b$ & 5.2\% & $\tilde{b}^*_2 b$ & 5.2\%\\
    \hline
\end{tabular} \\
\end{tabular}
\caption{\label{App:BRs-BC1} SUSY mass spectrum and BRs of the
benchmark scenario BC1 \cite{Allanach:2006st}. Only decays with a BR
of at least 1\% are shown. $R$-parity violating decays are in bold
face. }
\end{table*}

In Tab.~\ref{App:BRs-BC1} we show the supersymmetric mass spectrum and
the branching ratios (BRs) in the BC1 scenario. The non-vanishing
$\text{B}_3$ coupling at $M_{\rm GUT}$ is ${\lambda_{121}}|_{\rm
GUT}=0.032$.  The other mSUGRA parameters are $M_0=A_0=0$~GeV,
$M_{1/2}=400$~GeV, $\tan \beta=10$, and ${\rm sgn}(\mu)=+1$.

\mymed

The heavy part of the spectrum, \textit{e.g.} gluinos and
squarks, looks very similar to mSUGRA scenarios with a $\ninoone$
LSP, like SPS1a \cite{Allanach:2002nj}. In BC1, these sparticles
mainly decay via two-body decays mediated by the usual
$R$-parity conserving gauge interactions into a lighter sparticle and
a quark.

\mymed

Also the middle part of the mass spectrum, \ie~in
Tab.~\ref{App:BRs-BC1} masses of roughly $300-500$~GeV, is very
similar to many $\ninoone$ LSP mSUGRA scenarios. This includes
most of the charginos and neutralinos.  The heaviest neutralinos,
$\tilde{\chi}_3^0$ and $\tilde{\chi}_4^0$, and the heaviest chargino,
$\tilde{\chi}_2^+$, are Higgsino-like whereas the $\tilde{\chi}_2^0$
and $\tilde{\chi}_1^+$ are wino-like. These sparticles also mainly
decay via two-body decays involving gauge interactions.

\mymed 

However, the light part of the spectrum in Tab.~\ref{App:BRs-BC1}
looks very different than the $R$-parity conserving mSUGRA
scenarios with a $\ninoone$ LSP.  We now have a stau LSP and the
$\ninoone$ is only the next-to-next-to-next-to LSP (NNNLSP).  However,
it is nearly degenerate in mass with the slightly lighter
right-handed selectron, $\tilde{e}_R$ and smuon, $\tilde{\mu}_R$.

\mymed

We also observe that some of the lighter sparticles decay via
$R$-parity violating interactions, with BRs larger than
1\%. The stau LSP that can only decay via a $R$-parity violating
operator.  Because it does not directly couple to the dominant
$L_1 L_2 \bar E_1$-operator, it decays in BC1 via a four-body decay,
\cf~Tab.~\ref{tab_stau_LSP_sig_LLE}. The $\tilde{e}_R$ decays
mainly via $R$-parity violating interactions. This is because
it couples directly via $\lambda_{121}$ and can thus decay via
a two-body decay into two nearly massless SM particles. The competing
$R$-parity conserving decay mode is a three-body decay, namely $\tilde
{e}_R^-\rightarrow \tilde{\tau}_1^\pm \tau^\mp e^-$. The
$\ninoone$ NNNLSP is also unstable. It decays mainly into a tau plus
the stau LSP. See Refs.~\cite{Allanach:2006st,Allanach:2007vi} for
further details.

\subsection{Masses and Mass Differences}
\label{masses_stau}

\begin{figure*}[tbp!]
  \setlength{\unitlength}{1in}
  
    \subfigure[\label{Fig:gluino_mass}Gluino mass.]{ 
    \begin{picture}(3,2.3)
      \put(0.1,0){\epsfig{file=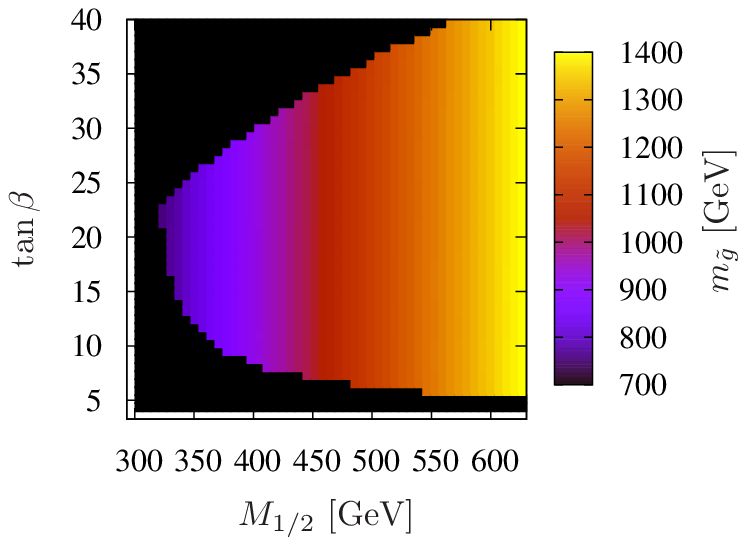,width=3.in}}
    \end{picture}
   }
   \subfigure[\label{Fig:squark_mass}$\tilde{d}_L$ squark mass.]{ 
        \begin{picture}(3,2.3)
      \put(0.1,0){\epsfig{file=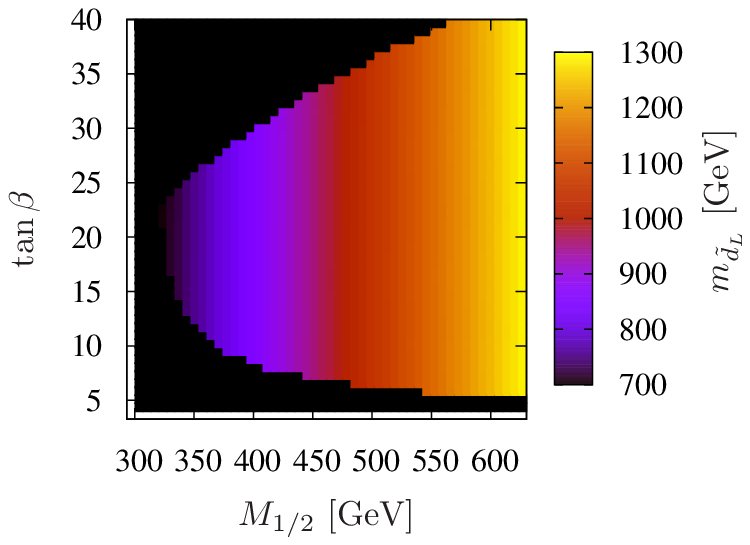,width=3.in}}
    \end{picture}
  }
  
  \subfigure[\label{Fig:neut1_mass}$\tilde{\chi}_1^0$ mass.]{
    \begin{picture}(3,2.3)
      \put(0.1,0){\epsfig{file=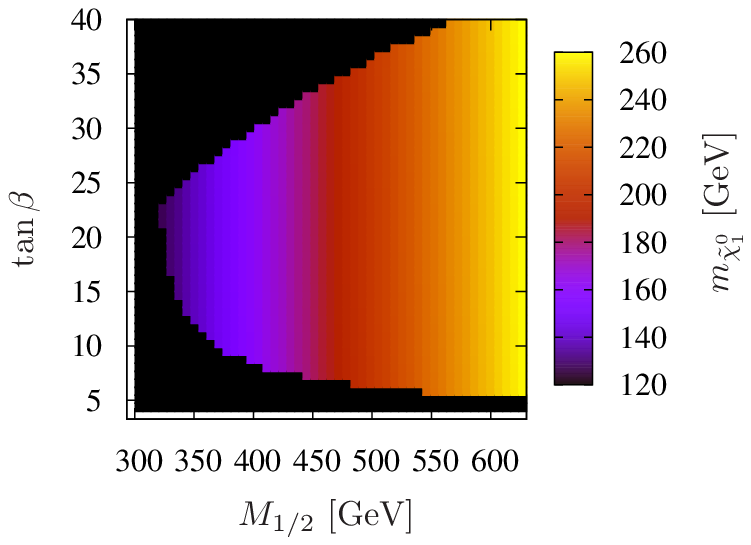,width=3.in}}
    \end{picture}
  }
  \subfigure[\label{Fig:stau_mass} $\tilde{\tau}_1$ LSP mass.]{
    \begin{picture}(3,2.3)
      \put(0.1,0){\epsfig{file=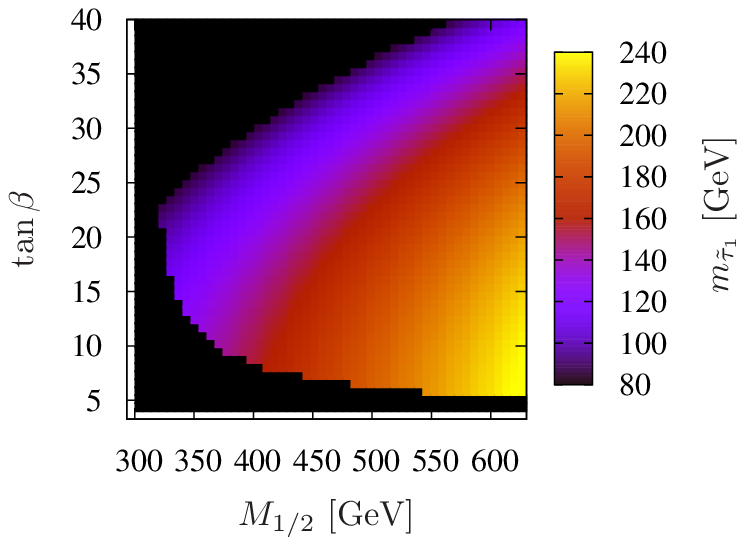,width=3.in}}
    \end{picture}
  }
    \hfill

  \subfigure[\label{Fig:mass_diff_stau1_neut1}
        Mass difference between the $\tilde{\chi}_1^0$ and the $\tilde{\tau}_1$ LSP.]{
    \begin{picture}(3,2.3)
      \put(0.1,0){\epsfig{file=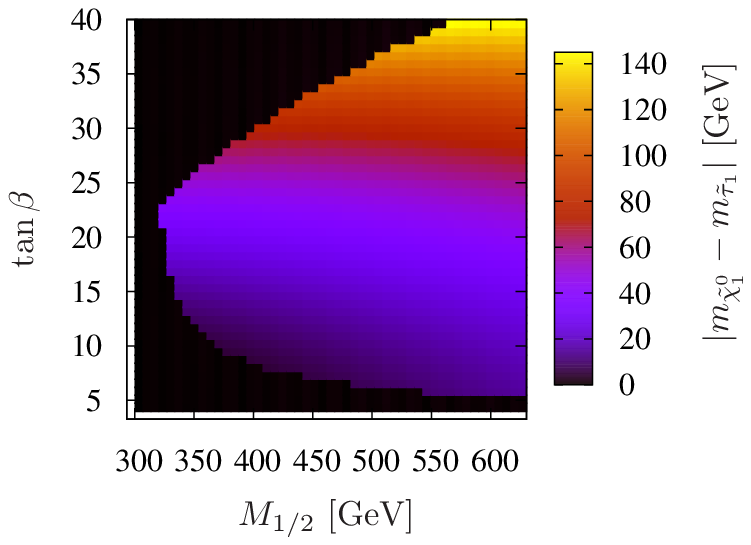,width=3.in}}
    \end{picture}
  }
     
    \caption{\label{Fig:M12_tanb_plane} Masses and mass differences in
    the $M_{1/2}$--$\tan \beta$ plane. The other mSUGRA parameter are
    that of BC1, \ie~$M_0=A_0=0$ GeV and $\text{sgn}(\mu)=+1$. The
    blackened out region is excluded due to tachyons or the LEP Higgs
    or $\tilde{\tau}_1$ mass bounds; see also
    Sec.~\ref{discovery_potential}.  The complete parameter space
    shown in the figures posses a stau LSP.}
\end{figure*} 

We review in Fig.~\ref{Fig:M12_tanb_plane} some sparticle masses and
mass differences in the $M_{1/2}$--$\tan \beta$ plane. The scan is
centered around BC1 ($M_{1/2}=400$~GeV and $\tan \beta=13$) and
corresponds to the parameter space for which we investigated the
discovery potential with early LHC data, \ie\ 
Figs.~\ref{Fig:parameterScan-BC1} and
\ref{Fig:parameterScan-BC1-minlumi}.

\mymed

We show in Figs.~\ref{Fig:gluino_mass}, \ref{Fig:squark_mass}, and
\ref{Fig:neut1_mass} the mass of the gluino, the left-handed down-type
squark, $\tilde{d}_L$, and the $\ninoone$, respectively. We observe
that these masses depend in first approximation only on $M_{1/2}$:
they increase with increasing $M_{1/2}$; see
Refs.~\cite{Drees:1995hj,Ibanez:1984vq} for further details.

\mymed 

However, the stau LSP mass, Fig.~\ref{Fig:stau_mass}, is different.
It also depends strongly on $\tan \beta$. For increasing $\tan \beta$
the stau LSP mass decreases. As already described in
Sec.~\ref{discovery_potential}, this is mainly due to the following
effects. On the one hand, $\tan \beta$ increases the tau Yukawa
coupling and thus its negative contribution to the RGE running of the
stau masses. On the other hand, a larger value of $\tan \beta$ leads
in general to a larger mixing between the left-handed and right-handed
stau. This than results in one lighter stau mass eigenstate;
\cf~\eg~Ref.~\cite{Dreiner:2008rv}.

\mymed

Therefore, we can change the mass difference between the
$\ninoone$ and the stau LSP by changing $\tan \beta$ as visualized
in Fig.~\ref{Fig:mass_diff_stau1_neut1}.

\section{Significance Definitions}
\label{diff_significances}
The expected signal significance for a given set of event selection
cuts depends on the significance definition used and the estimated
systematic uncertainties.  In many phenomenological publications the
simple $S/\sqrt{B}$ ratio is used, whereas more detailed experimental
analyses use other definitions. We compare some definitions in
Tab.~\ref{Tab:CutFlowBC1-detailed}, which make different assumptions
on the underlying statistical properties and give a very short review
of those. For details we refer to
Refs.~\cite{Cousins:2008,Cowan:2008sigcalc}.

\mymed

In all cases the $p$-value of a given number of observed events under
the background-only hypothesis with a certain number of expected
background events is related to the significance $Z$ by
\begin{equation}
    p = \int_Z^\infty \frac{1}{\sqrt{2\pi}} e^{-x^2/2} dx = 1 - \Phi(Z) \ ,
\end{equation}
where $\Phi$ is the cumulative distribution of the standard Gaussian.

\mymed

In all cases we use the so-called Asimov dataset, \ie~we estimate the
expected number of signal ($S$) and background ($B$) events by our
Monte Carlo estimate and the expected number of observed events by
$n_\text{obs} = S+B$.  Assuming Poisson statistics the $p$-value of
observing $n_\text{obs}$ events when $B$ background events are
expected is
\begin{equation}
    p_\text{P} = \sum_{n=n_\text{obs}}^\infty P_\text{P}(n; B) = \sum_{n=n_\text{obs}}^\infty \frac{B^n}{n!}e^{-B} \, ,
\end{equation}
where $P_\text{P}$ is the Poisson distribution. Using Wilks' theorem
the corresponding significance $Z_\text{P} = \Phi^{-1}(1-p_\text{P})$
can be approximated as
\begin{equation}
    Z_\text{W} = \sqrt{2(S+B) \ln(1+ S/B) -2S}
\end{equation}
in the limit of large statistics. In our case this approximation is
very good and $Z_\text{W}$ and $Z_\text{P}$ have nearly identical
values.  $Z_\text{W}$ itself can be approximated with the simple
$S/\sqrt{B}$ in the limit of $S \muchless B$ by expanding the
logarithm. This approximation is clearly not valid for our nearly
background free sample and is the reason for the discrepancy between
$S/\sqrt{B}$ and $Z_\text{P}$ in Tab.~\ref{Tab:CutFlowBC1-detailed}.

\mymed

The significance is reduced by uncertainties on the background estimate.
As the background is estimated from Monte Carlo itself, limited statistics of
the background Monte Carlo translates into uncertainties on the estimate of $B$, which
are not included in the previous definitions of significance. The ratio between
Monte Carlo and data luminosity $\tau = \frac{L^\text{MC}}{L^\text{data}}$ is used
to relate the $m$ Monte Carlo background events passing the cuts to the background
estimate $b=m/\tau$. The probability to observe $n = S+B$ events is therefore
\begin{eqnarray}
    P(n, m; S+B, \tau B) &=& P_\text{P}(n; S+B) \cdot P_\text{P}(m; \tau B)\nonumber \\
    &=& P_\text{P}(n+m; S+B+\tau B) \nonumber \\
   & & \cdot P_\text{Bi}(n| n+m; \rho) \, ,
\end{eqnarray}
with $\rho = \frac{S+B}{S+B+\tau B}$ and $P_\text{Bi}$ denoting the binomial distribution.
One can test the background only hypothesis ($S=0$) with a
standard frequentist binomial parameter test, giving the $p$-value
\begin{equation}
    p_\text{Bi} = \sum_{j=n}^{n+m} P_\text{Bi}\left(j| n+m; 1/(1+\tau)\right) \, .
    \label{eq:significance:binomial-p-value}
\end{equation}
The estimated uncertainty of estimating the true $b\tau$ by $m$ is
$\sqrt{m}$, which gives the correspondence $\sigma_b = \sqrt{m}/\tau$
and with the estimate of $b$ finally $\tau = b/\sigma_b^2$. This
relationship can be used to provide an ad-hoc background uncertainty
$\sigma_b$ to the Binomial $p$-value,
\cf~Eq.~(\ref{eq:significance:binomial-p-value}). We used this method
in our definition of $Z_0$ to set a fixed relative background
uncertainty of $f=$~50\%, \ie~in summary
\begin{eqnarray}
    \tau^\prime &=& 1/( b \cdot f^2 ) \\
    n_\text{off}^\prime &=& b\cdot \tau^\prime \\
    n_\text{on}^\prime &=& s +b \\
    p_0 &=& \sum_{k=n_\text{on}^\prime}^{n_\text{on}^\prime+n_\text{off}^\prime} P_\text{Bi}(k| n_\text{off}^\prime+n_\text{on}^\prime; 1/(1+\tau^\prime)) \\
    Z_0 &=& \Phi^{-1}(1-p_0)
    \label{eq:significance:z0}
\end{eqnarray}
For large number of background events, as in our
sample before cuts, this can lead to $p_0>0.5$, \ie~the significance is
not well-defined anymore.

\mymed

Finally we used a profile likelihood method to incorporate the
uncertainty on the Monte Carlo estimate of the background in the
significance. This has the advantage with respect to $Z_\text{Bi}$,
that different scale factors $\tau_i$ for different sub-samples $i$ of
the background can be treated correctly. A description of the method
can be found in the documentation of the \texttt{SigCalc} code
\cite{Cowan:2008sigcalc}, that was used for the calculation of
$Z_\text{PLH}$.

\begin{table*}
\begin{ruledtabular}
\begin{tabular}{l|*{2}{D{.}{.}{1}@{$\pm$}D{.}{.}{1}|}*{4}{d|}d}
cut &   \multicolumn{2}{c|}{all SM} & \multicolumn{2}{c|}{BC1} &  \multicolumn{1}{c|}{$S/\sqrt{B}$} &  \multicolumn{1}{c|}{$Z_0$} &  \multicolumn{1}{c|}{$Z_\text{PLH}$} &  \multicolumn{1}{c|}{$Z_\text{P}$} &  \multicolumn{1}{c}{$Z_\text{Bi}$} \\
\hline
                              before cuts  &  2~258~230 & 1392 &    282.8 & 2.8 &        0.2 & \multicolumn{1}{c|}{--}  &   0.1 &        0.2 &        0.1\\
      $\pT(1\text{st}\; \mu^\pm)>40 \GeV$ &    319~975 & 510   &      141.6 & 2.0 &        0.3 & \multicolumn{1}{c|}{--} &        0.2 &        0.2 &        0.2\\
      $\pT(1\text{st}\; e^\pm)>32 \GeV$ &        1~837 & 43    &      125.9 & 1.9 &        2.9 &  \multicolumn{1}{c|}{--}       &        2.1 &        2.9 &        2.1\\
      $\pT(2\text{nd}\; e^\pm)>7 \GeV$ &         184.9 & 14.8 &      113.7 & 1.8 &        8.4 &      0.7 &        5.5 &        7.6 &        5.5\\
      $\sum \pT^{\ell}>230 \GeV$ &               15.1 & 4.3 &       85.7 & 1.6 &       22.0 &        4.9 &        8.8 &       14.5 &        8.8\\
      $HT^\prime>200 \GeV$ &                      6.1 & 2.3 &       60.3 & 1.3 &       24.3 &        6.4 &        7.8 &       13.9 &        7.8\\
      $HT^\prime>300 \GeV$ &                      3.4 & 1.7 &       56.6 & 1.3 &       30.7 &        8.1 &        7.9 &       15.2 &        8.1\\
      $HT^\prime>400 \GeV$ &\multicolumn{2}{c|}{$\lesssim1.0$}&       52.6 & 1.2 &            &            &            &            &           \\
\end{tabular}
\end{ruledtabular}
\caption{\label{Tab:CutFlowBC1-detailed} Cut flow and different significance definitions for BC1. We assumed an integrated luminosity of $\int L = 1\ifb$
        at $\sqrt{s}=7\TeV$. The given uncertainties include statistical errors only. The significance $Z_0$ includes a systematic uncertainty of 50\%.
For the first three rows $Z_0$ is not well-defined.}
\end{table*}

\bibliographystyle{h-physrev}

\begin{thebibliography}{99}
\bibitem{Wjets}
We have checked that $W$ production
in association with one jets at parton level 
plays no role after our cuts in Tab.~\ref{Tab:CutFlowBC1} have been applied.
\bibitem{Peter}
The version of {\tt Herwig} used in this analysis was written 
by Peter Richardson and is available on request.
\end{thebibliography}

\end{document}